\documentclass[reqno]{amsart}

\usepackage{amssymb,amsfonts,latexsym,amsthm,amsmath,amscd}

\def\Ad{{\rm Ad}}
\def\Af{\vec A_{\sig}}
\def\Apar{A^\parallel_{\sig}}
\def\alg{{\mathfrak A}}

\def\bch{\bar\chi}
\def\Bf{\vec B_{\sig}}

\def\cref{c_2}

\def\cetaex{10 \, C_\Theta^2}
\def\cetaextw{20 \, C_\Theta^2}

\def\cdstrong{C_0}

\def\cuts{{\kappa_\sig}}
\def\cutsrho{{\kappa_{\rho^{-1}\sig}}}

\def\der{\partial}

\def\Dom{{\rm Dom}}

\def\e{\varepsilon}
\def\Egrd{{E(\vp,\sig)}}
\def\Egrdzero{{E(\vp,\sigzero)}}

\def\Erho{E_\rho}
\def\err{{\rm err}}

\def\Ez{{E}_\rho}

\def\Fpairs{{\mathfrak F}{\mathfrak P}}
\def\Fo{{\mathfrak F}}

\def\g{{\sqrt\alpha}}
\def\gs{\alpha}

\def\h{{\uw}}
\def\hfp{{\alpha}}
\def\H{{\mathcal H}}
\def\Hp{{\H_{\vp}}}
\def\Hps{H(\vp,\sig)}
\def\Hpszero{H(\vp,\sigzero)}

\def\ircut{\sigma}
\def\Iconst{100}
\def\Iconsttw{200}
\def\I{I_{\frac{1}{\Iconst}}}
\def\Ie{I}
\def\Ihfp{D_{\eta}}

\def\lTnl{{\lambda}}

\def\mat{{\rm Mat}}

\def\mirr{{M_{\vp}}}

\def\Omgrd{{\Psi_\vu(\vp,\sig)}}
\def\Omgrdzero{{\Psi_\vu(\vp,\sigzero)}}
\def\Omgrdzeropsi{{\Psi_\vu(\vp,\sigzero)}}

\def\op{{H_f,\Pf}}
\def\opp{\underline{\mathcal P}}

\def\pauli{\tau}
\def\Pf{\vec P_f}
\def\piop{\Upsilon}
\def\pol{{\vec\e}}
\def\pfp{{\beta}}
\def\Polpar{{(\e,\delta,\eta,\lTnl,\sig)}}
\def\Polparhat{{(\widehat\e,\widehat\delta,\widehat\eta,\widehat\lTnl,\widehat\sig)}}
\def\Polyd{{\mathfrak U}}
\def\Polyds{\Polyd^{(sym)}}
\def\Ppar{P_f^\parallel}
\def\Pperp{\Pf^\perp}
\def\puppbd{{\frac{1}{3}}}

\def\Ran{{\rm Ran}}

\def\remT{\delta T}
\def\ren{{\mathcal R}_\rho}
\def\renop{R_\rho^H}
\def\resc{{\mathcal S}_\rho}
\def\rotref{ {\bf Sym}$[\vp]$ }
\def\rotrefm{ {\rm\bf Sym}[\vp] }

\def\sbsrm{{\rm{\bf SR}}[\sig]}
\def\sbsr{ {\bf SR}$[\sig]$ }

\def\sha{\flat}

\def\sig{{\sigma}}
\def\sigzero{{\sig_0}}
\def\sind{{\rm{\bf sInd}}}
\def\spvar{{\underline{X}}}

\def\tuk{{\underline{\widetilde k}}}

\def\unull{{\underline{0}}}

\def\vnull{{\vec 0}}
\def\vac{\Omega_f}
\def\vacpsi{\Omega_{\vu}}
\def\rvar{x}

\def\xibd{1}

\def\z{r}
\def\zet{\widehat{\z}}

\def\bra{\big\langle}
\def\ket{\big\rangle}
\def\Bra{\Big\langle}
\def\Ket{\Big\rangle}

\def\C{{\mathbb C}}
\def\N{{\mathbb N}}

\def\R{{\mathbb R}}

\def\ua{{\underline{a}}}

\def\uk{{\underline{k}}}
\def\uv{{\underline{v}}}
\def\uw{{\underline{w}}}

\def\va{{\vec a}}

\def\vk{{\vec k}}

\def\vn{{\vec n}}
\def\vnull{{\vec 0}}
\def\vp{{\vec p}}

\def\vu{{\vec u}}
\def\vv{{\vec v}}
\def\vw{{\vec w}}
\def\vx{{\vec x}}

\def\vX{{\vec X}}

\def\cB{{\mathcal B}}
\def\cL{{\mathcal L}}

\def\Hspace{\underline{\mathfrak W}}

\def\Wspace{{\mathfrak W}}

\def\Tspace{{\mathfrak T}}

\def\1{{\bf 1}}

\def\eqnn{\begin{eqnarray*}}
\def\eeqnn{\end{eqnarray*}}
\def\eqn{\begin{eqnarray}}
\def\eeqn{\end{eqnarray}}

\def\bal{\begin{align}}
\def\eal{\end{align}}

\def\qedprf{\hspace*{\fill}\mbox{$\Box$}}

\theoremstyle{plain}
\newtheorem{theorem}{Theorem}[section]
\newtheorem{definition}[theorem]{Definition}
\newtheorem{proposition}[theorem]{Proposition}

\newtheorem{lemma}[theorem]{Lemma}

\newtheorem{remark}[theorem]{Remark}

\numberwithin{equation}{section}

\def\prf{\begin{proof}}
\def\endprf{\end{proof}}

\begin{document}

\bibliographystyle{plain}

\title[Infrared renormalization in QED]
{Infrared renormalization in non-relativistic QED and scaling criticality}

\author[T. Chen]{Thomas Chen}
\address{Department of Mathematics,
Princeton University, Fine Hall, Washington Road, Princeton,
NJ 08544, U.S.A.}
\email{tc@math.princeton.edu}

\subjclass[2000]{81Q10,81T16,81T08,81T17,81V10}

\begin{abstract}
We consider a spin-$\frac12$ electron in
a translation-invariant model of non-relativistic Quantum Electrodynamics (QED).
Let $H(\vp,\sig)$ denote the fiber Hamiltonian corresponding to the conserved total
momentum $\vp\in\R^3$ of the Pauli electron and the photon field,
regularized by a fixed ultraviolet cutoff in the interaction term,
and an infrared regularization parametrized by $0<\sig\ll1$ which we ultimately remove
by taking $\sig\searrow0$.
For $|\vp|<\puppbd$, all $\sig>0$, and all values of the finestructure constant  $\gs<\gs_0$,
with $\gs_0\ll1$ sufficiently small and {\em independent} of $\sig$,
we prove the existence of a ground state eigenvalue of multiplicity two at
the bottom of the essential spectrum. Moreover, we prove that
the renormalized electron mass satisfies $1<m_{ren}(\vp,\sig)<1+c\alpha$,
{\em uniformly} in $\sig\geq0$, in units where the bare mass has the value 1,
and we prove the existence of the renormalized mass in the limit $\sig\searrow0$.
Our analysis uses the isospectral renormalization group method
of Bach-Fr\"ohlich-Sigal introduced in \cite{bfs1,bfs2} and further developed in \cite{bcfs1,bcfs2}.
The limit $\sig\searrow0$ determines a scaling-critical renormalization group
problem of endpoint type, in which the interaction is strictly marginal
(of scale-independent size).
The main achievement of this paper is the development of a method that provides rigorous
control of the renormalization of a {\em strictly marginal} quantum field theory 
characterized by a {\em non-trivial scaling limit}.
The key ingredients entering this analysis include
a hierarchy of exact algebraic cancelation identities exploiting the
spatial and gauge symmetries of the model, and a combination of
the isospectral renormalization group method with the strong induction principle.
\end{abstract}

\maketitle

\tableofcontents

\parskip = 8 pt

\section{Introduction}

In this paper, we give a solution to the problem of infrared mass renormalization in 
non-relativistic Quantum Electrodynamics (QED),
the mathematical theory of non-relativistic quantum mechanical matter 
(electrons, positrons) interacting
with the quantized electromagnetic radiation field (light, photons).

We consider a Pauli electron of spin $\frac12$ in a translation-invariant model
of non-relativistic QED in $\R^3$. To make it mathematically well-defined,
we regularize the Hamiltonian with
a fixed  ultraviolet (high frequancy) cutoff in the interaction term which eliminates the
interaction of the electron with photons of high energy,
and an infrared (low frequency) regularization parametrized by $0<\sig\ll1$ which
we ultimately remove by letting $\sig\searrow0$.
Our aim is to characterize the particle spectrum of the Hamiltonian $H(\sig)$ of the regularized model,
to prove bounds on the  infrared renormalized mass  that are {\em uniform} in $\sig\geq0$,
and to establish its existence in the limit $\sig\searrow0$.


Since the model is translation invariant, we can study the fiber Hamiltonians $H(\vp,\sig)$
separately for different values of the conserved momentum $\vp\in\R^3$.
Our key aim is to control the regularity of the infimum,
$E(\vp,\sig)$, of the spectrum of $\Hps$  as a function of $|\vp|$,
in the limit   $\sig\searrow0$ ($\Egrd$ is a radial function of $\vp$).
For $|\vp|<\puppbd$ and $\sig>0$, we prove that $E(\vp,\sig)$ is an
eigenvalue of multiplicity two at the bottom of the essential spectrum of $\Hps$
(see also \cite{hisp1} for the degeneracy of the ground state energy).
All our results hold for sufficiently small values of the finestructure constant $\gs<\gs_0$,
where $\gs_0$ is {\em independent} of $\sig\geq0$.

We derive {\em uniform} upper and lower bounds on  the {\em renormalized electron mass}
\eqn
        m_{ren}(\vp,\sig) \;=\; \frac{1}{\partial_{|\vp|}^2E(\vp,\sig)}
        \label{eq:mren-HessE-def-1}
\eeqn
of the form (the "bare mass" (for $\gs=0$) has the value $1$ in our units)
\eqn
        1 \; < \; m_{ren}(\vp,\sig) \; < \; 1 \, + \, c_0 \, \gs \;,
        \label{eq:mren-bds-intro-1}
\eeqn
for $\gs<\gs_0$, where the constant $c_0$ is {\em independent} of $\sig\geq0$.
Moreover, we prove the existence of the renormalized mass
\eqn
		m_{ren}(\vp) \; := \; \lim_{\sigma\searrow0}m_{ren}(\vp,\sigma)
\eeqn
at fixed $\vp$ with $0\leq|\vp|<\puppbd$, and of the joint limit
\eqn
		m_{ren}(\vnull) \; = \; \lim_{(\vp,\sigma)\rightarrow(\vnull,0)}m_{ren}(\vp,\sigma) \;.
		\label{eq:mren-null-1}
\eeqn
The estimate (\ref{eq:mren-bds-intro-1}) plays a quintessential r\^ole in the construction of infraparticle
scattering theory and various related problems, \cite{chfr,pi2}. The Fourier multiplication
operator $e^{-it\Egrd}$ controls the dispersive behavior of the free time 
evolution of infraparticle states, and for stationary phase estimates, control of the
Hessian of $\Egrd$, as obtained from (\ref{eq:mren-HessE-def-1}) and (\ref{eq:mren-bds-intro-1}), is crucial.
Due to the absence of a gap separating $E(\vp,\sig)$ from the essential spectrum,
conventional perturbation theoretic approaches unavoidably produce
{\em divergent} results in the limit $\sig\searrow0$.
This is a manifestation of the {\em infrared problem} in non-relativistic QED.
For a discussion of the infrared problem in the operator-algebraic context,
we refer to \cite{chfr} and the references therein
(see also the remarks in Section {\ref{sec:mainthm-1}}).

In the joint work \cite{bcfs2} with V. Bach, J. Fr\"ohlich, and I.M. Sigal, 
analogous results are proven for the spin 0 model,
including bounds of the form (\ref{eq:mren-bds-intro-1}), for $0<|\vp|<\frac13$
and $\sig>0$, but for $\gs<\gs_0(\sig)$ where $\gs_0(\sig)\searrow0$ as $\sig\searrow0$.
For $|\vp|=0$ and under the hypothesis that the limits
\eqn
        \label{eq:derp-two-E-1}
        \lim_{\sig\searrow0}\lim_{\vp\rightarrow\vnull}m_{ren}(\vp,\sig)
        \; = \; \lim_{\vp\rightarrow\vnull}\lim_{\sig\searrow0}m_{ren}(\vp,\sig)
\eeqn
commute, bounds of the form (\ref{eq:mren-bds-intro-1})
on $m_{ren}(\vnull,\sig)$ are proven in \cite{bcfs2} for $\gs<\gs_0$
(independent of $\sig$) which are {\em uniform} in $\sig\geq0$.
In particular, an explicit, finite, convergent algorithm is constructed in
\cite{bcfs2} that determines $m_{ren}(\vnull,0)$ to any given precision, with rigorous
error bounds.
It is immediately clear that (\ref{eq:mren-null-1})
supplies \cite{bcfs2} with the condition (\ref{eq:derp-two-E-1}).
The present work is in many aspects a continuation of the analysis of \cite{bcfs2}, and
some familiarity with \cite{bcfs2} might be helpful for its reading.

The analysis in \cite{bcfs2} is based on the isospectral renormalization group method,
and shows for $0<|\vp|<\puppbd$ that, in the subcritical case $\sig>0$
and for the type of regularization used in \cite{bcfs2},
the interaction is driven to zero by scaling, at an {\em exponential}, $\sig$-dependent rate
under repeated applications of the renormalization map; the renormalization group problem
is of  is  {\em irrelevant} type.
In contrast, the case $\sigma=0$ is a problem of {\em endpoint type} in which
the interaction and the free Hamiltonian in $H(\vp,0)$
exhibit the same behavior under scaling.
In the context of renormalization group theory,
this defines a {\em  marginal} problem, and a priori, the following three scenarios are possible:
(1) The size of the interaction grows polynomially in the number of applications of the
renormalization map; the problem is {\em marginally relevant}.
(2) The size of the interaction decreases polynomially in the number of applications of the
renormalization map; the problem is {\em marginally irrelevant}.
(3) The size of the interaction neither diminishes nor increases under
repeated applications of the renormalization map; the problem is {\em strictly marginal}.

As we prove in the present work, the problem of infrared renormalization in non-relativistic QED
in the endpoint case $\sig=0$ is of {\em strictly marginal} type,
i.e. the size of the interaction is  {\em scale-independent}.
A main goal of the present work is to extend
the isospectral renormalization group method
of \cite{bcfs1,bcfs2},  based on the {\em smooth Feshbach map}, to models
in quantum field theory which are strictly marginal.
To prove uniform boundedness of the interaction, we invoke the strong induction principle,
and combine it with composition identities satisfied by the smooth Feshbach map.
Moreover, our method involves the use of hierarchies of
non-perturbative identities originating from
spatial and gauge symmetries of the model, which are used to control the precise
cancellations of terms in certain infinite sums.

The isospectral renormalization group produces a convergent series expansion of
$E(\vp,0)$ and $m_{ren}(\vp)$ in powers of $\gs$ in which the coefficients are $\gs$-dependent,
and {\em divergent} as $\gs\searrow0$
(see also \cite{pi}).
However, we emphasize that $E(\vp,0)$ and $m_{ren}(\vp)$ do {\em not} exist as ordinary power series
in $\gs$ (with $\gs$-independent coefficients), and are thus inaccessible to
more conventional methods of perturbation theory.


The contents of this paper are further developments of the work conducted in \cite{ch1}
which is available online, but unpublished.
All results and methods of \cite{ch1} are here fundamentally improved, optimized and extended. 
Some of the main differences comprise:
{\em(1)} The term of order $O(\gs)$ in the uniform upper bound
on the renormalized mass (\ref{eq:mren-bds-intro-1}) is optimal
in powers of the finestructure constant $\gs$.
In \cite{ch1}, the corresponding bound is  
of the form $O(\gs^{\delta})$, for some $\delta>0$.
{\em (2)} In contrast to \cite{ch1}, we include the electron spin here.
It is therefore necessary in our analysis to prove that the Zeeman
term (which involves the magnetic field operator) in  $\Hps$ is,
in renormalization group terminology, an {\em irrelevant} operator (it scales to zero).
The inclusion of electron spin has the consequence that the generalized 
Wick kernels in the present work are matrix-valued (in \cite{ch1}, they are scalar). 
For their analysis, the
spatial symmetries of the model enter in a more significant way in our proofs
than in \cite{ch1}.
{\em (3)} The existence of the renormalized mass in the limit
$\sigma\searrow0$ is proven here, but not in \cite{ch1}. 
{\em (4)} Most proofs are new or significantly improved.

A detailed introduction to the problem of infrared mass renormalization
in the context of the
isospectral renormalization group method is given in \cite{bcfs2}.
The uniform bounds on the infrared renormalized mass have important
applications, for example in infraparticle scattering theory \cite{chfr,pi2},
in certain approaches to the problem of enhanced binding \cite{cvv}
(see also \cite{hvv,hisp2} for enhanced binding), and a s noted above,
in algorithmic schemes for the computation of the renormalized mass \cite{bcfs2}.
Moreover, our results are used in \cite{bcv}.

In the present work, the ultraviolet cutoff $\Lambda$ is fixed.
Some important results related to the asymptotics of ground state
energies, binding, and thermodynamic limits are established in \cite{grlilo,lilo,lilo1,lilo2}
for arbitrary values of $\alpha$ and $\Lambda$, and without infrared cutoff.
For some recent works discussing the problem of ultraviolet mass renormalization,
which is not being addressed here, we refer to \cite{hasei,hiit,hisp}.
For a survey of recent developments in the mathematical study of non-relativistic
QED, we refer to \cite{sp1}.

\subsubsection*{Notations}

We use units in which the velocity of light $c$, Planck's constant
$\hbar$, and the bare electron mass $m$ have the values
$c=\hbar=m=1$.
\\
The letters $C$ or $c$ will denote various constants whose values
may change from one estimate to another.
\\
$\cB(\H_1,\H_2)$ denotes the bounded linear operators
$\H_1\rightarrow\H_2$ on Banach spaces $\H_i$.
\\
$I_c(a) \subset\R$ is the closed interval $[a-c,a+c]$, and
$I_c\equiv I_c(0)$.
\\
$\vv=(v_1,v_2,v_3)$ denotes a vector in $\R^3$.
\\
$\vv\cdot \vv'$ denotes the Euclidean scalar product, and
$\vv^2\equiv \vv\cdot \vv\equiv|\vv|^2$.
\\
$B_r(\vx)\subset\R^3$ is the closed ball of radius $r$ centered at
$\vx\in\R^3$, and $B_r\equiv B_r(\vnull)$.
\\
$\uv=(v_0,\vv)$ denotes a vector in $\R^4$.
\\
$\vec\pauli=(\pauli_1,\pauli_2,\pauli_3)$ denotes the vector of
Pauli matrices (cf. (\ref{eq:pauli-def-1})).

\section{Definition of the Model}

We consider a translation invariant model of non-relativistic QED
in $\R^3$ that describes a freely propagating, non-relativistic,
spin $\frac{1}{2}$ Pauli electron interacting with the quantized
electromagnetic radiation field.

The electron Hilbert space is given by
\eqn
        \H_{el} = L^2(\R^3) \otimes\C^2
\eeqn
where the factor $\C^2$ accounts for the electron spin.

The Hilbert space accounting for degrees of freedom of
the quantized electromagnetic field is given by the photon Fock space
\eqnn
        \Fo(L^2(\R^3)) &=& \bigoplus_{n\geq 0} \Fo_n(L^2(\R^3))  \;,\\
        \Fo_n(L^2(\R^3))&=&{\rm Sym}_n\Big(L^2(\R^3)\otimes\C^2
        \Big)^{\otimes n} \;,
\eeqnn
where the factors $\C^2$ accommodate the polarization of the photon in the
Coulomb gauge, and ${\rm Sym}_n$ fully symmetrizes the $n$ factors in the tensor product.
A vector $\Phi\in\Fo$ is a sequence
$$
        \Phi \; = \; \big(\Phi^{(0)},\Phi^{(1)},\dots,\Phi^{(n)},\dots\big)
        \; \; , \; \;
        \Phi^{(n)}\in\Fo_n  \;,
$$
where $\Phi^{(n)}=\Phi^{(n)}(k_1,\lambda_1,\dots,k_n,\lambda_n)$ is symmetric
in all $n$ variables $(k_j,\lambda_j)$.
$k_j\in\R^3$ is the momentum, and $\lambda_j\in\{+,-\}$
labels the two possible polarizations of the $j$-th photon.
For brevity, let $\Fo\equiv\Fo(L^2(X))$, $\Fo_n\equiv\Fo_n(L^2(X))$ if $X=\R^3$.
The scalar product on $\Fo$ is given by
\eqnn
        \bra \, \Phi_1 \, , \, \Phi_2 \, \ket \; = \; \sum_{n\geq0} \,
        \bra \, \Phi_1^{(n)} \, , \, \Phi_2^{(n)} \, \ket_{\Fo_n} \;.
\eeqnn
For $\lambda\in\{+,-\}$ and  $f\in L^2(\R^3)$,
we introduce annihilation operators
\eqn
        a(f,\lambda) \, : \, \Fo_n \, \rightarrow \, \Fo_{n-1} \;,
\eeqn
with
\eqn
        &&( a(f,\lambda)\Phi )^{(n-1)}( \vk_1,\lambda_1,\dots,\vk_{n-1},\lambda_{n-1} )
        \nonumber\\
        &&\hspace{1cm}= \; \sqrt n \;
        \int d^3\vk_n \, f^*(\vk_n) \,
        \Phi^{(n)}( \vk_1,\lambda_1,\dots,\vk_{n}, \lambda )
        \label{eq:ann-op-def-1}
\eeqn
and creation operators
\eqn
        a^*(f,\lambda):\Fo_n\rightarrow\Fo_{n+1}
        \;\;{\rm with}\;\;
        a^*(f,\lambda) \; = \; (a(f,\lambda))^*
\eeqn
which satisfy the canonical commutation relations
\eqn
      \big[ \, a(f,\lambda) \, , \, a^*(g, \lambda') \, \big] &=&
      \big( f \, , \, g \,\big)_{L^2} \; \delta_{\lambda, \lambda'}
        \;
      \nonumber\\
      &&
      \nonumber\\
      \big[ \, a^\sharp (f,\lambda) \, , \, a^\sharp (g, \lambda') \, \big] &=& 0
      \;\;\; , \;\;\; f,g\in L^2(\R^3) \;,
\eeqn
where $a^\sharp$ denotes either $a$ or $a^*$.
The {\em Fock vacuum}
\eqn
        \vac \; = \; ( 1 \, , \, 0 \, , \, 0 \, , \, \dots) \; \in \,\Fo
\eeqn
is the unique unit vector satisfying
\eqn
        a(f,\lambda) \, \vac = 0
\eeqn
for all $f\in L^2(\R^3)$.

Since $a(f,\lambda)$ is antilinear and $a^*(f,\lambda)$ is linear in $f$,
one can define operator-valued distributions $a^\sharp(\vk,\lambda)$ with
\begin{align}
        a(f,\lambda) \; = \; \int_{\R^3}d^3\vk \,  f^*(\vk) \, a(\vk,\lambda)
        \; \; \; , \; \; \;
        a^*(f,\lambda) \; = \; \int_{\R^3}d^3\vk \,  f(\vk) \, a^*(\vk,\lambda) \;,
\end{align}
satisfying
\eqn
        \big[ \, a(\vk',\lambda') \, , \, a^*(\vk,\lambda) \, \big] &=& \delta_{\lambda, \lambda'}
        \, \delta (\vk-\vk')
        \nonumber\\
        &&
        \nonumber\\
        \big[ \, a^\sharp (\vk',\lambda') \, , \, a^\sharp (\vk,\lambda) \, \big] &=& 0
\eeqn
for all $\vk, \vk'\in\R^3$ and $\lambda,\lambda'\in\{+,-\}$, and
\eqn
        a(\vk,\lambda)\,\vac \; = \; 0
\eeqn
for all $\vk$, $\lambda$.

We introduce the notation
\eqn
        K \; := \; (\vk,\lambda) \; \in \; \R^3\times\{+,-\}
        \; \; , \; \;
        \int dK \; := \; \sum_{\lambda=\pm} \, \int_{\R^3} d^3\vk \; ,
        \label{intdKdef}
\eeqn
for pairs of photon momenta and polarization labels.

The Hamiltonian and the momentum operator of
the free photon field are respectively given by
\eqn
        H_f &=& \int  dK \, |\vk| \, a^*(K) \, a (K)
        \nonumber\\
        &&
        \nonumber\\
        \Pf &=& \int  dK \, \vk \, a^*(K) \,  a (K) \;,
\eeqn
and are selfadjoint operators on $\Fo$.

The Hilbert space of states for the full system is given by the
tensor product Hilbert space
\eqn
        \H \; = \; \H_{el} \otimes \Fo \;.
\eeqn
The Hamiltonian of non-relativistic QED for the coupled system comprising
the electron and the quantized radiation field is given by
\begin{align}
        \label{uvham}
        H(\sig)  \; = \; \frac{1}{2} \,
        \big( i\nabla_{\vx}\otimes\1_{f} \, - \,
        \g \, \Af (\vx) \big)^{2} \, + \, \g\,\vec\pauli\cdot\Bf(\vx) \, + \,
        \1_{el}\otimes H_f   \;,
\end{align}
where $\vec\pauli=(\pauli^1,\pauli^2,\pauli^3)$
denotes the vector of Pauli matrices
\begin{align}
        \pauli_1
        \; = \; \left(
        \begin{array}{cc}0&1\\1&0\end{array}\right) \; \; , \; \;
        \pauli_2
        \; = \; \left(
        \begin{array}{cc}0&i\\-i&0\end{array}\right) \; \; , \; \;
        \pauli_3
        \; = \; \left(
        \begin{array}{cc}1&0\\0&-1\end{array}\right) \; \; ,
        \label{eq:pauli-def-1}
\end{align}
and  $\g$ is the bare electron charge, with $\alpha>0$ being the finestructure constant.
The operators
\eqn
         \Af (\vx)&=&
         \int
         \frac{dK}{\sqrt{|\vk|}}  \, \cuts (|\vk|) \,
         \big( \pol(K)e^{2\pi i\vk\cdot \vx}\otimes
         a(K) \, + \, h.c. \big)
        \label{Avdef}
        \nonumber\\
        \Bf(\vx)&=&
         \int
         \frac{dK}{\sqrt{|\vk|}}  \, \cuts (|\vk|) \,
         \big( i\vk\wedge \pol(K)e^{2\pi i\vk\cdot \vx}\otimes
         a(K) \, + \, h.c. \big)
\eeqn
stand for the quantized electromagnetic vector potential
and the magnetic field operator.
In agreement with the Coulomb gauge condition, the polarization vectors
$\pol(\vn_k,+)$, $\pol(\vn_k,-)\in \R^3$  (with $|\pol(K)|=1$)
form an orthonormal basis  together with $\vn_\vk:=\frac{\vk}{|\vk|}$ in $\R^3$,
for every $\vk\in\R^3\setminus\{0\}$.

The function $\cuts$
implements an ultraviolet cutoff
(comparable to the electron rest energy $mc^2=1$ in our units) and an
infrared regularization parametrized by $0<\sig\ll1$.
For technical reasons specific to our methods, we require that
$\cuts$ is non-zero for $0<x<\Lambda$, where we can assume $\Lambda=1$ for the ultraviolet cutoff.
The infrared regularization used in \cite{bcfs2} has the form $\cuts(x)=\chi(x<1)x^\sigma$, 
and softens the singularity of the photon form factor to $|\vk|^{-1/2+\sigma}$.

To study properties of the strictly marginal model in the scaling critical case $\sig\searrow0$,
it is more convenient to use an infrared regularization where $\cuts(x)=1$ for $x>\sig$.
For definiteness, we choose
\eqn
        \cuts (x):=\left\{
        \begin{array}{ll}
        (x/\sig)^K&{\rm for} \; x<\sig\\
        1&{\rm for}\;x\in[\sig,\frac12],\;{\rm and\;}C^\infty\;{\rm on}\;(\frac12,1)\\
        0&{\rm for}\;x>1\;,
        \end{array}
        \right.
        \label{cutsDef-1}
\eeqn
where $\sig>0$ is arbitrarily small, and which we will send to zero in the end. The exponent
$K>0$ is arbitrary. We will for simplicity assume that $K=1$, but everything discussed here
can be easily adapted to any $K>0$, or to any $\cuts$ which is smooth and monotonic
on $[0,\sig]$, with $\cuts(0)=0$ and $\cuts(\sig)=1$.

The operator of the total momentum of the electron and the quantized electromagnetic field
is given by
\eqn
        \vec P_{tot} \; = \; i\nabla_\vx\otimes\1_f \, + \, \1_{el}\otimes \Pf \;.
        \label{Ptotdef3}
\eeqn
The model is translation invariant,  $[\vec P_{tot},H(\sig)]=0$.
We write
\eqn
        \H &=& \int^{\oplus}_{\R^3} \, d^{3}\vp \; \Hp
\eeqn
in direct integral decomposition, where
\eqn
        \Hp&\cong&\C^2\otimes\Fo
\eeqn
denotes the fiber Hilbert space associated to the conserved
total momentum $\vp\in\R^3$.

The fibers $\Hp$ are invariant under $e^{-it H(\sig)}$.
It thus suffices to study the restriction of $H(\sig)$ to $\Hp$,
\eqn
        H(\vp,\sig) \; := \;  \frac{1}{2} \, \big( \vp \, - \, \Pf \, - \,
        \g \Af \big)^2  \, + \, \g \,\vec\pauli\cdot\Bf \, + \, H_f \;.
        \label{eq:Hps-def-1}
\eeqn
where
\eqnn
        \Af&:=&\Af(\vec0)
        \nonumber\\
        &&
        \nonumber\\
        \Bf&:=& i\big(\Pf\wedge\Af
        +\Af \wedge \Pf\big)    \;.
\eeqnn
$\Hps$ is the {\em fiber Hamiltonian} corresponding to the conserved momentum $\vp$.

\section{Main Theorem}
\label{sec:mainthm-1}

The main results of this paper characterize the infimum, $E(\vp,\sig)$, of the spectrum of $\Hps$,
for values $\gs<\gs_0$ where $\gs_0\ll1$ is effective, small, and {\em independent of } $\sig$.
For $0\leq|\vp|<\puppbd$,  we prove that  $E(\vp,\sig)$ is an eigenvalue of multiplicity two at the bottom of the
essential spectrum of $\Hps$, for any $\sig>0$.
In particular, we prove upper and lower bounds on the {\em renormalized electron mass}
which are {\em uniform} in $\sig\geq0$, and we establish the existence of the renormalized mass
in the limit $\sigma\searrow0$.
Using the results of this paper, it is shown in \cite{chfr} that
when $\sig=0$, $H(\vp,0)$ has no ground state in $\C^2\otimes\Fo$ if $|\vp|>0$; see
Theorem {\ref{thm-chfr-mainthm-1}} below, which quotes the main results of \cite{chfr}.

\begin{theorem}
\label{mainthm-1}
For $0\leq |\vp|<\puppbd$, there exists a constant $\gs_0>0$
(independent of $\sigma$)  such that for all $0<\gs<\gs_0$,
the following hold:
\begin{itemize}
\item
\underline{A.} For any $\sig>0$,
\eqn
        \Egrd \; := \; \inf {\rm spec}_{\C^2\otimes\Fo}\,\Hps
\eeqn
is an eigenvalue of multiplicity two at the bottom of the essential spectrum of $\Hps$.

\item
\underline{B.} There exists a constant $c_0$ independent of $\sig$ and $\gs$ such that
\eqn
        1- c_0 \gs\; < \;\partial_{|\vp|}^{2} \Egrd
        \; < \; 1 \;,
        \label{DerEgrBd-1}
\eeqn
and
\eqn
        \Big| \, \nabla_{\vp}E(\vp,\sig) \, - \, \vp \, \Big|&<&c_0 \gs |\vp|
        \nonumber\\
        \Big| \, E(p,\sig) \, - \, \frac{|\vp|^2}{2} \, - \,
        \frac{\gs}{2}\bra \,\vac\,,\,\Af^2 \, \vac \, \ket \, \Big|
        &<&\frac{c_0 \gs |\vp|^2}{2} \;.
\eeqn

\item
\underline{C.} The renormalized electron mass
\eqn
    m_{ren}(\vp,\sig) \; = \; \frac{1}{ \partial_{|\vp|}^2 E(\vp,\sig) }
\eeqn
is bounded by
\eqn
    1 \; < \; m_{ren}(\vp,\sig)  \; < \; 1 \, + \, c_0 \gs \;,
    \label{eq:mren-bds-mainthm-1}
\eeqn
uniformly in $\sig\geq0$.

\item
\underline{D.} For every $\vp$ with $0\leq|\vp|<\puppbd$, 
there exists a sequence $\{\sigma_n\}_{n\in\N}$ converging to zero
such that
\eqn
	m_{ren}(\vp) \; := \; \lim_{n\rightarrow\infty}m_{ren}(\vp,\sigma_n)  
\eeqn
exists. 
Moreover,  
\eqn
	\widetilde m_{ren}(\vnull) \; := \; \lim_{\sigma\searrow0} m_{ren}(\vnull,\sigma)
\eeqn
exists (for the proof, see \cite{bcfs2}), and 
\eqn
	m_{ren}(\vnull) \; = \;  \lim_{|\vp|\searrow0}m_{ren}(\vp) \; = \; 
	\widetilde m_{ren}(\vnull)  \,, 
	\label{eq:mren-commute}
\eeqn
i.e., the limits $|\vp|\searrow0$ and $\sigma\searrow0$ commute.
\end{itemize}
\end{theorem}

\subsection{Remarks} $\;$

{\em (1)}
For the proof of Theorem {\ref{mainthm-1}}, we can invoke
many constructions and results from \cite{bcfs1}, and especially from \cite{bcfs2}.
As noted in the introductory section, it is established in \cite{bcfs2}
that there exists $\gs_0(\sig)>0$ for any $\sig>0$
such that for all $\gs<\gs_0(\sig)$,
the statements {\em\underline{A.}} - {\em\underline{C.}} of Theorem {\ref{mainthm-1}} hold.
However, the bound derived in \cite{bcfs2} is such that
$\gs_0(\sig)\searrow0$ as $\sig\searrow0$. The key purpose of the present
paper is to prove $\sig$-independent estimates.

For $|\vp|=0$, bounds of the form (\ref{eq:mren-bds-mainthm-1})
are proved in \cite{bcfs2} for $\gs<\gs_0$, with $\gs_0$ independent of $\sig$, under the assumption that
the limits
\eqn
        \lim_{\sig\rightarrow0}\lim_{|\vp|\rightarrow0} \partial_{|\vp|}^2 E(\vp,\sig)
        \; = \;
        \lim_{|\vp|\rightarrow0}\lim_{\sig\rightarrow0}\partial_{|\vp|}^2 E(\vp,\sig)
        \label{eq:limEexch-1}
\eeqn
commute.
In particular,  an explicit, finite,
and convergent algorithm is constructed in \cite{bcfs2} which determines $m_{ren}(\vnull,0)$
to {\em any arbitrary given precision, with rigorous error bounds}.
It is clear that (\ref{eq:mren-commute})
supplies \cite{bcfs2} with the condition  (\ref{eq:limEexch-1}).

{\em (2)}
The uniform bounds (\ref{DerEgrBd-1}) have important applications in the construction
of dressed one-electron states in the operator-algebraic framework,
and in infraparticle scattering theory.
Let
$\alg_\rho:=\cB(\C^2\otimes\Fo_\rho)$ denote the $C^*$-algebra of bounded
operators on the Fock space $\Fo_\rho(L^2(\R^3\setminus B_\rho))$.
Then, we define the $C^*$-algebra
\eqn
    \alg \; := \; \overline{\bigvee_{\rho>0}\alg_\rho}^{\|\,\cdot\,\|_{op}} \;,
\eeqn
where the closure is taken with respect to the operator norm.
Ground state eigenvectors belonging to $\Egrd$ are parametrized by $\vu\in S^2$,
\eqn
    \Omgrd \in\C^2\otimes\Fo \;\;\;{\rm with} \;\;\;
    \bra \, \Omgrd  \, , \, \vec\pauli \, \Omgrd \, \ket \; = \; \vu
\eeqn
and $\|\Omgrd\|_{\C^2\otimes\Fo} = 1$.
For fixed $\vu\in S^2$, $\Omgrd$ defines a normalized, positive state
\eqn
    \omega_{\vp,\sig}(A) \; := \; \bra\Omgrd\,,\,A\Omgrd\ket \;\;\;,\;\;\;A\in\alg \;,
\eeqn
on $\alg$, referred to as a {\em dressed one-electron state} or an
{\em infraparticle state}.
In physical terms, it accounts for an electron in a bound state with an
infinite number of low frequency (soft) photons of small total energy.

The following results are proved in \cite{chfr}:

\begin{theorem} (C-Fr\"ohlich, \cite{chfr})
\label{thm-chfr-mainthm-1}
Assume Theorem {\ref{mainthm-1}}. Then, with $\Omgrd$ and $\omega_{\vp,\sigma}$
as defined above, the following hold independently of $\vu\in S^2$:
\begin{itemize}
\item
Let
\eqn
    N_f \; = \; \int \, dK \, a^*(K) \, a(K)
\eeqn
denote the photon number operator.
Then,
\eqn
    \Big(-c\gs + c'\gs |\nabla_{\vp}E(\vp,\sig)|^2\log\frac1\sig \Big)_+&<&
    \bra \, \Omgrd \, , \, N_f \, \Omgrd \, \ket
    \\
    &<&C \, \gs \, + \, C'\,\gs\, |\nabla_{\vp}E(\vp,\sig)|^2 \, \log\frac1\sig \;,
    \nonumber
\eeqn
for positive constants $c$, $C$, $c'<C'$, and $r_+:=\max\{r,0\}$.
That is, the expected photon number in the ground state  diverges logarithmically
in the limit $\sig\searrow0$ if $|p|>0$ (since $\vp\neq\vnull\Leftrightarrow\nabla_{\vp}E(\vp,\sig)\neq\vnull$).
\item
Every sequence $\{\omega_{\vp,\sig_n}\}$ with $\sig_n\searrow0$ as $n\rightarrow\infty$
possesses a subsequence $\{\omega_{\vp,\sig_{n_j}}\}$ which converges
weak-* to a state $\omega_{\vp}$ on $\alg$ as $j\rightarrow\infty$.
The state $\omega_{\vp}$ restricted to $\alg_\rho$ is normal for any $\rho>0$.
\item
The state $\omega_{\vp,\sigma}$ satisfies
\eqn
    \int dK\,\Big|\,\omega_{\vp,\sigma }( a(K)^*  a(K))
    -|\omega_{\vp,\sigma }(a(K))|^2\,\Big|
    &\leq& c\gs \;,
\eeqn
uniformly in $\sig\geq 0$.
\item
Let $\pi_{\vp}$ denote the representation of $\alg$,
$\H_{\omega_{\vp}}$ the Hilbert space, and $\Omega_{\vp}\in\H_{\omega_p}$
the cyclic vector corresponding to $(\omega_{\vp},\alg)$ by the GNS construction,
(with $\omega_{\vp}(A)=\bra\Omega_{\vp}\,,\,\pi_{\vp}(A)\Omega_{\vp}\ket$, for all $A\in \alg$).
Moreover, let
\eqn
    v_{{\vp},\sigma,\lambda}(k )&:=&-\g \, \e(\vk,\lambda)\cdot\nabla_{\vp} E(\vp,\sig) \,
    \frac{\cuts(|\vk|)}{|\vk|^{\frac12}}\frac{1}{|\vk|-\vk\cdot\nabla_{\vp} E(\vp,\sig)} \;,
    \label{eq:vpkernel-def}
\eeqn
and
\eqn
    v_{\vp,\lambda}(\vk ) \; := \; \lim_{\sigma\searrow0}v_{\vp,\sigma,\lambda} (\vk) \;.
\eeqn
Then,  $\pi_\vp$ is quasi-equivalent to $\pi_{Fock}\circ\alpha_\vp$ (where
$\pi_{Fock}$ is the Fock representation of $\alg$), and
$\alpha_\vp$ is the *-automorphism of $\alg$
determined by
\eqn
    \alpha_\vp(a_\lambda^\sharp(\vk)) \; = \;
    a_\lambda^\sharp(\vk) \, + \, v_{\vp,\lambda}^\sharp(\vk) \;.
    \label{eq:coh-state-1}
\eeqn
%
\item
The following relations between the Fock representation and $\pi_\vp$ hold:

(i) If $\vp=\vec{0}$
\eqn
    | \, \lim_{\sigma\searrow0} \omega_{\vec{0},\sigma}(N_f) \, | \; < \; c \, \gs \;,
    \label{eq:num-lim-2}
\eeqn
and $\pi_{\vec{0}}$ is (quasi-)equivalent to $\pi_{Fock}$.

(ii) If $\vp\neq\vec0$, $\pi_\vp$ is unitarily inequivalent to the Fock representation, and
\eqn
    \lim_{\sigma\searrow0} \omega_{\vp,\sigma}(N_f) \; = \; \infty \;,
    \label{eq:num-lim-1}
\eeqn
but $\omega_\vp$ has a "local Fock property":
\begin{itemize}
\item[(a)]
For every $\rho>0$,
the restriction of $\omega_\vp$ to $\alg_\rho$ determines a GNS representation which
is quasi-equivalent to the Fock representation.
\item[(b)]
For every bounded
region $B$ in physical $\vx$-space, the restriction of $\omega_\vp$ to the local algebra
$\alg(B)$ determines a GNS representation which is quasi-equivalent to the
Fock representation of $\alg(B)$.
\end{itemize}
\end{itemize}
\end{theorem}

A key ingredient in the proof is
the uniform bound (\ref{DerEgrBd-1}) on the renormalized electron mass.

Theorem {\ref{thm-chfr-mainthm-1}} provides a crucial ingredient (the correct coherent
transformation in the construction of the scattering state) for the
construction of infraparticle scattering states in non-relativistic QED,
extending recent results of Pizzo \cite{pi2} for the Nelson model,
see \cite{chfr}.

{\em (3)}
Theorem {\ref{mainthm-1}} can be straightforwardly extended to Nelson's model.
It is defined on the Hilbert space
\eqn
        \H \; = \; L^2(\R^3)\otimes \Fo_{bos}
        \; \; \; , \; \; \;
        \Fo_{bos} \; := \; \bigoplus_{n\geq0} \, \big(L^2(\R^3)\big)^{\otimes_s n} \;,
\eeqn
with $\Fo_{bos}$ a Fock space of scalar bosons. Introducing creation- and annihilation
operators $a^\sharp(\vk)$,
the Nelson Hamiltonian is given by
\eqn
        H_{Nelson}(\sig)\;=&-&\frac{1}{2} \, \Delta_{\vx}\otimes\1_{bos} \, + \, \1\otimes H_{bos}
        \nonumber\\
        &+&g\int_{\R^3}d^3\vk\,v_\sig(\vk)\big(
        e^{-i\langle \vk,\vx\rangle}\otimes a^*(\vk)+
        e^{i\langle \vk,\vx\rangle}\otimes a(\vk)\big)\;,
\eeqn
where $v_\sig(\vk):=\frac{\cuts(|\vk|)}{|\vk|^{\frac12}}$
and where $g$ is a small coupling constant.
$$
    H_{bos} \; = \; \int dK \, |\vk| \, a^*(K) \, a(K) \;\;\;\;,\;\;\;\;
    \vec P_{bos} \; = \; \int dK \, \vk \, a^*(K) \, a(K)
$$
are the Hamiltonian and momentum operator of the free boson field.
Due to translation invariance,
it again suffices to consider the restriction of $H_{Nelson}(\sig)$
to a fiber Hilbert space
$\H_\vp$ corresponding to the conserved total momentum $\vp\in\R^3$,
\eqn
        H_{Nelson}(\vp,\sig) \; = \; \frac{1}{2} \,
        \big( \vp \, - \, \vec P_{bos} \big)^2 \, + \, H_{bos} \, + \, g a^*(v_\sig) \, + \, g a(v_\sig)  \;.
\eeqn
Applying a {\em Bogoliubov transformation},
\eqn
        H_{Nelson}(\vp,\sig) \; \mapsto \; \alpha_{Bog,\sig} (\, H_{Nelson}(\vp,\sig) \, ) \;,
\eeqn
which acts on creation- and annihilation operators by way of
\eqn
	 \alpha_{Bog,\sig} (\, a^\sharp(\vk) \, ) \; = \; a^\sharp(\vk)-\frac{v_\sig(\vk)}{|\vk|} \, ,
\eeqn
the Nelson Hamiltonian at fixed conserved total momentum $\vp$ is transformed into
\begin{align}
        H_{B-N}(\vp,\sig) \; = \; \frac{1}{2} \,
        \big( \vp \, - \, \vec P_{bos} \, - \, g a(\vw_\sig) \, - \,
    g a^*(\vw_\sig) \big)^2 \, + \, H_{bos} \;,
\end{align}
where $\vw_\sig(\vk):=v_\sig(\vk)\frac{\vk}{|\vk|}$ is a
radially directed, vector-valued function in
the boson momentum space.

The Bogoliubov transformation can be implemented by
\eqn
	\alpha_{Bog,\sig}( \, A \, ) \; = \; U_{Bog,\sig} \, A \, U_{Bog,\sig}^* \, ,
\eeqn	 
where $U_{Bog,\sig}$ is unitary if $\sig>0$, but which is not unitary 
in the limit $\sig\searrow0$. 

The Nelson model
admits {\em soft boson sum rules} that are very similar to the soft photon
sum rules introduced in section {\ref{spsrsect}} for the QED model
(although the Nelson model has no gauge symmetry), \cite{bcfs2}. The
only difference is that the photon polarization vector $\pol(K)$ appearing
in (\ref{eq:sbsr-fib-1}),(\ref{eq:sbsr-fib-2-1}),
(\ref{eq:sbsr-fib-2-2}) is replaced by the radial unit vector $\frac{\vk}{|\vk|}$.
The results of this paper
can be straightforwardly extended to the Nelson model.

{\em (4)}
We remark that the upper  bound $|\vp|<\puppbd$ in Theorem  {\ref{mainthm-1}} has a purely
technical origin. While we make no attempt to optimize it here,
we note that it cannot be improved beyond a critical value $p_c<1$.
It is expected that the eigenvalue $\Egrd$ dissolves into the continuous
spectrum as $|\vp|\nearrow p_c$,
while a resonance appears (a phenomenon similar to Cherenkov radiation).
An analysis of this problem is beyond the scope of this work.

{\em (5)}
Due to the absence of creation or annihilation of particles (electrons) 
or antiparticles (positrons) in non-relativistic
QED, there is, in contrast to relativistic QED (see \cite{bjdr}), 
no renormalization of the finestructure constant $\gs$.

{\em (6)}
Our proof uses the isospectral, operator-theoretic renormalization
group method pioneered by V. Bach, J. Fr\"ohlich, and I. M. Sigal in \cite{bfs1,bfs2},
and further developed in \cite{bcfs1,bcfs2}.
We apply and further extend the formulation based on the "smooth Feshbach map" of \cite{bcfs1,bcfs2}.
In the limit $\sig\searrow0$, the interaction in $H(\vp,\sig)$ is
{\em purely marginal}.
The main goal in the present work is the development of a method to
control the size of the interaction in a purely marginal theory.

\section{Wick ordering and symmetries}

In this section, we discuss three properties of $\Hps$
which play a crucial r\^ole in a more general context later:
\begin{itemize}
\item
The fiber Hamiltonian $\Hps$ can be written in {\em generalized Wick ordered normal form},
i.e., as a sum of Wick-ordered (all creation-operators stand on the left
of all annihilation operators)  monomials of creation- and annihilation operators
which are characterized
by {\em operator-valued} integral kernels (referred to as {\em generalized Wick kernels}).
\item
$\Hps$ is symmetric under rotations and reflections with respect to
a plane perpendicular to $\vp$ containing the origin. We observe that the non-interacting
Hamiltonian in $\Hps$ is a scalar a multiple of $\1_2$.
We prove in Lemma {\ref{rotreflinvlemma}} below that any ${\rm Mat}(2\times2,\C)$-valued
generalized Wick monomial $f[\op]$ of degree zero that admits these symmetries is
necessarily a multiple of $\1_2$.
\item
Moreover, $\Hps$ admits {\em soft photon sum rules},
which are a generalization of the differential {\em Ward-Takahashi identities} of QED.
Those are hierarchies of non-perturbative, exact identities which originate
from $U(1)$ gauge invariance.
\end{itemize}

\subsection{Generalized Wick ordered normal form}

The {\em generalized Wick ordered normal form} of
the fiber Hamiltonian $H(\vp,\sig)$ is given by
\eqn
        H(\vp,\sig) \; = \; E[\vp]+T[\op,\vp] \, + \,  W_1+  W_2  \;,
\eeqn
where
\eqn
        E[\vp] \; := \; \frac{\vp^2}{2} \, + \, \frac\alpha2 \,
        \bra\,\vac \, , \, \Af^2 \, \vac \, \ket \;\; \in \R_+ \;.
\eeqn
The {\em free Hamiltonian}
\eqn
        T[\op,\vp] \; := \; H_f \, - \, \vp\cdot\Pf \, + \, \frac{\Pf^2}{2}
\eeqn
commutes with $\op$.
The {\em interaction Hamiltonian} is a sum
\eqn
        W_L \; = \; \sum_{M+N=L}W_{M,N} \; \; , \; \;
        W_{M,N} \; = \;  W_{N,M}^* \; \; , \; \; L=1,2 \;,
\eeqn
where the operators $W_{M,N}$ are the following {\em generalized Wick monomials}:
\eqn
        W_{0,1} \; = \; \int \frac{dK \cuts(|\vk|)}{|\vk|^{1/2}}
        \, w_{0,1}[\op,\vp;K] \,a(K) \; = \; W_{1,0}^* \;.
\eeqn
The integral kernel
\begin{align}
        w_{0,1}[\op,\vp;K] \; := \; - \, \g \, \big( \vp \, - \, \Pf \big)\cdot\pol(K) \, + \,
        \g\,\vec\pauli\cdot(i\vk\wedge \pol(K))
\end{align}
is a ${\rm Mat}(2\times2,\C)$-valued operator-function of $K$,
which commutes with $\op$. We shall refer to it as the
{\em generalized Wick kernel} of order $(0,1)$, and $w_{1,0}=w_{0,1}^*$.
Furthermore, we have the Wick monomials
\eqn
        W_{1,1}&=&\int \frac{dK d K' \cuts(|\vk|)\cuts(| \vk'|)}{(|\vk||\vk'|)^{1/2}}
        \;a^*(K) \, w_{1,1}[\op;K, K'] \, a(K') \;,
        \nonumber\\
        W_{0,2}&=&\int \frac{dK d K' \cuts(|\vk|)\cuts(|\vk'|)}{(|\vk||\vk'|)^{1/2}}
        \;w_{0,2}[\op;K,K'] \, a(K) \, a(K') \;,
    \nonumber\\
        W_{2,0}&=&\int \frac{dK d K' \cuts(|\vk|)\cuts(|\vk'|)}{(|\vk||\vk'|)^{1/2}}
        \;a^*(K) \, a^*(K') \, w_{0,2}[\op;K,K'] \;,
\eeqn
with generalized Wick kernels
\eqn
        w_{1,1}[\op;K,K']&=&2 \gs\, \pol(K)\cdot\pol(K') \;,
        \nonumber\\
        w_{0,2}[\op;K, K']&=&\;\;\gs\, \pol(K)\cdot\pol(K') \;,
        \nonumber\\
        w_{2,0}[\op;K, K']&=&\;\;\gs\, \pol(K)\cdot\pol(K')
\eeqn
of orders $(1,1)$, $(0,2)$, and $(2,0)$, respectively.
In case of $\Hps$, the number of Wick monomials is evidently finite;
we will later study classes of Hamiltonians where the interaction
part is a norm-convergent, infinite series of Wick monomials.

\subsection{Rotation and reflection invariance}
\label{rotreflinvsect}

We let $U_{R}^\Fo: \Fo\rightarrow\Fo$
denote the unitary representation of $SO(3)$ defined by
\begin{align}
        (U_{R}\Phi)_n(\vk_1,\lambda_1,\cdots,\vk_n,\lambda_n)
        \; = \;  \Phi_n(R\vk_1,\lambda_1,\cdots,R\vk_n,\lambda_n)
    \;\;\;\;,\;\;R\in SO(3) \;.
        \label{eq:U-R-def-1}
\end{align}
We denote the representation $SU(2) \rightarrow SO(3)$ by
$R_{\bullet}:h\mapsto R_h$, and
\eqn
        \Ad_{U_{R_h}^\Fo}[A] \; = \; U_{R_h}^\Fo A (U_{R_h}^\Fo)^* \;,
\eeqn
with $A$ defined on $\Fo$.
Then, clearly,
\begin{align}
        \Ad_{U_{R_h}^\Fo}[H_f] \; = \; H_f
        \;\;\; , \;\;\;
        \Ad_{U_{R_h}^\Fo}[\vp] \; = \;  \vp
        \;\;\; , \;\;\; \Ad_{U_{R_h}^\Fo}[\Pf] \; = \;  R_h\Pf \;,
        \nonumber\\
        \Ad_{U_{R_h}^\Fo}[\vp\,\wedge\,\Pf] \; = \; R_h(\vp\,\wedge\,\Pf)
        \;\;\; , \;\;\;
        \Ad_{U_{R_h}^\Fo}[\Af] \; = \; R_h\Af \;.
\end{align}
Moreover, conjugating the vector of Pauli matrices
$\vec\pauli=(\pauli_1,\pauli_2,\pauli_3)$ by $h$ yields
$$
        h \, \vec \pauli \, h^* \; = \; R_h \, \vec\pauli \;,
$$
and
\eqn
        U_h \; := \; h\otimes U_{R_h}^\Fo
\eeqn
defines a unitary representation of $SU(2)$ on $\C^2\otimes\Fo$.

It is easy to see that
\eqn
        U_h \, H(R_h \vp,\pauli) \, U_h^* \; = \;   \Hps \;,
\eeqn
i.e.  $\Hps$ is rotation invariant.

Let $\vn_\vp=\frac{\vp}{|\vp|}$.
We consider the unitary reflection operator on $\Fo$
\begin{align}
        U_{ref,\vp}^\Fo \; := \; \exp\Big[ \, \frac{i\pi}{2}
        \int_{\R^3\times\{+,-\}} dK \, \big( a^*(\vk,\lambda)a(\vk,\lambda) \, - \,
        a^*(\vk,\lambda)a(\mirr \vk,\lambda) \big) \, \Big]\; ,
    \label{eq:def-refl-op-1}
\end{align}
where $\mirr \vk:=-k^\parallel \vn_\vp+\vk^\perp$, with
$k^\parallel:=\vk\cdot \vn_\vp$ and $\vk^\perp:=(\vk-k^\parallel \vn_\vp)$.
Clearly, $\mirr^2=\1$.
We point out the similarity of (\ref{eq:def-refl-op-1}) to the parity inversion
operator in relativistic QED, see for instance \cite{bjdr}.

One can straightforwardly verify that
\eqn
        U_{ref,p}^\Fo \, a^\sharp(\vk,\lambda)
        \, (U_{ref,p}^\Fo)^* \; = \; a^\sharp(\mirr \vk,\lambda) \;,
\eeqn
and correspondingly with $\vk$ and $\mirr\vk$ exchanged.
Hence, $\Hps$ is invariant under reflection with
respect to a plane perpendicular to $\vp$ containing the origin.

Under conjugation by $U_{ref,\vp}^\Fo$,
\eqn
        \Ad_{U_{ref,\vp}^\Fo}[H_f] \; = \; H_f
        \; , &
        \Ad_{U_{ref,\vp}^\Fo}[\vp] \; = \; \vp
        & , \; \;
        \Ad_{U_{ref,\vp}^\Fo}[\Ppar] \; = \; - \, \Ppar
        \nonumber\\
        \Ad_{U_{ref,\vp}^\Fo}[\Af^\parallel] \; = \; - \, \Af^\parallel
        \; , &
        \Ad_{U_{ref,\vp}^\Fo}[\Pperp] \; = \; \Pperp
        & , \; \;
        \Ad_{U_{ref,\vp}^\Fo}[\Af^\perp] \; = \; \Af^\perp \;,
        \nonumber\\
        \Ad_{U_{ref,\vp}^\Fo}[\vp \, \wedge \, \Pf] \; =
        &\Ad_{U_{ref,\vp}^\Fo}[\vp \, \wedge \, \Pperp]&= \; \vp \, \wedge \,
        \Pperp \;\; = \;\; \vp \, \wedge \, \Pf
        \label{UrefId-1}
\eeqn
while under conjugation by $\pauli^\parallel=\vec\pauli\cdot\vn_{\vp}$,
\eqn
        \pauli^\parallel\rightarrow\pauli^\parallel
        \; \; , \; \;
        \vec\pauli^\perp\rightarrow-\vec\pauli^\perp \;,
        \label{UrefId-2}
\eeqn
where
\eqn
        \vec\pauli^\perp \; = \; \vec\pauli \, - \, {\rm diag}(\vn_{\vp})\cdot\vec\pauli  \;.
\eeqn
For
$$
        U_{ref,\vp} \; := \; \pauli^\parallel\otimes U_{refl,\vp}^\Fo \;,
$$
it follows that
$$
        U_{ref,\vp} \, H(-\vp,\pauli) \, U_{ref,\vp}^* \; = \; \Hps \;,
$$
i.e. $\Hps$ is reflection invariant.

An important ingredient in our analysis is the fact that
any reflection and rotation invariant
${\rm Mat}(2\times2,\C)$-valued function
of $H_f$, $\Pf$ and $\vp$ is a scalar operator (i.e. a multiple of $\1_2$).

\begin{lemma}\label{rotreflinvlemma}
Let $A$ denote a ${\rm Mat}(2\times2,\C)$-valued
Borel function of $H_f,\Pf,\vp$, satisfying
\eqn
        U_h \, A(H_f,\Pf,R_h \vp) \, U_h^* &=&  A(H_f,\Pf,\vp)
        \label{eq:RotInv-1}
        \\
        U_{ref,p} \, A(H_f,\Pf,-\vp) \, (U_{ref,\vp})^* &=& A(H_f,\Pf,\vp)
        \label{eq:RefInv-1}
\eeqn
for all $h\in SU(2)$. Then,
\eqn
        A(H_f,\Pf,\vp) \; = \; a_0(H_f,\Pf,\vp) \, {\bf 1}_2 \;,
\eeqn
for a Borel function $a_0:\R_+\times\R^3\times\R^3\rightarrow\C$, where
\eqn
        U_h  a_0(H_f,\Pf,R_h \vp) U_h^* &=&  a_0(H_f,\Pf,\vp)
        \label{eq:RotInv-2}
        \\
        U_{ref,p} \, a_0(H_f,\Pf,-\vp) \, (U_{ref,\vp})^* &=& a_0(H_f,\Pf,\vp) \;.
        \label{eq:RefInv-2}
\eeqn
Hence, $A(H_f,\Pf,\vp)$ transforms according to the trivial representation of $SU(2)$.
\end{lemma}

\prf
Representing $A\in {\rm Mat}(2\times2,\C)$ in the basis of Pauli matrices
$\{\pauli_0,\pauli_1,\pauli_2,\pauli_3\}$,
\eqn
        A \; = \; \left({a_{11}\atop a_{21}}\; {a_{12}\atop a_{22}}\right)
        &=&\frac{a_{11}+a_{22}}{2}\,\pauli_0
        \, + \,
        \frac{a_{11}-a_{22}}{2}\,\pauli_3
        \nonumber\\
        &+&\frac{a_{12}+a_{21}}{2}\,\pauli_1
        \, + \, \frac{a_{12}-a_{21}}{2i}\,\pauli_2\;,
\eeqn
we write
\eqnn
    A \; = \; a_0 \, \pauli_0
        \, + \, \va \cdot \vec \pauli \;,
\eeqnn
with $\vec a=(a_1,a_2,a_3)$, and $\vec\pauli=(\pauli_1,\pauli_2,\pauli_3)$.
We will refer to $a_0$ as the {\em scalar}, and $\vec a$ as the
{\em vector} part of $A$ (which are in general $\C$-valued).

Since
\eqn
        U_h \, A(H_f,\Pf,R_h \vp) \, U_h^*&=&
        a_0(H_f,R_h\Pf,R_h \vp) \, \pauli_0
        \nonumber\\
        &&+\;\va(H_f,R_h\Pf,R_h \vp)\cdot(R_h \vec\pauli)\;,
\eeqn
assumption (\ref{eq:RotInv-1}) implies that
\eqn
        a_0(H_f,R_h\Pf,R_h \vp)
        &=&a_0(H_f,\Pf, \vp)\nonumber\\
        \vec a(H_f,R_h\Pf,R_h \vp)&=&R_h \,\vec a(H_f,\Pf,\vp)\;.
\eeqn
We write $\vec a$ in the basis $\vp$, $\Pf$, $\vp\,\wedge\,\Pf$,
\begin{align}
        \vec a(\op,\vp)=b_1(\op,\vp)\vec p+b_2(\op,\vp)\Pf
        +b_3(\op,\vp)\vp\,\wedge\,\Pf\;,
\end{align}
where
\eqn
        b_j(H_f,R_h\Pf,R_h \vp)
        \; = \; b_j(H_f,\Pf, \vp) \; \; \; , \; j=1,2,3\;,
\eeqn
are scalar functions of $\op,\vp$.
By rotation invariance, $a_0,b_1,b_2,b_3$ are functions of the rotation
invariant combinations $\vp^2$, $\vp\cdot\Pf$, $\Pf^2$ only. Hence,
$$
        U_{ref,\vp}^\Fo \, a_0(H_f,\Pf, -\vp) \, (U_{ref,\vp}^\Fo)^*
        \; = \; a_0(H_f,\Pf, \vp) \;,
$$
and likewise for $b_j$.
However, $\vec\pauli\cdot \vp$, $\vec\pauli\cdot\Pf=\pauli^\parallel\Ppar
+\vec\pauli^\perp\cdot\Pf^\perp$, and $\vec\pauli\cdot(\vp\,\wedge\,\Pf)
=\vec\pauli^\perp\cdot(\vp\,\wedge\,\Pf)$ change
their signs under conjugation by $U_{ref,\vp}$, see (\ref{UrefId-1})
and (\ref{UrefId-2}). Therefore, the
conditions (\ref{eq:RotInv-1}) and (\ref{eq:RefInv-1})
can only be simultaneously satisfied if $b_1=b_2=b_3=0$.
\endprf

\subsection{Gauge invariance and soft photon sum rules}

An important property of the model under consideration is that on all levels of the
renormalization group analysis, the corresponding {\em effective Hamiltonians}
(introduced in Section {\ref{sec:RG-effHam}}) satisfy
{\em soft photon sum rules}, \cite{bcfs2}, which
can be considered as a generalization of the differential Ward-Takahashi identities in QED.
For the fiber Hamiltonian $\Hps$, they correspond to the following relations.

Let $\vn\in\R^3$, $|\vn|=1$. It is easy to see that
\eqn
        \g\pol(\vn,\lambda)\cdot\nabla_{\Pf}T[\opp,\vp]
        &=&-\lim_{\rvar\rightarrow0}
        w_{0,1}[\opp;(\rvar\vn,\lambda)] \nonumber\\
        &=&-\lim_{\rvar\rightarrow0}
        w_{1,0}[\opp;(\rvar\vn,\lambda)]
        \label{eq:sbsr-fib-1}
\eeqn
holds for any choice of $\vn$.
Furthermore,
\eqn
        \g\pol(\vn,\lambda)\cdot\nabla_{\Pf}w_{0,1}[\opp,\vp;\widetilde K]
        &=&-2\lim_{\rvar\rightarrow0}w_{0,2}[\opp;(\rvar\vn,\lambda) , \widetilde K]
        \nonumber\\
        &=&
        -\lim_{\rvar\rightarrow0}w_{1,1}[\opp;(\rvar\vn,\lambda) , \widetilde K]
        \; ,
        \label{eq:sbsr-fib-2-1}
\eeqn
and likewise,
\eqn
        \g\pol(\vn,\lambda)\cdot\nabla_{\Pf}w_{1,0}[\opp,\vp;\widetilde K]&=&
        -2\lim_{\rvar\rightarrow0}w_{2,0}[\opp;(\rvar\vn,\lambda) , \widetilde K]
        \nonumber\\
        &=&
        -\lim_{\rvar\rightarrow0}w_{1,1}[\opp;(\rvar\vn,\lambda) , \widetilde K]
        \; .
        \label{eq:sbsr-fib-2-2}
\eeqn
(\ref{eq:sbsr-fib-1}), (\ref{eq:sbsr-fib-2-1}) and (\ref{eq:sbsr-fib-2-2}) correspond  to  the
{\em soft photon sum rules} on the most basic level.

\subsection{Organization of the proof}

For an introductory exposition of the
isospectral renormalization group method, and a discussion of problems connected to
infrared mass renormalization in non-relativistic QED, we refer to \cite{bcfs1,bcfs2}.
The proof of Theorem {\ref{mainthm-1}} is essentially organized as follows:


\begin{itemize}
\item
In Section {\ref{sec:smFesh-1}}, we introduce the isospectral smooth Feshbach map,
and recall some of its key properties from \cite{bcfs1,bcfs2}.
\item
In Section {\ref{sec:RG-effHam}}, we introduce effective Hamiltonians
belonging to a subclass of the bounded operators on the reduced Hilbert space $\H_{red}=\C^2\otimes\1[H_f<1]\Fo$,
which are reflection and rotation symmetric, and satisfy soft photon sum rules.
Moreover, we introduce a Banach space of generalized
Wick kernels $\Hspace_{\geq0}$ which parametrize the effective Hamiltonians.
\item
In Section {\ref{Rentrsf}}, we define an isospectral renormalization map
$\ren$ on a polydisc $\Polyd\subset\Hspace_{\geq0}$, given by the composition of the smooth Feshbach map
with a rescaling transformation, and  a renormalization of a spectral parameter.
We then state the main technical results of this work:
\begin{itemize}
\item
Theorem {\ref{thm:codim2contrthm}}
asserts that $\ren$ is codimension-3 contractive on $\Polyd$, and that
it is marginal on a subspace of dimension 3 (after explicitly projecting out a one-dimensional
subspace of relevant perturbations). However, no control on the growth
of the marginal interactions under repeated applications of $\ren$ is provided at this point.
$\ren$ is shown to preserve reflection and rotation symmetry, and the soft photon sum rules.
\item
We introduce a strong induction assumption
$\sind[n]$ which asserts that the marginal interactions admit an $n$-independent
upper bound after $n$ applications of $\ren$.
Theorem {\ref{thm:strong-induct}} asserts that $\sind[n-1]$ implies $\sind[n]$
for any $n$.
\end{itemize}
\item
In Section {\ref{sect:codim2-contr-1}}, we prove Theorem {\ref{thm:codim2contrthm}}.
We use the soft photon sum rules
to reduce the number of a priori independent marginal operators, and the
spatial symmetries of the model to prove that the operators originating from
the Zeeman term in $\Hps$ (the term proportional to the magnetic field operator $\Bf$)
are irrelevant.
\item
In Section {\ref{sect:sind-thm-prf-1}}, we prove Theorem {\ref{thm:strong-induct}}.
To establish the strong induction step
$\sind[n-1]\Rightarrow\sind[n]$, we combine Theorem {\ref{thm:codim2contrthm}}
with composition identities satisfied by the smooth Feshbach map.
\item
In Section {\ref{sec:main-thm-proof-1}}, we prove the uniform bounds
on the renormalized mass asserted in Theorem {\ref{mainthm-1}}.
This is accomplished by relating $m_{ren}(\vp,\sig)$ to the
renormalization group flow of one of the operators contained in
the effective Hamiltonians.
\item
In Section {\ref{sec:mren-limits-1}}, we prove the existence of the renormalized
mass in the limit $\sigma\searrow0$ for $\vp$ with $0\leq|\vp|<\puppbd$.
\end{itemize}
For the proof of Theorem {\ref{mainthm-1}}, we will invoke many constructions and
results from \cite{bcfs1,bcfs2}.

\section{The Smooth Feshbach Map}
\label{sec:smFesh-1}

In this section, we introduce the smooth Feshbach map and the associated intertwining
maps, mostly quoting results from \cite{bcfs1,bcfs2}.

\subsection{Definition of the smooth Feshbach map}

Let $\H$ be a separable Hilbert space, and
let $0\leq\chi\leq1$ denote a positive, selfadjoint operator on $\H$.
Introducing $\bar\chi:=\sqrt{1-\chi^2}$, we obtain the partition of unity
$\chi^2+\bar\chi^2=\1$ on $\H$.

We let $P_\chi$, $P_{\bar\chi}$ denote the
orthoprojectors associated to the subspaces
$\Ran(\chi)$, $\Ran(\bar\chi)\subset\H$, respectively, and
let $P^\perp_\chi= \1-P_\chi$ and $P_{P_{\bar\chi}}^\perp=\1-P_{\bar\chi}$,
their respective complements.
It is clear that the spaces $\Ran(\chi)$ and $\Ran(\bar\chi)$ are
mutually complementary if and only if $\chi$ is a projector.

\begin{definition}\label{Feshbtripledef}
A pair of closed operators $(H,\tau)$ acting on $\H$
is  a Feshbach pair corresponding to $\chi$ if:
\begin{itemize}
\item
The domains of $H$ and $\tau$ coincide, and are invariant under
$\chi$ and $\bar\chi$. Moreover, $[\chi,\tau]=0=[\bar\chi,\tau]$.

\item
Let
\eqn
        H_{\bar\chi}&:=&\tau+\bar\chi \omega\bar\chi
    \nonumber\\
    \omega&:=&H -\tau \;.
\eeqn
The operators $\tau$, $H_{\bar\chi}$
are bounded invertible on $\Ran(\bar\chi)$.

\item
Let
\eqn
        \bar R \; := \; H_{\bar\chi}^{-1}\;\;on\;\;\Ran(\bar\chi) \;,
\eeqn
and let $H_{\bar\chi}=U|H_{\bar\chi}|$ denote the polar decomposition of
$H_{\bar\chi}$ on $\Ran\bar\chi$. Then,
\eqn
    \big\|\bar{R}\big\|_{\H\rightarrow\H}&<&\infty
    \;,
    \nonumber\\
    \big\|\big|\bar{R}\big|^{\frac{1}{2}}U^{-1}
    \bar\chi\omega\chi\big\|_{\Ran(\chi)\rightarrow\H}
    &,&
    \big\|\chi \omega\
    \bar\chi\big|\bar{R}\big|^{\frac{1}{2}}
    \big\|_{\H\rightarrow\Ran(\chi)} <\infty \; .
    \label{Feshboundopasscond}
\eeqn
\end{itemize}
We write
\eqn
        \Fpairs(\H,\chi)
\eeqn
for the set of Feshbach pairs on $\H$ corresponding to $\chi$.
\end{definition}

The {\em smooth Feshbach map} is defined by
\eqn
        F_\chi:\Fpairs(\H,\chi)&\rightarrow&\cL(\H)
        \nonumber\\
        (H,\tau)&\mapsto& \tau + \chi\,
        \omega \,\chi -
        \chi  \,\omega \,\bar\chi\, \bar{R}\, \bar\chi \, \omega \,\chi \; ,
        \label{FchiHtaudef}
\eeqn
where $\cL(\H )$  denotes the linear operators $\H \rightarrow\H $.
Furthermore, we introduce the {\em intertwining maps}
\eqn
        Q_\chi : \Fpairs(\H,\chi)&\rightarrow&\cB(\Ran(\chi),\H)
        \nonumber\\
        (H,\tau)&\mapsto&\chi \,- \,
        \bar\chi\, \bar{R} \,\bar\chi \,\omega \,\chi
        \nonumber\\
        \nonumber\\
        Q^\sharp_\chi  :
        \Fpairs(\H,\chi)&\rightarrow&\cB(\H,\Ran(\chi))
        \nonumber\\
        (H,\tau)&\mapsto&\chi \,- \,\chi\, \omega \,\bar\chi\,
        \bar{R}\,\bar\chi \;.
\eeqn
We note that the mutually complementary subspaces $\Ran(\chi)$, $\Ran(\chi)^\perp\subset\H$ are invariant
under $F_\chi(H,\tau)$. On $\Ran(\chi)^\perp$, $F_\chi(H,\tau)$
equals $\tau$, while it is a bounded operator on $\Ran(\chi)$.

\subsection{Isospectrality}

The smooth Feshbach map, combined with the intertwining operators, implements a
non-linear, isospectral correspondence between closed operators on $\H$
and ones on the Hilbert subspace $\Ran(\chi)\subset\H$, according to
the following main theorem.

\begin{theorem}\label{FeshIsoThm} (Feshbach isospectrality theorem)
Let $(H,\tau)\in\Fpairs(\H,\chi)$.  Then, the following hold:
\begin{itemize}
\item
The operator $H$ is bounded invertible on $\H$ if and only if
$F_{\chi}(H,\tau)$ is bounded invertible on $\Ran(\chi)\subset\H$.
If $H$ is invertible,
\eqn
    F_\chi(H,\tau)^{-1} \; = \; \chi H^{-1}\chi \, + \, \bar\chi \tau^{-1}\bar\chi
\eeqn
and
\eqn
    H^{-1} \; = \; Q_\chi(H,\tau)F_\chi(H,\tau)^{-1}Q_\chi^\sharp(H,\tau)
    \, + \, \bar\chi\bar R\bar\chi \;.
\eeqn

\item
Let $\psi\in\H$. Then, $H\psi=0$ on $\H$ if and only if
$F_{\chi}(H,\tau) \chi \psi = 0$ on $\Ran(\chi)\subset\H$.

\item
Let $\phi\in \Ran(\chi)$. Then, $F_{\chi}(H,\tau)\zeta=0$ on $\Ran(\chi)\subset\H$
if and only if $H Q_{\chi}(H,\tau)\zeta=0$ on $\H$.
\end{itemize}
\end{theorem}

We furthermore quote the following lemma from \cite{bcfs2}.

\begin{lemma}
Let $(H,\tau)\in\Fpairs(\H,\chi)$. Then, the following identities hold.
\eqn
        \chi F_\chi(H,\tau)&=&HQ_\chi(H,\tau)\nonumber\\
        F_\chi(H,\tau)\chi&=&Q^\sharp_\chi(H,\tau) H \;,
        \label{FHQid}
\eeqn
and
\eqn
        Q^\sharp_\chi(H,\tau)HQ_\chi(H,\tau) \; = \; F_\chi(H,\tau)
        -F_\chi(H,\tau)\bar\chi\tau^{-1}\bar\chi F_\chi(H,\tau) \;.
        \label{QHQid}
\eeqn

\end{lemma}

\subsection{Derivations}
Consider a Hilbert space $\H$ with a dense subspace $\mathcal{D}\subset\H$,
and let $\mathcal{L}(\mathcal{D},\H)$ denote the space of linear (not
necessarily bounded) operators from $\mathcal{D}$ to $\H$.

A derivation $\der$ is a linear map $\Dom(\der)
\rightarrow \mathcal{L}(\mathcal{D}, \H)$, defined on a subspace
$\Dom(\der) \subset \mathcal{L}(\mathcal{D}, \H)$, which obeys
Leibnitz' rule. That is, for $A, B \in \Dom(\der)$,
$\Ran( B) \subseteq \mathcal{D}$, and $A \, B \in
\Dom(\der)$,
$$
        \der[A\,B]=\der[A]B + A\der[B]\;.
$$
Let $(H,\tau)\in\Fpairs(\H,\chi)$, and
assume that $H,\tau \in \mathcal{L}(\mathcal{D}, \H)$, where
$\mathcal{D} := \Dom(H) = \Dom(\tau)$ and that $H, \tau, \chi, \bar\chi$
and the composition of operators in the definition of
$F_{\chi}(H,\tau)$ are contained in $\Dom(\der)$.

\begin{theorem}
\label{FeshDerThm}
Assume that $\der[\bar\chi]$, $\bar\chi$ are bounded operators which leave
$\mathcal{D}$ invariant, and which commute with $\tau$ and with one another.
Then, under the assumptions stated above, and writing $Q^{(\sharp)}\equiv
Q^{(\sharp)}_\chi(H,\tau)$,
\eqn
    \der [F_{\chi}(H,\tau)]
    &=&
    \der[\tau] \, + \,  \chi \omega \bar\chi \bar{R}
    \der[\tau]
    \bar{R} \bar\chi \omega \chi
    \, + \, Q^\sharp \der[\omega] Q\;
    \nonumber\\
    &+&\der[\chi]H Q + Q^\sharp H \der[\chi]
    \nonumber\\
    &-&2 \chi \omega (\tau^{-1}\der[\bar\chi] \, - \, \bar{R}
    \bar\chi \omega  \tau^{-1}\der[\bar\chi])\,\tau \,
    \bar{R}\bar\chi\omega\chi \; .
        \label{FeshDer-1}
\eeqn
and
\eqn
        \der [Q ]&=&- \, \bar\chi \bar R
        \bar\chi \der [H]  Q
        \nonumber\\
        \der [Q^\sharp]&=&-
         Q^\sharp
         \der [H] \bar\chi\bar R\bar\chi \;.
        \label{QDer-1}
\eeqn
In particular (\ref{FeshDer-1}) reduces to
\eqn
        \der [F_{\chi}(H,\tau)] \; = \; Q^\sharp \der [H] Q \;.
        \label{FeshDer-2}
\eeqn
in the special case where
\eqn
        [\der [\chi],\bar\chi] \; = \; 0 \; = \; \der [\tau]
\eeqn
is satisfied.
\end{theorem}

\subsection{Composition identities}

For two subsequent applications of the smooth Feshbach map, the
following concatenation rule holds.

\begin{theorem} \label{FeshCompThm}
Let $0 \leq \chi_1 , \chi_2 \leq 1$ be a pair of mutually commuting,
selfadjoint operators, and $\bar\chi_j := (\1 - \chi_j^2)^{\frac12}$.
We assume that $\chi_1 \chi_2 = \chi_2
\chi_1=\chi_{2}$, such that $\Ran(\chi_2)\subseteq \Ran(\chi_1) \subset \H$.
Let
\eqn
        (H,\tau_1)&\in&\Fpairs(\H,\chi_1)
        \nonumber\\
        (H,\tau_{2})&\in&\Fpairs(\H,\chi_2)
        \nonumber\\
        (F_1,\tau_{12})&\in&\Fpairs( \Ran(\chi_1),\chi_2)
\eeqn
with $F_1 := F_{\chi_1}(H,\tau_1)$, where $\tau_1$,
$\tau_{12}$ commute with
$\chi_j, \bar\chi_j$.

Then,
\eqn
        F_{\chi_{2}}(H,\tau_2) &=& F_{\chi_2}(F_1,\tau_{12}) \; ,
        \nonumber\\
        Q_{\chi_{2}}(H,\tau_2) &=& Q_{\chi_1}(H,\tau_1)
        \, Q_{\chi_2}(F_1,\tau_{12}) \; ,
        \nonumber\\
        Q^\#_{\chi_{2}}(H,\tau_{2}) &=& Q^\#_{\chi_2}(F_1,\tau_{12})
        \, Q^\#_{\chi_1}(H,\tau_1) \; ,
        \label{FeshComp-1}
\eeqn
if and only if $\tau_2=\tau_{12}$. Furthermore,
\eqn
        A \, Q_{\chi_{2}}(H,\tau_{2}) &=&  A \, Q_{\chi_2}(F_1,\tau_{12})
        \nonumber\\
        Q^\sharp_{\chi_{2}}(H,\tau_{2}) \, A
        &=&  Q^\sharp_{\chi_2}(F_1,\tau_{12})  \, A \; ,
        \label{FeshComp-2}
\eeqn
for all operators $A$ acting on $\H$ that satisfy $A \bar\chi_1 = \bar\chi_1 A = 0$.
\end{theorem}

\subsection{Grouping of overlap terms}
\label{overlapopsubsubsec}

The Feshbach pairs $(H,\tau)\in\Fpairs(\H,\chi)$ considered in this paper
have the property that $H=T+W$ with $T\neq\tau$, $[T,\chi]=0=[T,\tau]$ and
$[W,\chi],[W,\tau]\neq0$, and where the operator $W$ has a small relative bound
with respect to $T$.

For the resolvent expansions in powers of $W$ instead of
$\omega=H-\tau=T-\tau+W$ (which is in general {\em not} small),
we regroup the terms in the smooth Feshbach map
to manifestly separate the contributions from $T$ and $W$ contained in $\omega$.
For this purpose, we introduce the operator $\piop_\chi(T,\tau)$ in (\ref{piopdef}).
Notably, it differs from the identity operator only on the spectral support of
$\chi\bar\chi$ where the smooth cutoff operators overlap.

\begin{lemma}\label{piopdeflemma}
Let $(H,\tau)\in\Fpairs(\H,\chi)$, and
assume that
$H=T+W$, where $[T,\chi]=[T,\tau]=0$. Let
\eqn\label{barR0def}
        T':=T-\tau
        \; \; {\rm and} \; \;
        \bar{R}_0(T,\tau):=(\tau+\bch T'\bch)^{-1}
\eeqn
on $\Ran(\bch)$.
Moreover, let
\eqn
    \piop_\chi(T,\tau) \; := \; \1 \, - \, \bch T'\bch \bar{R}_0
    \; = \;
    P_{\bch}^\perp \, + \, P_{\bch} \tau \bar{R}_0 \;
    \label{piopdef}
\eeqn
on $\Ran(\chi)$, where
$\Ran(\piop_\chi(T,\tau)-\1)=\Ran(\chi\bch)$,
and where $\piop_\chi(T,\tau)$ commutes with $\tau,\chi,\bch$ and $T$.
Then,
\eqn
        F_{\chi}(H,\tau)\;=\;\tau &+& \chi T' \piop_\chi(T,\tau) \chi
        \nonumber\\
        &+& \chi \piop_\chi(T,\tau) (W - W \bch\bar{R}\bch W) \piop_\chi(T,\tau)\chi\;,
\eeqn
and in particular, $\piop_\chi\equiv\1$ if and only if $\tau=T$.

Moreover,
\eqn
        \piop_\chi(T_1,\tau_1)-\piop_\chi(T_2,\tau_2)
        &=&\bar\chi^2(T_2-T_1)\bar R_{0}(T_2,\tau_2)\piop_\chi(T_1,\tau_1)
        \label{eq:piop-diff-ed-1}\\
        &&\hspace{2cm}-\,\bar\chi^2 (\tau_2-\tau_1)T_1 \bar R(T_1,\tau_1) \bar R(T_2,\tau_2)
        \nonumber
\eeqn
where $(T_i,\tau_i)$, $i=1,2$, satisfy the same assumptions as $(T,\tau)$.
\end{lemma}

\prf
We only verify (\ref{eq:piop-diff-ed-1}); all other statements were proved in \cite{bcfs2}.
let $\bar R_{0,i}:=\bar R(T_i,\tau_i)$. We have
\eqn
        \piop_\chi(T_1,\tau_1)-\piop_\chi(T_2,\tau_2)&=&
        \bar\chi^2(T_2-T_1) \bar R_{0,2}-\bar\chi^2 T_1(\bar R_{0,1}-\bar R_{0,2})
        \nonumber\\
        &=&\bar\chi^2(T_2-T_1)(\bar R_{0,2} - \bar\chi^2 T_1\bar R_{0,1} \bar R_{0,2})
        \nonumber\\
        &&\hspace{2cm}
        -\bar\chi^2 T_1 \bar R_{0,1} \bar R_{0,2} (\tau_2-\tau_1)
        \nonumber\\
        &=&\bar\chi^2(T_2-T_1) \bar R_{0,2}\piop_\chi(T_1,\tau_1)
        -\bar\chi^2 T_1 \bar R_{0,1} \bar R_{0,2} (\tau_2-\tau_1)
\eeqn
using
\eqn
        \bar R_{0,1} - \bar R_{0,2}&=&\bar R_{0,1}  \bar R_{0,2}
        \big(\tau_2+\bar\chi^2 T_2' - (\tau_1+\bar\chi^2 T_1')\big)
        \nonumber\\
        &=&\bar R_{0,1}  \bar R_{0,2}
        \big(\tau_2-\tau_1+\bar\chi^2( T_2 - T_1)\big) \;,
\eeqn
where $T'_i=T_i-\tau_i$. This establishes (\ref{eq:piop-diff-ed-1}).
\endprf

\section{Isospectral renormalization group: Effective Hamiltonians}
\label{sec:RG-effHam}

In this section, we introduce a space of {\em effective Hamiltonians}.
While the basic constructions are similar or equal to those in \cite{bcfs2},
some significant modifications will be formulated in later sections.

We introduce the "reduced" Hilbert space
\eqn
        \H_{red} \; := \;  \C^2\otimes {\bf 1}(H_f<1)\Fo
        \; \subset \; \C^2\otimes\Fo \;,
\eeqn
and choose a smooth cutoff function
\eqn
        \chi_1[x] \; := \; \sin[\frac\pi2 \Theta(x)]
        \label{Thetadef}
\eeqn
on $[0,1)$, with
\eqn
        \Theta \, \in \, C_0^\infty([0,1);[0,1])
\eeqn
and
\eqn
        \Theta \; = \; 1 \; \; {\rm on } \; [0,\frac34] \;.
\eeqn
Together with
$$
        \bar\chi_1[x] \; := \; \sqrt{ \, 1 \, - \, \chi_1^2[x] \,}\;,
$$
we obtain the selfadjoint cutoff operators $\chi_1[H_f]$ and $\bar\chi_1[H_f]$
on $\H_{red}$ (and on $\Fo$).

We introduce the notation
\eqn
    \opp \; := \; (\op) \;,
       \label{spvardef}
\eeqn
with associated spectral variables
\eqn
    \spvar \; := \; (X_0,\vX) \, \in \, [0,1]\times B_1 \;.
\eeqn
We introduce a class of bounded operators on $\H_{red}$,
referred to as {\em effective Hamiltonians}, of the form
\eqn
        H \; = \;  E[\vp]  \chi_1^2[H_f] \, + \, T[\opp;\vp] \, + \, \chi_1[H_f] W[\vp] \chi_1[H_f] \;,
        \label{effhamWickexpdef}
\eeqn
parametrized by the conserved momentum $\vp\in\R^3$. $E[\vp]\in\R$ is a spectral parameter.
The operator $T[\opp;\vp]$ is referred to as the {\em free}, or the {\em non-interacting term}
in the effective Hamiltonian, and the function $T[\,\cdot\,;\vp]:[0,1]\times B_1\rightarrow\R$ has the form
\eqn
        T[\spvar;\vp] \; = \; X_0 \, + \, T'[\spvar;\vp]
        \; \; \; , \; \; \;
        T'[\spvar;\vp] \; = \; \chi_1^2[X_0]\widetilde T[\spvar;\vp] \;,
        \label{eq:T-def-structure-1}
\eeqn
with
\eqn
    \partial_{X_0}^a\widetilde T[\unull;\vp]\;=\;0
    \; \; \; {\rm and} \; \; \;
    \partial_{X_0}^a T'[\unull;\vp]\;=\;0 \; \; , \; a \; = \; 0,1 \;.
\eeqn
Clearly, $T[\unull;\vp]=0$, and $T[\opp;\vp]$ commutes with
every component of $\opp$.
The detailed list of assumptions imposed on $\widetilde T[\spvar;\vp]$
is presented in Section {\ref{subsect:T-structure-detail-1}}.

The operator $W$ in the effective Hamiltonian is referred to
as its {\em interaction term},
\eqn
        W=\sum_{M+N=1}^\infty W_{M,N}\;,
\eeqn
where the operator $W_{M,N}$ is a {\em generalized Wick monomial} of degree $(M,N)$ of the form
\eqn
        W_{M,N}&\equiv&W[w_{M,N}]
        \nonumber\\
        &=&\int
        d \mu_\ircut(K^{(M,N)}) \, a^*(K^{(M)})\, w_{M,N}\left[\opp;\vp;K^{(M,N)}\right]\,
        a^\sharp(\widetilde{K}^{(N)}) \;,
        \label{Wmon}
\eeqn
where we introduce the notation (recalling that $K =  (\vk , \lambda)    \in   B_1\times\{+,-\}$)
\eqn
        K^{(M)} &:=& (K_1,\dots,K_M)
        \nonumber\\
        \widetilde K^{(N)}&:=& (\widetilde K_1,\dots,\widetilde K_N)
        \nonumber\\
        K^{(M,N)} &:=&  (K^{(M)} , \widetilde K^{(N)}  )
        \nonumber\\
        a^\sharp (K^{(M)})&:=&\prod_{i=1}^M
        a^\sharp (K_i)
        \nonumber\\
        \Sigma[\uk^{(N)}]&:=& \uk_1+\dots+\uk_N
        \label{multiinddef}
\eeqn
for $M,N\geq 0$, and $a^\sharp=a$ or $a^*$.

The integration measure $d\mu_\ircut$ on $(B_1\times\{+,-\})^{M+N}$ is given by
\begin{align}
        d \mu_\ircut (K^{(M,N)})   \;  :=  \;
        \prod_{i=1}^M \prod_{j=1}^N
        \frac{d K_i \,  \max\{1,\cuts(|\vk_i|)\}  }{|\vk_i|^{1/2}}
        \frac{d\widetilde K_j \, \max\{1, \cuts(|\widetilde \vk_j|) \} }{|\widetilde \vk_j|^{1/2}}
        \;,
        \label{dmudef-leq1}
\end{align}
We note that hereby, the cutoff function
$\cuts$ is incorporated into the
integration measures $d\mu_\sig$ if $\sig\leq1$, and absorbed
into the generalized Wick kernels $w_{M,N}$ if $\sig>1$. Moreover, we note
that for $\sigma>1$ and $|k|\leq1$, we have $\cuts(|k|)=\frac{|k|}{\sigma}$
with $\cuts$ given in (\ref{cutsDef-1}).

For $M+N\geq1$,  the {\em generalized Wick kernels} $w_{M,N}$ are $\mat(2\times2,\C)$-valued functions of
$\spvar$, $K^{(M,N)}$, and $\vp$, of the form
\eqn\label{genwMNdefspinor}
        w_{M,N} &:=& w_{M,N}^0 \1_2 + \vec\pauli\cdot\vw_{M,N}
        \nonumber\\
        &&
        \nonumber\\
        &=&\left(
        \begin{array}{cc}w_{M,N}^0+w_{M,N}^3&
        w_{M,N}^1+iw_{M,N}^2\\
        w_{M,N}^1-iw_{M,N}^2&w_{M,N}^0-w_{M,N}^3\end{array}\right)
\eeqn
in the basis of Pauli matrices $\vec\pauli=(\pauli_1,\pauli_2,\pauli_3)$.

We shall refer to
\eqn
        w^0_{M,N}&&{\rm and}
        \nonumber\\
        \vw_{M,N}&:=&( w^1_{M,N} \, , \, w^2_{M,N} \, , \, w^3_{M,N} )
\eeqn
as the scalar, and the vector component of $w_{M,N}$, respectively.
Every component of $w_{M,N}$ is separately fully symmetric with
respect to $K_1,\dots,K_M$ and $\widetilde K_1,\dots,\widetilde K_N$.

For $M+N=0$,
\eqn
        w_{0,0} \;  = \;  w_{0,0}^0 \, \1_2 \;\;\;\;\;\;\;\;( \vw_{0,0} \; \equiv \; \vnull )
\eeqn
is assumed to be purely scalar.

\subsection{The Banach space of generalized Wick kernels}

We recall that
\eqn
        \spvar \; = \; (X_0,\vX) \, \in \, [0,1]\times B_1
\eeqn
denotes the quadruple of spectral variables corresponding to $\opp=(\op)$.
Let
\eqn
        \ua \; := \; (a_0,\va)
        & , &
        \va \; := \; (a_1,a_2,a_3)
        \nonumber\\
        |\va| \; := \; \sum_{j=1}^3 a_j
        & , &
        |\ua| \; := \; |a_0| \, + \, |\va|
\eeqn
with $a_i\in\N_0$,
\eqn
        \partial_{\spvar}&:=&(\partial_{X_0},\nabla_{\vX})
        \nonumber\\
        \nabla_{\vX}&:=&(\partial_{X_1},\partial_{X_2},\partial_{X_3})
\eeqn
and
\eqn
        \partial_{\spvar}^\ua&:=&\prod_{j=0}^3\partial_{X_j}^{a_j}
        \nonumber\\
        \nabla_{\vX}^\va&:=&\prod_{j=1}^3\partial_{X_j}^{a_j}  \;.
\eeqn
For $M=N=0$,  we introduce the norms
\eqn
        \big\|w_{0,0}\big\|_{0,0}&:=&
        \sup_{|\vX|\leq X_0\in I }
        \big| w_{0,0}\big|
\eeqn
and
\eqn
        \big\|w_{0,0}\big\|_{0,0}^\sha&:=&
        \sum_{0\leq|\ua| \leq2}
        \big\|\partial_{\spvar}^\ua   w_{0,0}\big\|_{0,0}
        \,+\,\sum_{ |\ua| =0,1}
        \big\| \partial_{|\vp|} \partial_{\spvar}^\ua w_{0,0}\big\|_{0,0}
\eeqn
(by definition, the vector part of $w_{0,0}$ is zero).
Writing
\eqn
        \|A\|_{{\rm Mat}(2\times2,\C)}:=\sqrt{ \, {\rm Tr}A^*A \, }\;,
\eeqn
we define
\eqnn
        \big\|w_{M,N}\big\|_{M,N} &:=&(2\pi^{\frac12})^{M+N}
        \sup_{|\vX|\leq X_0\in I}\sup_{K^{(M,N)}}
        \big\|w_{M,N}[\spvar;K^{(M,N)}]\big\|_{{\rm Mat}(2\times2,\C)}
        \; ,
\eeqnn
and
\eqn
        \big\| w_{M,N}\big\|_{M,N}^\sha & := & \sum_{0\leq|\ua|\leq2}
        \big\|\partial_{\spvar}^\ua \, w_{M,N}\big\|_{M,N}
        \, + \, \sum_{|\ua|=0,1}
        \big\|\partial_{|\vp|}\partial_{\spvar}^\ua \, w_{M,N}\big\|_{M,N}
        \nonumber\\
        &&\hspace{1cm}+\,
        \sum_{a=0,1}\sup_{(\vk,\lambda)\in K^{(M,N)}}
        \big\|\partial_{|\vp|}^a\partial_{|\vk|} \, w_{M,N}\big\|_{M,N}
        \label{WMNnormdefinition}
\eeqn
for $M+N\geq1$.

We note the following differences from \cite{bcfs2}:
\begin{itemize}
\item
The kernels $w_{M,N}$ in \cite{bcfs2} are scalar, and
the norms used in \cite{bcfs2} do not contain second order
derivatives with respect to $X_0$, or mixed derivatives in $|\vp|$, $|\vk|$.
\item
In \cite{bcfs2}, different norms are introduced for $|\vp|=0$, and for $|\vp|>0$.
\item
In \cite{bcfs2}, the infrared regularization $\cuts$ is attributed to the generalized Wick kernels
$w_{M,N}$, and not to the integration measure $d\mu_\sigma$.
Therefore, the corresponding norm in \cite{bcfs2} depends on $\sig$,
while here, it does not.
\end{itemize}

We define the Banach spaces
\eqn
        \Wspace_{0,0}^\sharp&=&\Big\{\; w_{0,0} \; \Big| \,
        \|w_{0,0}\|_{0,0}^\sha \, < \, \infty\Big\}
        \nonumber\\
        &&
        \nonumber\\
        \Wspace_{M,N}^\sharp&:=&\Big\{ w_{M,N}\Big| \,
        \big\|w_{M,N}\big\|_{M,N}^\sha \, < \, \infty  \Big\} \;,
\eeqn
of generalized Wick kernels of degree $(M,N)$ with $M+N\geq0$.

\begin{lemma}\label{WMNwMNrelboundslemma3}
Let  $M,N\in\N_0$, and $M+N\geq1$. Let $\|w_{M,N}\|_{M,N}<\infty$,
and $W_{M,N}:=W_{M,N}[w_{M,N}]$ as in (\ref{Wmon}). Then,
the operator norm $\|\cdot\|_{op}$ of $W_{M,N}$ on $\H_{red}$ is bounded by
\eqnn
        \big\| W_{M,N}\big\|_{op} &\leq&
        \big\|(P_{\vac}^\perp H_f)^{-\frac M2}W_{M,N}
        (P_{\vac}^\perp H_f)^{-\frac N2}\big\|_{op}\\
        &\leq&  \Big(\frac{1}{M}\Big)^{\frac M2}
        \Big(\frac{1}{N}\Big)^{\frac N 2}\big\|w_{M,N}\big\|_{M,N} \;.
\eeqnn
$P_{\vac}^\perp:={\bf 1}-|\vac\rangle\langle\vac|$ is the projection
onto the complement of the subspace spanned by the Fock vacuum in $\Fo$.
\end{lemma}

The proof is given in \cite{bcfs1}.

In order to accommodate infinite sums of Wick monomials, we define the spaces
\eqn
        \Hspace_{k}^\sharp \; := \; \bigoplus_{M+N=k}\Wspace_{M,N}^\sharp
\eeqn
for $k\geq1$ and
\eqn
        \h_k \; := \; (w_{M,N})_{M+N=k} \;,
\eeqn
with
\eqn
        \|\h_k\|^{\sha }_\xi \; := \; \xi^{-k}\,\sum_{M+N=k}\|w_{M,N}\|_{M,N}^\sha  \;.
\eeqn
Moreover, for $0<\xi<1$, we introduce the Banach space
\eqn
        \Hspace_{\geq k}^\sharp \; := \;  \bigoplus_{m\geq k}
        \Hspace_{m}^\sharp \; \; \; \; , \; \; k\geq1\;,
\eeqn
of sequences of generalized Wick kernels
\eqn
        \h_{\geq k} \; := \; (\h_{m})_{m\geq k}
\eeqn
for which
\eqn
        \|\h_{\geq k}\|_{\xi}^\sha  \; := \; \sum_{m\geq k}
        \|\h_{m}\|_{\xi}^\sha
\eeqn
is finite.

In the case $M=N=0$
(recalling again that $w_{0,0}=w_{0,0}^{0}\1_2$ is scalar),
\eqn
        w_{0,0}[\spvar;\vp] \; = \; w_{0,0}[\underline{0};\vp] +
        (w_{0,0}[\spvar;\vp]-w_{0,0}[\underline{0};\vp])
\eeqn
induces the decomposition
$$
        \Wspace_{0,0} \; = \; \R \, \oplus \, \Tspace^\sharp
$$
with
\eqnn
        \Tspace^\sharp  \; := \;
        \Big\{T: \bigcup_{r\in[0,1)}\{r\}\times B_r\rightarrow\R
        &\Big|&\|T\|_{\Tspace}^\sha<\infty \;,\;T[\underline{0};p]=0\;,\;\;\;\;
        \nonumber\\
        &&T[X_0,RX;p]=T[\spvar;R^{-1}p]
        \;\;\forall\; R\in O(3) \Big\} \;,
\eeqnn
and
\eqn
        \|T\|^\sha_{\Tspace}  \; := \;\|T\|_{0,0}^\sha  \;.
        \label{Tnormsharpdef}
\eeqn
The pair $(\Tspace^\sharp,\|\,\cdot\,\|_{\Tspace}^\sha)$ is a Banach space.

We introduce the Banach space
\eqn
        \Hspace_{\geq0}^\sharp \; := \;
        \R\oplus\Tspace^\sharp\oplus\Hspace_{\geq1}^\sharp
\eeqn
endowed with the norm
\eqn
        \|\h\|_{\xi}^\sha
        \; := \; \sum_{a=0,1}|\partial_{|\vp|}^a w_{0,0}[\underline{0};\vp]|
        \, + \|T\|_{\Tspace}^\sha
        \, + \|\h_{\geq 1}\|_{\xi}^\sha
\eeqn
for $\h\in\Hspace_{\geq0}^\sharp$.
To a sequence of generalized Wick kernels
$$
        \h:=\big(E,T,\{w_{M,N}\}_{M+N\geq1}\big)\in\Hspace_{\geq0}^\sharp \;,
$$
we associate the effective Hamiltonian
\eqn
        H[\h] \; = \;  E[\vp] \, \chi_1^2[H_f] \, + \, T[\opp;\vp] \, + \,
        \chi_1[H_f] W[\h] \chi_1[H_f]
        \label{phimapdef}
\eeqn
with
\eqn
        W[\h]&:=&\sum_{M+N\geq1} W_{M,N}[w_{M,N}] \;,
\eeqn
which is of the form (\ref{effhamWickexpdef}).

The following result corresponds to Theorem 3.3 of \cite{bcfs1}.

\begin{lemma}
\label{lm:H-embedd-Hspace-1}
The map
\eqn
        H:\Hspace_{\geq0}^\sharp&\rightarrow&\cB(\H_{red})
        \nonumber\\
        \h&\mapsto& H[\h] \; = \; (\ref{phimapdef})
        \label{H-Hsp-cB-map-def-1}
\eeqn
is an injective embedding of $\Hspace_{\geq0}^\sharp$
into the bounded operators on $\H_{red}$.

Moreover,
\eqn
        \|H[\h]\|_{op} \; \leq \; \|\h\|^\sha_{\xi }
\eeqn
for $0<\xi<\xibd$ and $\h\in\Hspace_{\geq0}^\sharp$, and more generally,
\eqn
        \|H[\h_{\geq k}]\|_{op} \; \leq \; \xi^k\|\h_{\geq k}\|^\sha_{\xi }
\eeqn
for $\h_{\geq k}\in\Hspace_{\geq k}^\sharp$.
\end{lemma}

\section{Isospectral renormalizaton group: Renormalization map}
\label{Rentrsf}

In this section, we introduce the isospectral renormalization map.
While the structure of the exposition is similar as in
\cite{bcfs1,bcfs2}, the constructions themselves are significantly more subtle.
The strictly marginal type of the problem under consideration now enters the
constructions in a very essential manner
(see also Remarks {\ref{rem:hfp-choice-1}} and {\ref{rem:hfp-reason}} below).

\subsection{Definition of the isospectral renormalizaton map}
\label{sect:ren-def}

We consider families
of effective Hamiltonians parametrized by $\h[\z]$ which
depend differentiably on a real-valued spectral
parameter $\z\in\I:=[-\frac{1}{\Iconst} \, , \, \frac{1}{\Iconst}]$.
Let $\Hspace_{\geq0}$ denote the Banach space of
$\Hspace_{\geq0}^\sharp$-valued differentiable functions
on the interval $\I$, endowed with the norm
\eqn
        \|T\|_{\Tspace} \; := \; \sup_{\z\in\I}\|T[\z]\|_{\Tspace}^\sharp
        \;\;\;,\;\;\;
        \|\h\|_{\xi} \; := \;
        \sup_{\z\in\I}\|\h[\z]\|_{\xi}^\sharp \;,
\eeqn
where
\eqn
    \|T[\z]\|_{\Tspace}^\sharp&:=&\|T[\z]\|_{\Tspace}^\sha \, + \, \| \partial_{\z} T[\z] \|_{0,0}
\eeqn
and
\eqn
        \|\h\|_{\xi}^\sharp
        \; := \; \sum_{a+b=0,1}\sup_{\z\in\I}
        | \, \partial_{|\vp|}^a \, \partial_{\z}^b \,  w_{0,0}[\z;\underline{0};\vp] \, |
        \, + \|T\|_{\Tspace}^\sharp
        \, + \|\h_{\geq 1}\|_{\xi}^\sharp \;,
\eeqn
with
\eqn
    \|\h_{ k}[\z]\|_{\xi}^\sharp&:=&\|\h_{k}[ \z]\|_{\xi}^\sha \, + \, \xi^{-k} \,
    \sum_{M+N= k} \| \partial_{\z} \, w_{M,N}[\z] \|_{M,N}
    \nonumber\\
    \|\h_{\geq k}[\z]\|_{\xi}^\sharp&:=&\sum_{j\geq k}\|\h_{j}[\z]\|_{\xi}^\sharp \;.
\eeqn
The statements of Lemma {\ref{lm:H-embedd-Hspace-1}} also hold in the case where
$\|\,\cdot\,\|_\xi^\sha$ is replaced by  $\|\,\cdot\,\|_\xi^\sharp$, as can be easily seen.
Let  $H[\Hspace_{\geq0}]$ denote the Banach space of differentiable families of effective Hamiltonians
\eqn
    \I\rightarrow H[\Hspace_{\geq0}^\sharp]
    \;\;\;\;,\;\;\;\;
    \z\mapsto H[\h[\z]] \;,
\eeqn
in $\cB(\H_{red})$.

\begin{remark}
The spectral parameter $\z$ is chosen real, and not in $\C$
as in \cite{bcfs2}, due to certain technical advantages related to strict marginality,
see Remark {\ref{rem:real-spec-var}} below.
\end{remark}

\begin{remark}
Henceforth, we will frequently omit $\vp$ from the notation;
it is always understood that the effective Hamiltonians depend on $\vp$.
\end{remark}

Let $\h\in\Hspace_{\geq0}$ and let $0<\rho\leq\frac12$, which will be fixed later.
The renormalization transformation
$\ren$ is defined by the composition of the following three operations:

\begin{enumerate}

\item[{\bf (F)}]
The smooth
Feshbach map $F_{\chi_\rho[H_f]}$ is applied to the Feshbach pair
$$
    (H[\h[\z]],\hfp[\z] H_f) \in \Fpairs(\H_{red},\chi_\rho[H_f])\;.
$$
Thereby, the degrees of freedom in the range of photon field
energies in $[\rho,1]$ are "eliminated ({\em decimated})".
The scalar $\hfp[\z]\in \R$ with $|\hfp[\z]-1|\ll1$ is determined
by the implicit condition (\ref{eq:hfp-def-1}) below, and
\eqnn
        \chi_\rho[H_f]:=\chi_1[H_f/\rho] \;,
\eeqnn
where $\chi_1$ is defined in (\ref{Thetadef}).

\item[{\bf (S)}] A unitary rescaling transformation $\resc$ with
$$
    \resc[A] \; = \; \alpha[\z]^{-1}\rho^{-1} \Gamma_\rho A \Gamma_\rho^* \;,
$$
where $\Gamma_\rho$ implements unitary dilation by a factor $\rho$ on $\Fo$
(see Section {\ref{resc-def-subsubsect-1}} for detailed definitions).

\item[{\bf (E)}] A transformation $\Ez$ of the
spectral parameter $\z\in\I$ in $\h[\z]$.
\end{enumerate}

\subsubsection{The operation {\bf (F)}}

The detailed constructions are presented in Sections {\ref{subsec:ref-theory}}
and {\ref{polyd-subsect-1-1}}.
The main steps can be summarized as follows.

We first verify for $\h$ in a {\em polydisc} $\Polyd\Polpar\subset\Hspace_{\geq0}$
(the definition is given in Section {\ref{polyd-subsect-1-1}}),
$|\z|<\frac{1}{\Iconst}$, and $|\hfp-1|<\eta$  sufficiently small, that
\eqn
        \big( H[\h[\z]] \, , \, \hfp H_f \big)\in\Fpairs( \H_{red} \, , \, \chi_\rho[H_f] ) \;,
\eeqn
i.e. $(H[\h[\z]],\hfp H_f)$ is a Feshbach pair corresponding to $\chi_\rho[H_f]$.

We then choose the coefficient $\hfp$ in $\tau=\hfp H_f$ to be
given by the unique solution of the implicit equation
\eqn
        \hfp[\z] \; = \;
        \Bra\vac\,,\,\partial_{H_f}F_{\chi_\rho[H_f]}( H[\h[\z]] \, , \, \hfp[\z] H_f )\vac\Ket
        \label{eq:hfp-def-1}
\eeqn
for  $\hfp[\z]$, with $|\hfp[\z]-1| <\eta$. The existence and uniqueness
of this solution are proved in Proposition {\ref{Polydlemma}}.
The correct choice of $\hfp[\z]$ is crucial for the convergence
of the renormalization group recursion in later sections.
The reasons are outlined in Remark {\ref{rem:hfp-choice-1}} below.

\subsubsection{The operation {\bf (S)}}
\label{resc-def-subsubsect-1}

The rescaling transformation $\resc$ is
obtained from unitarily scaling the photon momenta by a factor $\rho$, followed by
multiplication with a scalar factor $(\hfp[\z]\rho)^{-1}$,
\eqn
        \resc[A] =\frac{1}{\hfp[\z]\rho } \, \Gamma_\rho \, A \, \Gamma_\rho^* \;,
        \label{eq:resc-def-1}
\eeqn
where $\Gamma_\rho$ is the unitary dilation operator on $\Fo$. It satisfies
\eqn
        \;\;\;\;
        \resc[a^*(K^{(M)})a(\widetilde K^{(N)})]= \hfp[\z]^{-1 } \rho^{-1-\frac32(M+N)}
        a^*(\rho^{-1}K^{(M)})a(\rho^{-1}\widetilde K^{(N)}) \;,
        \label{Gammarhodef}
\eeqn
where we write
\eqn
    \rho^{-1}K \; := \; (\rho^{-1}k,\lambda)
    \;\;\;\;,\;\;\;\;
    \rho^{-1}K^{(M)} \; := \; (\rho^{-1}K_1,\dots,\rho^{-1}K_M)
\eeqn
for $K\in\R^3\times\{+,-\}$.

To determine the action of rescaling on the generalized Wick kernels,
we first observe that under the scaling $\vk_i\rightarrow \rho\vk_i$,
$\widetilde \vk_j\rightarrow \rho \widetilde \vk_j$,
the integration measures $d\mu_\sigma(K^{(M,N)})$ in (\ref{dmudef-leq1})
produce a factor $\rho^{\frac{5}{2}(M+N)}$.
The cutoff function is modified by $\cuts\rightarrow\cutsrho$.


As a convention, we attribute the scaling factors $\rho^{\frac52(M+N)}$ from the
integration measures
to $w_{M,N}$.
In addition, when $\sig>1$,  the cutoff function $\cuts$,  given by $\cuts(|k|)=\frac{|k|}{\sig}$,
is absorbed into the generalized Wick kernel $w_{M,N}$.

Then, restricted to $H[\Wspace^\sharp_{\geq0}]\subset\cB[\H_{red}]$, $\resc$
induces a rescaling map $s_\rho$ on $\Hspace_{\geq0}$ by
\eqn
        \resc[H[\h]] \; =: \; H[s_\rho[\h]] \; =: \; H[(s_\rho[w_{M,N}])_{M+N\geq0}]\;,
\eeqn
where
\eqn
        \label{rescwMNdef}
        \lefteqn{
        s_\rho[w_{M,N}][\spvar;K^{(M,N)}]
        }
        \nonumber\\
        &&\; = \;
        \left\{
        \begin{array}{rl}
        \hfp[\z]^{-1}\rho^{M+N-1}
        w_{M,N}[\rho\spvar;\rho K^{(M,N)}]
        &{\rm if} \; \sig\leq1 \\
        \hfp[\z]^{-1}\rho^{2(M+N)-1}
        w_{M,N}[\rho\spvar;\rho K^{(M,N)}]
        &{\rm if} \; \sig>1\;.
        \end{array}
        \right.
\eeqn
The powers of $\rho$ are obtained as follows.
A factor $\rho^{\frac52(M+N)}$ enters from the scaling of the integration measure
$d\mu_\sig(K^{(M,N)})$.
For $\sig>1$, an additional factor $\rho^{ M+N }$ enters from
the scaling of $\cuts(|k|)=\frac{|k|}{\sig}$ (one factor $\rho$ for each of the $M+N$ momentum variables;
if $K\neq1$ in (\ref{cutsDef-1}), the factor is $\rho^{ K(M+N) }$).
In addition, there is a factor $\rho^{-\frac32(M+N)}$ from the unitary scaling of $M$ creation-
and $N$ annihilation operators, see (\ref{Gammarhodef}).
Finally, an overall factor $\rho^{-1}$ is produced by
multiplicative factor $(\hfp[\z]\rho)^{-1}$ in the definition of $\resc$.

This implies the bounds
\eqn
        \| s_\rho[w_{M,N}] \|_{M,N} &\leq& | \hfp[\z] |^{-1}\rho^{M+N-1}
        \| w_{M,N} \|_{M,N} \;\;\;{\rm if} \;\sig\leq1 \;.
        \label{srhowMNaprioribd}
\eeqn
and
\eqn
        \| s_\rho[w_{M,N}] \|_{M,N} &\leq& | \hfp[\z] |^{-1}\rho^{2(M+N)-1}
        \| w_{M,N} \|_{M,N} \;\;\;{\rm if}\;\sig>1 \;.
        \label{srhowMNaprioribd-1}
\eeqn
Thus,  if $\sig\leq1$, all $\|w_{M,N}\|_{M,N}$ with $M+N\geq2$ are
contracted by a factor $\leq\rho$; they are therefore {\em irrelevant} in the renormalization
group terminology.
The generalized Wick kernels with $M+N=1$ do not scale with
any positive power of $\rho$; this property is referred to as {\em marginality}.

In the case $\sig>1$, $\|w_{M,N}\|_{M,N}$ is
contracted by a factor $\leq\rho$,  for all $M+N\geq1$;
that is, all generalized Wick kernels are irrelevant if $\sig>1$.
When we speak of marginal interactions, it is understood that we refer
to the case $\sig<1$.
\\

\subsubsection{The operation {\bf (E)}}

Given $\h\in\Hspace_{\geq0}$ with $E[\z]:=w_{0,0}[\z;\underline{0}]$,
we define
\eqn
        {\mathcal U}[\h]:=\Big\{
        \z\in\I\Big|\,|E[\z]|\leq\frac{\rho}{\Iconst}\Big\} \;,
\eeqn
and consider the map
\eqn
        \Ez:{\mathcal U}[\h] \; \rightarrow \; \I
        \;\;\;,\;\;\;
        \z \; \mapsto \; (\hfp[\z]\rho)^{-1} E[\z] \;.
        \label{eq:Erho-def-1}
\eeqn
$\Ez$ is a bijection, and ${\mathcal U}[\h]$ is close to the
interval $I_{\frac{\rho}{\Iconst}}$, provided that $\h$ is close to the non-interacting
theory defined in Section {\ref{subsec:ref-theory}} below.
\\

\subsubsection{The renormalization transformation}

Composing the rescaling transformation $\resc$, the transformation
of the spectral parameter $\Ez$, and the smooth Feshbach map, we now define the
{\em renormalization transformation} $\ren$.

We recall from Lemma {\ref{lm:H-embedd-Hspace-1}} that
the map $H:\h\mapsto H[\h]$ injectively embeds $\Hspace_{\geq0}$
into the bounded operators on $\H_{red}$.
$\Dom(\ren)$, the domain of $\ren$, is defined by those elements $\h\in\Hspace_{\geq0}$ for which
\eqn
        \renop\big[\, H[\h] \,\big] [\zet]  &:=& \resc\Big[\;F_{\chi_\rho[H_f]}
        \big(H[\h[\z]] \,,\,\hfp[\z] H_f\big) \;\Big]
        \label{renHh-def-1}
\eeqn
is well-defined and in the domain of $H^{-1}$, where $\zet\in\I$ and
\eqn
        \z=\Ez^{-1}[\zet]\in\I\;.
\eeqn
The real number $\hfp[\z]\in 1+\Ihfp$ is defined by
(\ref{eq:hfp-def-1}). The map
$\renop:\cB(\H_{red})\rightarrow\cB(\H_{red})$ is referred to as the
{\em renormalization map acting on operators}.

\begin{remark}
\label{rem:hfp-choice-1}
The definition (\ref{eq:hfp-def-1}) of $\hfp[\z]\in 1+\Ie_\eta$ ensures that no operator
proportional to $H_f\chi_1^2[H_f]$ is generated by $\ren$.
We note that with an additional term of the form
$c \, H_f\chi_1^2[H_f]$ in the non-interacting part $T$ of the effective
Hamiltonian,  it cannot be ruled out that $|c|$ becomes large under repeated applications
of $\ren$. Once $|c|\geq1$, the operator $H_f + c \, H_f\chi_1^2[H_f]$ may develop
spurious zero spectrum in the vicinity of $H_f=1$ which strongly complicates the analysis. This
phenomenon is suppressed by the choice of $\hfp[\z]$ stated in (\ref{eq:hfp-def-1}).
\end{remark}

\begin{remark}
\label{rem:hfp-reason}
Using the factor $\hfp[\z]^{-1}$ in $\resc$, the coefficient of the operator $H_f$ in 
$H[\ren[\h]]$ is normalized to have the value 1 (as in $H[\h]$).
In the absence of this factor, the coefficient of $H_f$ increases to
$O(\log\frac{1}{\sigzero})$ under repeated applications of
the renormalization map ($\sigzero$ is the infrared cutoff in the
fiber Hamiltonian $H(\vp,\sigzero)$).
In contrast, it is not necessary in \cite{bcfs2} to keep the coefficient of $H_f$
fixed because there, its smallness is ensured by the $\sigzero$-dependent bounds on the
finestructure constant (in contrast, the results proved here
hold for $\gs<\gs_0$ with $\gs_0$ independent of $\sigzero$).
\end{remark}

Given $\renop$, we define the {\em renormalization map acting on generalized Wick
kernels}
\eqn
        \ren \; := \; H^{-1}\circ \renop \circ H
\eeqn
on $\Dom(\ren)\subset\Hspace_{\geq0}$.
It is shown in Section {\ref{polyd-subsect-1-1}} that the intersection of the
domain and range of $\ren$ contains a family of polydiscs.

\subsection{Choice of a reference theory}
\label{subsec:ref-theory}

We compare
$\h\in\Dom(\ren)$ to a reference family of non-interacting
theories parametrized by $\h_0^{(\vp;\lTnl)}\in\Dom(\ren)$,
which we introduce here (of the same form as in \cite{bcfs2}).
A central task of our analysis is to prove
that $\|\h-\h_0^{(\vp;\lTnl)}\|_{\xi}$ remains small under interations
of $\ren$.

We choose comparison kernels of the form
\eqn
         \h_0^{(\vp;\lTnl)}[\z] \; = \; E[\z]\oplus T_0^{(\vp;\lTnl)}[\z;\spvar]
        \oplus\underline{0}_{\geq1}\;,
        \label{h0plTnl-def-1}
\eeqn
where
\eqn
        T_0^{(\vp;\lTnl)}[\z;\opp]
        \; = \;  H_f \, + \, \chi_1^2[H_f] \,
        \piop_{\chi_1}^{(\vp;\lTnl)}[\opp] \, ( - \, |\vp|\Ppar \, + \, \lTnl P_f^2 ) \;,
        \label{h0plambdadef}
\eeqn
with
\eqn
        \piop_{\chi_1}^{(\vp;\lTnl)}[\z;\opp] \; := \; \piop_{\chi_1[H_f]}
        \big( E[\z] \chi_1^2[H_f]+T_0^{(\vp;\lTnl)}[\opp] \, , \, H_f \big)  \;,
        \label{eq:piop-ref-def-1}
\eeqn
see (\ref{piopdef}).

Therefore,
\eqn
        H[\h_0^{(\vp;\lTnl)}[\z]]
        \; = \; H_f & + & \chi_1^2[H_f]\big(E[\z]-|\vp|\Ppar+\lTnl P_f^2\big)
        \nonumber\\
        &-&\left.  \frac{\big(|p|\Ppar -\lTnl P_f^2\big)^2
        \chi_1^2[H_f]\bar\chi_1^2[H_f]}{H_f+\bar\chi_1^2[H_f]
        \big(E[\z]-|p|\Ppar +\lTnl P_f^2\big)}
        \right|_{\Ran(\bar\chi_1[H_f])} \;. \;\;\;\;\;\;
        \label{fixedpt-1}
\eeqn
In the limit $E[\z],\lTnl\rightarrow0$,  the operator
\begin{align}
        \lim_{\z\rightarrow0}\lim_{\lTnl\rightarrow0}H[\h_0^{(\vp;\lTnl)}[\z]]
        =H_f- |\vp|\Ppar\chi_1^2[H_f]
        - \left.\frac{(|\vp|\Ppar)^2
        \chi_1^2[H_f]\bar\chi_1^2[H_f]}{ H_f-\bar\chi_1^2[H_f]
        |\vp|\Ppar }\right|_{\Ran(\bar\chi_1[H_f])}
\end{align}
defines a {\em fixed point} of the renormalization transformation $\ren$.


\subsection{Detailed structure of $T$}
\label{subsect:T-structure-detail-1}

It is necessary to impose more detailed requirements on the
structure of $T$ than those formulated in  (\ref{eq:T-def-structure-1}).

We recall from (\ref{eq:T-def-structure-1}) that
\eqn
        T[\z;\opp;\vp] \; = \; H_f \, + \, \chi_1^2[H_f] \, \widetilde T[\z;\opp;\vp] \;.
        \label{eq:T-def-structure-2}
\eeqn
We require that $\widetilde T$ has the form
\eqn
        \widetilde T[\z;\opp]
        \; = \;
        \big( \pfp[\z;\vp] \Ppar \, + \, \lTnl_T \Pf^2 \, + \,
        \remT[\z;\opp;\vp] \big)\,\widetilde\piop[\z;\opp;\vp] \;,
        \label{eq:T-def-structure-3}
\eeqn
where:
\begin{itemize}
\item
The scalar $\pfp[\z;\vp]\in\R$ is $C^1$ in $\z\in\I$ and $\vp$, with
\eqn
        |\pfp[\z;\vp]+\vp|\,,\,|\partial_{|\vp|}\pfp[\z;\vp]+1|\;\ll\;1
        \;\;\;,\;\;\;{\rm for\;all}\;\z\in\I \;.
\eeqn
\item
The parameter $\lTnl_T$ is a real number independent of $\z$ and $\vp$, and $0\leq\lTnl_T\leq\frac12$.
\item
The operator $\widetilde\piop[\z;\opp;\vp]$ is
close to $\piop_{\chi_1}^{(\vp;\lTnl)}[\z;\opp;\vp]$,
\eqn
        \|\widetilde\piop -\piop_{\chi_1}^{(\vp;\lTnl)} \|_{\Tspace}\;\ll\;1\;.
\eeqn
Moreover,
\eqn
        \chi_\rho[H_f]\widetilde\piop[\z;\opp;\vp]\;=\;0 \;.
        \label{eq:chirho-tildpiop-annih-1}
\eeqn
\item
The function $\remT[\z;\spvar;\vp]$ satisfies
\eqn
        \partial_{\spvar}^{\ua}\Big|_{\spvar=\unull}\remT[\z;\spvar ]&=&0
        \;\;\;\;{\rm for}\; 0\leq|\ua|\leq1 \;,
        \nonumber\\
        \|\,\remT\,\|_{\Tspace}&\ll&1\;.
        \label{eq:T-def-structure-4}
\eeqn
It is a  small error term, and $O(|\spvar|^2)$
in the limit $|\spvar|\rightarrow 0$.
\end{itemize}

\subsection{The domain of $\ren$}
\label{polyd-subsect-1-1}

We next prove that the domain of $\ren$
contains a {\em polydisc} of the form
\eqn
        \Polyd\Polpar :=\Big\{\; \h=(E,T,\h_{\geq1})\in\Hspace_{\geq 0} \,\Big|
        \hspace{1.2cm}
        \|\h_{1}\|_{\xi}&<&\eta   \; ,
        \nonumber\\
        \|\h_{\geq2}   \|_{\xi} &<&\e \; ,
        \label{eq:Polyd-def-hgeq2-1}\\
        {\rm with}\;
        T\;{\rm as\;in\;}(\ref{eq:T-def-structure-2})\,\sim\,(\ref{eq:T-def-structure-4}),\;
        {\rm where}\hspace{0.7cm}&&
        \nonumber\\
        \lTnl_T&=&\lTnl \;,
        \label{eq:Polyd-def-lTnl-1}\\
        \|\,\remT\|_{\Tspace}
        &<& \delta \;,
        \label{eq:Polyd-def-remT-1}\\
        \|\widetilde\piop -\piop_{\chi_1}^{(\vp;\lTnl)} \|_{\Tspace}
        &<&K_\Theta \, \delta \;  ,
        \label{eq:Polyd-def-piop-1}\\
        {\rm and\;for}\;a =0,1 \;, \hspace{3.5cm}
        &&
        \nonumber\\
        \sup_{\z\in\I } |\,\partial_{|\vp|}^a (\pfp[\z;\vp]+|\vp|) \, |
        &<&\frac\delta2\;,\;
        \label{eq:Polyd-def-pfp-1}\\
        \sup_{\z\in\I } \{ \, | \, \partial_{\z}  \pfp[\z;\vp]  \, |
        \; , \;
        |\, \partial_{\z}^a( E[\z;\vp]-\z )  \,| \; , \;
        |\,\partial_{|\vp|}   E[\z;\vp] \,| \, \}
        &<&\eta
        \;\; \Big\} \;, \;\;\;\;
        \label{eq:Polyd-def-Ez-1}
\eeqn
for parameters
\eqn
    0 \; \; \leq \; \; |\vp| &<& \puppbd
    \nonumber\\
    0 \; \; \leq \; \; \e \; \leq \; \eta&\ll& 1
    \nonumber\\
    0 \; \; \leq \; \;\delta &\ll&1
    \nonumber\\
    0 \; \; \leq \; \; \lTnl &\leq&    \frac12 \;,
    \label{eq:Polpar-choice-1}
\eeqn
with
\eqn
    \xi \; = \; \frac{1}{10}
\eeqn
fixed. The constant $K_\Theta >2$ only depends on the
smooth cutoff function $\Theta$ introduced in (\ref{Thetadef}), and is
determined in (\ref{eq:F0-piop0-est-1}) below.
%
%
The parameter $\e$ measures the size of the projection of the polydisc to a codimension 3 subspace of
irrelevant perturbations, and is referred to
as an {\em irrelevant parameter}. On the other hand, $\delta$ and $\eta$ measure
the projection of the polydisc to a dimension 3 subspace of operators which are strictly marginal
in the limit $\sig\searrow0$, and are therefore referred to as {\em marginal parameters}.

We remark that $\h\in\Polyd\Polpar$ implies that
\eqn
        \|T- T^{(\vp;\lTnl)} \|_{\Tspace} &<&K_\Theta' \, \delta \;,
\eeqn
where the constant $K_\Theta'$ only depends on $\Theta$. This is discussed in detail
in Section {\ref{subsubsec:T-bounds-1}} below.

Accordingly, one can verify that
\eqnn
        \Big\{\h\in\Hspace_{\geq0}\Big|\,\|\h-\h_0^{(\vp;\lTnl)}\|_{\xi}\leq\eta\Big\}
        &\subseteq&\Polyd\Polpar
        \\
        &\subseteq&
        \Big\{\h\in\Hspace_{\geq0}\Big|\,\|\h-\h_0^{(\vp;\lTnl)}\|_{\xi}\leq
        2\delta+2\eta\Big\} \;,
\eeqnn
see  \cite{bcfs1}.
Hence, $\Polyd\Polpar$ is comparable to an $(\delta,\e,\eta)$-ball around
$\h_0^{(\vp;\lTnl)}$.

\begin{lemma}
\label{Ezlemma}
Let $0<\xi<\xibd $, $\sig>0$ and $0<\rho<\frac{1}{2}$. Then,
\eqn
        \Ie_{\frac{\rho}{200}}\subseteq {\mathcal U}[\h]\subseteq \Ie_{\frac{3\rho}{\Iconsttw}}
        \label{eq:Ezlemma-eq-1}
\eeqn
for all $\h\in\Polyd\Polpar$ with $\eta<\frac{\rho}{200}$, and
\eqn
        |\, \rho \, \hfp[\z;\vp] \, \partial_{\z}\Ez[\z] \, - \,1 \,| \; \leq \; 2 \, \eta \;,
        \label{eq:Ezlemma-eq-2}
\eeqn
for all $\z\in{\mathcal U}[\h]$. Then, $\Ez:{\mathcal U}[\h]\rightarrow
\I$ is a bijection.
\end{lemma}

\prf
By  definition of $\Polyd\Polpar$, we have   $|E[\z]-\z|<\eta$, and since
$\z\in{\mathcal U}[\h]=\{\z\in\I\,|\,|E[\z]|<\frac{\rho}{\Iconst}\}$, one infers
that
\eqn
        \Big|\,|\z|-|E[\z]|\,\Big| \; < \;|E[\z]-\z| \; < \; \eta\;.
\eeqn
Hence, (\ref{eq:Ezlemma-eq-1}) holds for $\eta<\frac{\rho}{200}$.

To prove (\ref{eq:Ezlemma-eq-2}), we note that
\eqn
        \sup_{\z\in{\mathcal U}[\h]}|\partial_{\z}(E[\z]-\z)|
        \; \leq \; \sup_{\z\in \Ie_{\frac{3\rho}{\Iconsttw} }}|\partial_{\z}(E[\z]-\z)|
        \; \leq \; \eta
        \label{eq:der-E-z-bds-2}
\eeqn
from the definition of $\Polyd\Polpar$.
Using Proposition {\ref{Polydlemma}} below, we find
\eqn
        |\,\rho\,\hfp[\z;\vp]\,\partial_{\z} \, \Ez[\z] \, - \, 1 \,| &\leq&
        \Big|\frac{\partial_{\z}\hfp[\z;\vp]}{\hfp[\z;\vp]}\Big|\,
        |E[\z]| + |\partial_{\z}E[\z]-1|
        \nonumber\\
        &\leq& c\eta^2 \, + \, \eta \; < \; 2 \, \eta
\eeqn
for $\eta$ sufficiently small.
\endprf


\begin{proposition}\label{Polydlemma}
Assume that $0\leq|\vp|<\puppbd$, $0< \rho< \frac{1}{K_\Theta} $,
$0<\xi<\xibd $, and $\h\in\Polyd\Polpar$.
Then,
\eqn
        (H[\h[\z;\vp]],\hfp H_f)\in\Fpairs(\H_{red},\chi_\rho[H_f])
    \label{eq:ren-F-pair-1}
\eeqn
for $\z\in\I$.
That is, $(H[\h[\z;\vp]],\hfp H_f)$ defines a Feshbach pair corresponding to $\chi_\rho[H_f]$,
for all $\hfp\in1+\Ihfp$, and all $\z\in{\mathcal U}[\h]$.

Moreover, there is a unique solution $\hfp[\z;\vp]$ of
\eqn
        \hfp[\z;\vp]=\Bra\vac\,,\,\partial_{H_f}F_{\chi_\rho[H_f]}(H[\h[\z;\vp]],\hfp[\z;\vp]H_f)\vac\Ket
        \label{eq:hfp-def-2}
\eeqn
which satisfies
\eqn
        | \, \hfp[\z;\vp] \, - \, 1 \, |&<&\frac{c\eta^2}{\rho^3}
        \nonumber\\
        | \, \partial_{\z} \, \hfp[\z;\vp] \, | \,,\,
        | \, \partial_{|\vp|} \, \hfp[\z;\vp] \, |&<&\frac{c\eta^2}{\rho^3}
    \label{eq:hfp-der-1}
\eeqn
and in particular,
\eqn
        \Big|\frac{\partial_{|\vp|}\hfp[\z;\vp]}{\hfp[\z;\vp]}\Big|&<&\frac{c\eta^2}{\rho^3} \;.
\eeqn
The constants  are independent of $\rho$, $\eta$.
\end{proposition}

\prf
To verify (\ref{eq:ren-F-pair-1})
for $\z\in\I$, $|\vp|<\puppbd$, and all $\hfp\in1+\Ihfp$,
one can straightforwardly adopt the corresponding results from  \cite{bcfs2}.

To prove that (\ref{eq:hfp-def-2}) has a unique solution in $1+\Ihfp$, we note first that
\eqn
        \;\;\;
        F_{\chi_\rho[H_f]}(H[\h[\z;\vp]],\hfp[\z;\vp]H_f)
        \; = \; E[\z]\chi_\rho^2[H_f] \, + \, (I) \, + \, (II)
\eeqn
with
\eqn
        (I)&:=&\hfp[\z;\vp]H_f \, + \, \chi_\rho^2 [H_f]\, \piop_\rho \,
        \big(T[\z;\opp;\vp] \, - \, \hfp[\z;\vp]H_f\big)
\eeqn
and
\eqn
        (II)&:=&\chi_\rho[H_f]\piop_{\rho}  \, W \, \piop_{\rho} \, \chi_\rho[H_f]
       \\
        &&\hspace{1cm}
        - \, \chi_\rho[H_f]\piop_{\rho}  \, W \, \bar\chi_\rho[H_f]
        \, \bar R[\h[\z;\vp]] \, \bar\chi_\rho[H_f] \,W  \, \piop_{\rho}  \, \chi_\rho[H_f] \;,
        \nonumber
\eeqn
where we introduce the abbreviated notations
\eqn
    W  &\equiv&W[\z;\vp]\;\equiv\; W[\h[\z;\vp]] \; = \; \sum_{M+N\geq1}W_{M,N}[\h[\z;\vp]]
    \nonumber\\
    \piop_\rho&\equiv&\piop_\rho[\z;\opp;\vp] \; \equiv \;
    \piop_{\chi_\rho}(T[\z;\opp;\vp] \, , \, \hfp[\z;\vp]H_f)  \;.
    \label{piop-def-1}
\eeqn
For $(I)$, we note that
\eqn
        &&T[\z;\opp;\vp]-\hfp[\z;\vp]H_f
        \; = \; (1-\hfp[\z;\vp])H_f \, + \, T'[\z;\opp;\vp] \;,
\eeqn
where
\eqn
        \bra \, \vac \, , \, \partial_{H_f}T'[\z;\opp;\vp] \, \vac \, \ket \; = \; 0
\eeqn
(see the definition of $T$ in (\ref{eq:T-def-structure-1})). Therefore,
\eqn
        \bra \, \vac \, , \, \partial_{H_f}(I) \, \vac \, \ket \; = \; 1  \;.
\eeqn
Next, we consider $(II)$.
Using
\eqn
        \partial_{H_f}\chi_\rho[H_f] \, \vac &= & 0
        \nonumber\\
        \partial_{H_f}\piop_\rho[\z;\opp;\vp] \, \vac & = & 0
        \nonumber\\
        \bra\vac \, , \, \partial_{H_f}W[\z;\vp] \, \vac\ket & = & 0 \;,
\eeqn
we get
\begin{align}
        \bra \, \vac \, , \, \partial_{H_f}(II)\vac \, \ket \; = \; - \, \Bra \, \vac \, , \, \partial_{H_f}
        \big(W \bar\chi_\rho[H_f]
        \bar R[\h[\z;\vp]]\bar\chi_\rho[H_f]W\big)\vac \, \Ket \;.
        \label{eq:part-Hf-II-id-1}
\end{align}
From Lemma {\ref{lm:res-aux-bds}} below, we find
\eqn
        \| \partial_{H_f}^a W[\z;\vp] \|_{op}&\leq&\| \h_{\geq1} \|_\xi
        \; < \; \eta \, + \, \e\; < \; 2\eta
        \nonumber\\
        \| \partial_{H_f}^a\bar R[\h[\z;\vp]] \|_{op}&\leq&\frac{C_\Theta}{\rho^{1+a}}
        \nonumber\\
        \| \partial_{H_f}\chi_\rho[H_f] \|_{op}&\leq&\frac{C_\Theta}{\rho}
        \label{eq:hfp-aux-est-1}
\eeqn
for $a=0,1$. Indeed, let
\eqn
        \bar R_0[\z;\opp;\vp] \; = \; \big(\hfp[\z;\vp]H_f \, + \,
        \chi_\rho^2[H_f]\widetilde T [\z;\opp;\vp]\big)^{-1}
        \;,
        \label{bar-R-0-def-1}
\eeqn
denote  the free resolvent on $\Ran(\bar\chi_\rho[H_f])$,
and $\widetilde T[\z;\opp;\vp]=T[\z;\opp;\vp]-\hfp[\z;\vp]H_f$.
From the resolvent identity
\begin{align}
    \bar R[\h[\z;\vp]] \; = \; \bar R_0[\z;\opp;\vp]
    \, - \, \bar R_0[\z;\opp;\vp] \bar\chi_\rho\chi_1[H_f] W \chi_1\bar\chi_\rho[H_f] \bar R[\h[\z;\vp]]
\end{align}
and Lemma {\ref{lm:res-aux-bds}}, we find
\eqn
    \|\bar R\|_{op} \; \leq \; (1-\|\bar R_0\|_{op})^{-1}\|W\|_{op}\|\bar R_0\|_{op}
    \; \leq \; \frac{C_\Theta}{\rho} \;.
\eeqn
Moreover,
\eqn
    \| \partial_{H_f}\bar R \|_{op} &\leq&
    \big( \|\partial_{H_f}\bar R_0\|_{op}\,+\,\|\partial_{H_f} W\|_{op} \big)
    \sum_{L\geq1} L  \big( \|W\|_{op} \|\bar R_0\|_{op} \big)^{L-1}
    \nonumber\\
    &\leq&\Big(\frac{C_\Theta}{\rho^2}+\eta\Big) \sum_{L\geq1}    L
    \Big(\frac{C_\Theta\eta}{\rho}\Big)^{L-1}
    \; \leq \; \frac{2C_\Theta }{\rho^2}
    \label{eq:derHf-barR-est-1}
\eeqn
by Lemma {\ref{lm:res-aux-bds}}, and $\eta\ll\rho$.

Consequently, one finds
\eqn
        | \, \hfp[\z;\vp] \, - \, 1 \, | \; = \; | \, \bra \, \vac \, ,
        \, \partial_{H_f}(II) \, \vac \, \ket \, |
        \; \leq \; \frac{c\eta^2}{\rho^2} \;
\eeqn
for a constant independent of $\rho$.
This implies that $r.h.s.\; of\; (\ref{eq:hfp-def-2})=1+O(\eta^2)$ for $\hfp[\z;\vp]\in1+\Ihfp$.
Consequently, there exists a solution of (\ref{eq:hfp-def-2}) in $1+\Ihfp$.

To prove uniqueness, we note that only
$\bar R[\h[\z;\vp]]$ in (\ref{eq:part-Hf-II-id-1}) depends on $\hfp[\z;\vp]$.
Similarly as in (\ref{eq:derHf-barR-est-1}), one finds
\eqn
        \| \partial_{H_f}^a \, \partial_{\hfp[\z;\vp]}^b \, \bar R[\h[\z;\vp]] \|_{op}
        \; \leq \;
        \frac{10 C_\Theta}{\rho^{3}}
        \;\;\;,\;\;\; a,b=0,1 \;,
        \label{eq:derHf-barR-est-2}
\eeqn
and a straightforward calculation shows that
\begin{align}
        \sup_{\alpha\in1+\Ihfp}\Big|\partial_\alpha\Bra\vac\,,\,
        F_{\chi_\rho[H_f]}\big(H[\h[\z;\vp]] \, , \, \alpha H_f\big)
        \vac\Ket\Big|\;<\;\frac{c\eta^2}{\rho^3} \;  \ll \; 1\;
        \label{eq:deralpha-hfp-aux-1}
\end{align}
for a constant $c$ independent of $\rho$.
This implies that (\ref{eq:hfp-def-2})
has a unique solution.

The estimates in (\ref{eq:hfp-der-1}) are obtained from
\eqn
    \partial_{\z} \, \hfp[\z;\vp]&=&\bra \, \vac \, , \,
    \partial_{\z} \, \partial_{H_f}(II) \, \vac \, \ket
    \nonumber\\
    \partial_{|\vp|} \, \hfp[\z;\vp]&=&\bra \, \vac \, , \,
    \partial_{|\vp|} \, \partial_{H_f}(II) \, \vac \, \ket \;,
\eeqn
and a straightforward calculation using
\eqn
        \| \partial_{\z}^a \, \partial_{|\vp|}^b \, \partial_{H_f}^c \, W \|_{op}
        &\leq&c \|\h_{\geq1}\|_\xi \;<\;c \, ( \eta+\e)\; < \; c' \, \eta
        \nonumber\\
        \| \partial_{\z}^a \, \partial_{|\vp|}^b \, \partial_{H_f}^c \, \bar R[\h[\z;\vp]] \|_{op}&\leq&
        \frac{ c}{\rho^3}
        \nonumber\\
        \| \partial_{\z}^a \, \partial_{|\vp|}^b \, \partial_{H_f}^c \, \chi_\rho[H_f] \|_{op}&=&0
    \label{eq:alph-aux-est-1}
\eeqn
for $a+b=1$ and $c=0,1$,
similarly as in  (\ref{eq:derHf-barR-est-1}) and (\ref{eq:derHf-barR-est-2}).
%
\endprf

\begin{lemma}
\label{lm:res-aux-bds}
Assume that $\h\in\Polyd\Polpar$ and $\z\in\I$.
There is a constant $C_\Theta$ only depending on the
smooth cutoff function $\Theta$ in (\ref{Thetadef}) such that
\begin{align}
    \|\partial_{\z}\bar R_0\|_{op}
    +\sum_{0\leq|\ua|\leq2}\rho^{1+|\ua|}\|\partial_{\opp}^{\ua}\bar R_0\|_{op}
    +\sum_{0\leq |\ua|\leq 1}\rho^{2+|\ua|} \|\partial_{|\vp|}\partial_{\opp}^{\ua}\bar R_0\|_{op}
    \; \leq \; C_\Theta \;.
    \label{eq:barR0-norm-est-aux-1}
\end{align}
Moreover,
\begin{align}
    \|\partial_{\z}W[\h]\|_{op}
    +\sum_{0\leq|\ua|\leq2} \|\partial_{\opp}^{\ua}W[\h]\|_{op}
        +\sum_{0\leq |\ua|\leq 1}  \|\partial_{|\vp|}\partial_{\opp}^{\ua}W[\h]\|_{op}
    \; \leq \; \eta+\e \;.
    \label{eq:W-norm-est-aux-1}
\end{align}
\end{lemma}

\prf
For $\h\in\Polyd\Polpar$, we have
\eqn
    |\bar R_0[\z;\opp]| \; < \; c \, H_f^{-1}
\eeqn
on $\Ran(\bar\chi_\rho[H_f])$ (see \cite{bcfs2}), and thus in particular
$\|\bar R_0\|_{op}\leq \frac{c}{\rho}$.
The estimate (\ref{eq:barR0-norm-est-aux-1}) follows from the fact that its left hand side can be
bounded by
\eqn
    {\rm l.h.s.\;of\;}(\ref{eq:barR0-norm-est-aux-1})
    &\leq&\Big(\sum_{0\leq|\ua|\leq2}\rho^{1+|\ua|}\| \bar R_0\|_{op}^{1+|\ua|}
        +\sum_{0\leq |\ua|\leq 1}\rho^{2+|\ua|} \| \bar R_0\|_{op}^{2+|\ua|}\Big)\|T\|_{\Tspace}
        \nonumber\\
    &\leq&c_\Theta \| T\|_{\Tspace}
\eeqn
where the constant $c_\Theta$ only depends on the smooth cutoff function $\Theta$, and where
$\|T\|_{\Tspace}<c$ follows from the definition of $\Polyd\Polpar$.

The estimate (\ref{eq:W-norm-est-aux-1}) is an immediate consequence of
Lemma {\ref{lm:H-embedd-Hspace-1}} and  the definition of $\Polyd\Polpar$.
\endprf

\subsection{Generalized Wick ordering}

The next step in determining $\widehat\h=\ren[\h]$,
consists of finding the {\em generalized Wick ordered} normal
form of the right hand side of (\ref{renHh-def-1})
(we suppress $\vp$ in the notation).

We note that
\eqn
        \| \piop_\rho[\z;\opp] \|_{op}&\leq&c \;,
\eeqn
where the constant is independent of $\rho$, and that
\eqn
        \| W[\z] \|_{op}&\leq&\eta \, + \, \e \; < \; 2\eta
        \label{TWbounds}
\eeqn
for $\h\in\Polyd\Polpar$ (recalling that $\e<\eta$ by (\ref{eq:Polpar-choice-1})).

Recalling the expression for the smooth Feshbach map given in Lemma {\ref{piopdeflemma}},
the resolvent expansion in powers of $W[\z]$ yields
\eqn
        F_{\chi_\rho[H_f]}\big(H[\h[\z]],\hfp[\z]H_f\big)
        \; = \; E[\z]\chi_\rho^2[H_f] \, + \, A_0 \, + \, \sum_{L=1}^\infty (-1)^{L-1}  A_L
        \label{resolvexp}
\eeqn
where
\eqn
        A_0 \; := \;\hfp[\z]H_f \, + \,
        \chi_\rho^2[H_f] \, \piop_{\rho}[\z;\opp] \, \big(T[\z;\opp]-\hfp[\z] H_f \big)
\eeqn
and
\eqn
        A_L & := &
        \chi_\rho[H_f] \, \piop_{\rho}[\z;\opp] \;
        \nonumber\\
        &&\hspace{1cm}
        \Big[ W[\h[\z]]\chi_1^2 [H_f]   \;
        \bar\chi_\rho^2[H_f]\bar R_0[\z,\opp] \Big]^{L-1}   \;
        \nonumber\\
        &&\hspace{4cm}
        W[\h[\z]] \piop_{\rho}[\z;\opp]\, \chi_\rho[H_f] \;.
\eeqn
From (\ref{TWbounds}),
\eqn
        \|A_L\|_{op} \; < \; C_\Theta^L \, \rho^{-L+1} \, \eta^L \;.
\eeqn
Hence, the series $\sum_{L=1}^\infty (-1)^{L-1}  A_L$
is norm convergent when $\eta$ is sufficiently small.

We introduce the operators
\eqn
        &&W_{p,q}^{m,n}[\h\big|\spvar;K^{(m+p,n+q)}]
        \;:=\; P_{red}\int_{B_1^{p+q}}
        d\mu_{\sig}(Q^{(p,q)}) a^*(Q^{(p)})
        \nonumber\\
        &&\hspace{3cm}
        w_{m+p,n+q}[\opp+\spvar;Q^{(p)},K^{(m)};\widetilde Q^{(q)},\widetilde K^{(n)}]
        a(\widetilde Q^{(q)})P_{red} \;.
        \label{Wmnpqdef}
\eeqn
The generalized Wick ordered form of the $L$-th term in the
resolvent expansion (\ref{resolvexp}) is given as follows.

\begin{lemma}
\label{Wickorderthm}
For $\h=(w_{M,N})_{M+N\geq1}\in\Hspace_{\geq1}^\sharp$, let
$W_{M,N}:=W_{M,N}[w_{M,N}]$, $W=\sum_{M+N\geq1}W_{M,N}$, and let $F_0,\dots,F_L\in
\Wspace_{0,0}$. Moreover, let $S_M$ denote the $M$-th symmetric group. Then,
\eqnn
        F_0 W F_1 W\cdots W F_{L-1} W F_L = H[\widetilde\h] \;,
\eeqnn
where $\widetilde \h = (\widetilde w_{M,N})_{M+N\geq0}\in \Hspace_{\geq0}^\sharp$
is determined by the symmetrization with respect to $K^{(M)}$ and $\widetilde K^{(N)}$,
\eqn
        \widetilde w_{M,N}[\spvar;K^{(M,N)}]
        ={\rm Sym}_{M,N}\widetilde w_{M,N}'[\spvar;K^{(M,N)}] \;,
\eeqn
with
\eqnn
        &&{\rm Sym}_{M,N} w_{M,N}[\spvar;K^{(M,N)}]
        \nonumber\\
        &&\hspace{1cm}\;=\;
        \frac{1}{M!N!}\sum_{\pi\in S_M}\sum_{\widetilde\pi\in S_N}
        w_{M,N}[\spvar;K_{\pi(1)},\dots,K_{\pi(M)};\widetilde K_{\widetilde\pi(1)},
        \dots,\widetilde K_{\widetilde\pi(N)}]\;,
        \nonumber
\eeqnn
and
\eqn
        \lefteqn{
        \widetilde w_{M,N}'[\spvar;K^{(M,N)}]
        \; = \;
        \sum_{ m_1+\cdots+m_L=M \atop n_1+\cdots+n_L=N}
        \sum_{ p_1,q_1,\dots,p_L,q_L \atop m_\ell+p_\ell+n_\ell+q_\ell\geq1}
        \prod_{\ell=1}^L
        {m_\ell+p_\ell\choose p_\ell}
        {n_\ell+q_\ell\choose q_\ell}
        }
        \nonumber\\
        &&\Bra \, \vac \, , \,F_0[\spvar+\widetilde\spvar_0]\,
        \widetilde W_1[\spvar+\spvar_1;K_1^{(m_1,n_1)}]
        \nonumber\\
        &&\hspace{1cm}F_1[\opp+\spvar+\widetilde\spvar_1;
        K_2^{(m_2,n_2)}]
        \cdots\cdots F_{L-1}[\opp+\spvar+\widetilde\spvar_{L-1}]
        \\
        &&\hspace{4cm}
        \widetilde W_L[\spvar+\spvar_L;K_L^{(m_l,n_L)}]
        F_L[\spvar+\widetilde\spvar_L]\, \vac\,\Ket \;.
        \nonumber
\eeqn
\end{lemma}

Here we are using the definitions
\eqn
        \widetilde W_\ell[\spvar+\spvar_\ell;K^{(m_\ell,n_\ell)}_\ell]
        &:=&W^{m_\ell,n_\ell}_{p_\ell,q_\ell}[\h\big|\spvar+\spvar_\ell;
        K^{(m_\ell,n_\ell)}_\ell] \;,
    \label{eq:tilde-W-def-1}
\eeqn
\eqnn
        K^{(M,N)} \; = \; (K_1^{(m_1,n_1)},\dots,K_L^{(m_L,n_L)})
        & , &
        K_\ell^{(m_\ell,n_\ell)} \; := \; (K_\ell^{(m_\ell)},\widetilde K_\ell^{(n_\ell)}) \;,
\eeqnn
and
\eqn
        \spvar_\ell&:=& \Sigma\big[\tuk_1^{(n_1)}\big]+\cdots+
        \Sigma\big[\tuk_{\ell-1}^{(n_{\ell-1})}\big]
        +\Sigma\big[\uk_{\ell+1}^{(m_{\ell+1})}\big]+\cdots
        +\Sigma\big[\uk_L^{(m_L)}\big]
        \nonumber\\
        \widetilde \spvar_\ell&:=&\Sigma\big[\tuk_1^{(n_1)}\big]+\cdots+
        \Sigma\big[\tuk_\ell^{(n_\ell)}\big]
        +\Sigma\big[\uk_{\ell+1}^{(m_{\ell+1})}\big]+\cdots+\Sigma\big[\uk_L^{(m_L)}\big]
        \;,
    \label{eq:tilde-X-def-1}
\eeqn
where $\Sigma[\uk_j^{(n_j)}]$ is defined in (\ref{multiinddef}).

Next, we apply rescaling, and transform the spectral parameter, thus obtaining
\eqn
        H[\widehat\h[\zet]]&=&\renop[H[\h[\z]]]
        \nonumber\\
        &=&
        \resc(F_{\chi_\rho}[H_f](H[\h[\z]],\hfp[\z]H_f))
        \label{hath-renopH-id-1}
\eeqn
(see (\ref{renHh-def-1})) with $\z=\Ez^{-1}[\zet]$.

The renormalized generalized Wick kernels $\widehat \h[\zet]$ have the following
explicit form.

\begin{lemma}
\label{hatwMNformalseriesthm}
Let $\zet\in \I$, and $\z:=\Ez^{-1}[\zet]\in{\mathcal U}[\h]$.
Then, one obtains $\widehat\h=(\widehat\h_{M,N})_{M+N\geq0}$ from (\ref{hath-renopH-id-1})  with
\eqn
        \lefteqn{
        \widehat w_{M,N}[\zet;\spvar;K^{(M,N)}]\; =\;
        \rho^{M+N-1}\, \frac{1}{\hfp[\z]} \,
        {\rm Sym}_{M,N}\sum_{L=1}^\infty
        (-1)^{L-1}
        }
        \nonumber\\
        && \sum_{m_1+\cdots+m_L=M \atop n_1+\cdots n_L=N}
        \sum_{p_1,q_1,\dots,p_L,q_L: \atop m_\ell+p_\ell+n_\ell+q_\ell\geq1}
        \prod_{\ell=1}^L
        {m_\ell+p_\ell\choose
                p_\ell}
        {n_\ell+q_\ell\choose
                q_\ell}
        \nonumber\\
        &&\hspace{0.5cm}
        \Bra\,\vac\,,\,
        \piop_{\rho}[\z;\opp+\rho (\spvar+\widetilde\spvar_0)]
        \widetilde W_1[\z;\rho(\spvar+\spvar_1);\rho K^{(m_1,n_1)}_1]
        \nonumber\\
        &&\hspace{1cm} \;\;\;\;
        (\bar\chi_\rho^2\chi_1^2 \bar R_0)[\z;\opp+\rho(\spvar+\widetilde\spvar_1)]
        \widetilde W_2[\z;\rho(\spvar+\spvar_2);\rho K^{(m_2,n_2)}_2]
        \\
        &&\hspace{1.5cm} \;\;\;\;
        (\bar\chi_\rho^2\chi_1^2\bar R_0)[\z;\opp+\rho(\spvar+\widetilde\spvar_2)]
        \cdots\cdots
        \nonumber\\
        &&\hspace{2cm} \hspace{3.5cm}\cdots\cdots
        (\bar\chi_\rho^2\chi_1^2\bar R_0)
        [\z;\opp+\rho(\spvar+\widetilde\spvar_{L-1})]
        \nonumber\\
        &&\hspace{2.5cm} \;\;\;\;
        \widetilde W_L[\z;\rho(\spvar+\spvar_L);\rho K_L^{(m_L,n_L)}]
        \piop_{\rho}[\z;\opp+\rho (\spvar+\widetilde\spvar_L)]\, \vac\,\Ket
        \nonumber
\eeqn
for $M+N\geq1$, and
\eqn
        \widehat w_{0,0}[\zet;\spvar]&=&\frac{1}{\hfp[\Ez^{-1}[\zet]]}
        \Big\{\ren[E[\,\cdot\,]\oplus w_{0,0}\oplus\unull_1]
       \\
        &+&\rho^{-1}\sum_{L=2}^\infty(-1)^{L-1}
        \sum_{p_1+q_1\geq1}\cdots\sum_{p_L+q_L\geq1}\piop_{\rho}^2[\z;\spvar]
        \nonumber\\
        && \; \Bra \,\vac\,,\, W_{p_1,q_1}[\h[\z]\big|\rho\spvar]
        (\bar\chi_\rho^2\chi_1^2\bar R_0)[\z;\opp+\rho\spvar]
        W_{p_2,q_2}[\h[\z]\big|\rho\spvar]
        \nonumber\\
        &&\hspace{3cm}
        \cdots (\bar\chi_\rho^2\chi_1^2\bar R_0)[\z;\opp+\rho\spvar]
        W_{p_L,q_L}[\h[\z]\big|\rho\spvar]
        \, \vac\,\Ket \Big\}
        \nonumber
\eeqn
for $M=N=0$.
\end{lemma}

The statements of Lemmata {\ref{hatwMNformalseriesthm}}  and {\ref{Wickorderthm}} are purely
algebraic, and the proofs can be adopted straightforwardly from \cite{bcfs1,bcfs2}.

\subsection{Spatial symmetries}

We shall require that the effective Hamiltonians possess the spatial symmetries
of the fiber Hamiltonian $H(\vp,\sig)$ in Section {\ref{rotreflinvsect}}.

\begin{definition}
\label{def:rotref-1}

Let the operators $U_h$ and
$U_{ref,\vp}$ be defined as in Section {\ref{rotreflinvsect}}.
We say that the effective Hamiltonian $H=H[\opp;\vp]\in\cB(\H_{red})$
in (\ref{effhamWickexpdef}) satisfies property \rotref
if
$$
        U_h \, H[\opp;R_h \vp] \, U_h^* \; = \; H[\opp;\vp]
$$
for all $h\in SU(2)$, and
$$
        U_{ref,\vp} \, H[\opp;-\vp] \, U_{ref,\vp}^* \; = \;  H[\opp;\vp] \;.
$$
(Invariance under rotations and
under reflections with respect to a plane
orthogonal to $\vp$.)
\end{definition}

\subsection{Soft photon sum rules}
\label{spsrsect}

The generalized Wick kernels $w_{M,N}$ are all mutually linked
by a hierarchy of non-perturbative identities,
referred to as the {\em soft photon sum rules}.
For the scalar model, which neglects the spin of the electron,
they were introduced in \cite{bcfs2}.
For the model including the spin of the electron, 
the generalized Wick kernels $w_{M,N}$ are ${\rm Mat}(2\times2,\C)$-valued
(for $M+N\geq1$; we recall that $w_{0,0}$ is scalar), but
the formal expressions for the identities remain unchanged.
For our construction, the quintessential property of the soft photon sum rules
is the fact that they are {\em preserved
by the renormalization map}, see Section {\ref{spsrproofsssect}}.

\begin{definition}
Let $\vn\in\R^3$, $|\vn|=1$, be an arbitrary unit vector, and let $\pol(\vn,\lambda)$
denote the photon polarization vector orthonormal to $\vn$
labeled by the polarization index $\lambda$.
We say that the sequence of generalized Wick kernels $\h\in\Hspace_{\geq0}$
satisfies the
{\em soft photon sum rules} \sbsr  if  the identity
\eqn
        &&\g  \pol(\vn,\lambda)\cdot\nabla_{\vX} \,
        w_{M,N}[\spvar;K^{(M,N)}]        
        \label{sbsr}\\
        &&\hspace{3cm}=\mu(\sig)(M+1)\lim_{\rvar\rightarrow0} w_{M+1,N}
        [\spvar;K^{(M+1,N)}]\Big|_{K_{M+1} =(\rvar \vn,\lambda)}
        \nonumber\\
        &&\hspace{3cm}=\mu(\sig)(N+1)\lim_{\rvar\rightarrow0}  w_{M,N+1}
        [\spvar;K^{(M,N+1)}]\Big|_{\widetilde K_{N+1}=(\rvar \vn,\lambda)}
        \nonumber
\eeqn
holds for all $M,N\geq0$, and any choice of $\vn$. The factor $\mu(\sig)$ is
given by  $\mu(\sig)=1$ if $\sig\leq1$, and satisfies $\mu(\rho^{-1}\sig)=\rho \mu(\sig)$
if $\sig>1$, for $0<\rho<1$.
\end{definition}

We remark that for $K\neq1$ in (\ref{cutsDef-1}), one would have
$\mu(\rho^{-1}\sig)=\rho^K \mu(\sig)$ instead.
The recursive application of (\ref{sbsr}), rooted at $M,N=0$, and in the order indicated by
\eqn
        \begin{array}{llllllll}
        &&&&&&\cdots\\
        &&&&&\nearrow& \\
        &&&&w_{2,0}&&\\
        &&&\nearrow&&\searrow&\\
        &&w_{1,0}&&&&\cdots\\
        &\nearrow&&\searrow&&\nearrow&\\
        w_{0,0}&&&&w_{1,1}&&\\
        &\searrow&&\nearrow&&\searrow&\\
        &&w_{0,1}&&&&\cdots\\
        &&&\searrow&&\nearrow&\\
        &&&&w_{0,2}&&\\
        &&&&&\searrow&\\
        &&&&&&\cdots\\
        \end{array}
\eeqn
links all generalized Wick kernels to one another.

In QED, the soft photon sum rules can be interpreted as a generalization of the
differential Ward-Takahashi identities.
A more detailed discussion is given in \cite{bcfs2}.

\subsection{Codimension-3 contractivity of $\ren$ on a polydisc}
\label{Codimcontrsubsect}

Let
\eqn
        \Polyds\Polpar:=\Big\{\h\in\Polyd\Polpar&\Big|&
        \h\;{\rm satisfies\;} \sbsrm
        \\
        &&
        {\rm \; and \; the \; symmetries \;}\rotrefm\Big\}
        \nonumber
\eeqn
denote the subset of elements in the polydisc $\Polyd\Polpar$ (defined
in Section {\ref{polyd-subsect-1-1}}),
which are {\em rotation and reflection symmetric} according to Definition {\ref{def:rotref-1}},
and which {\em satisfy the soft photon
sum rules} (\ref{sbsr}).

Our first main result states that the renormalization map is codimension-3 contractive
on sufficiently small polydiscs of this type.

\begin{theorem}
\label{thm:codim2contrthm}
The renormalization map $\ren$
is codimension-3 contractive on the polydisc
$\Polyds\Polpar$:

Assume that $0\leq|\vp|<\puppbd$, and let
$\rho$ and $\xi$ be given as in (\ref{eq:rho-xi-parmchoice-1}).
Then, there exist constants  $\e_0$, $\delta_0$  (small and independent of $\sig$)
such that for
all $0\leq\e  \leq \e_0 $ and $0\leq\delta \leq \delta_0+2\e_0 $,
\eqn
        \ren: \Polyds\Polpar\rightarrow
        \Polyds(\widehat\e,\widehat\delta,\widehat\eta,\widehat\lTnl,\widehat\sig) \;,
        \label{codim2contr}
\eeqn
where
\eqn
        \left.\begin{array}{rcl}
        \widehat\e&\leq&\frac{\e}{4} + \frac\eta4
        \\
        \widehat\delta&\leq&\delta+\frac\eta2
        \\
        \widehat\eta&=&\cetaex \, \g \, \xi^{-1} \, (1+|\vp|+\widehat\delta) \, + \, \frac{\eta}{2}
        \\
        \widehat\lTnl&=&\rho\lTnl
        \\
        \widehat\sig&=&\rho^{-1}\sig
        \end{array}\;\;\right\}\;{\rm if}\;\sig\leq1
        \label{eq:Polparhat-leq1}
\eeqn
and
\eqn
        \left.\begin{array}{rcl}
        \widehat\e&\leq&\frac{\e}{4}+\frac\eta4
        \\
        \widehat\delta&\leq&\delta+\frac\eta2
        \\
        \widehat\eta&=&\frac{\eta}{2}
        \\
        \widehat\lTnl&=&\rho\lTnl
        \\
        \widehat\sig&=&\rho^{-1}\sig
        \end{array}\;\;\right\}\;{\rm if}\; \sig>1\;.
        \label{eq:Polparhat-geq1}
\eeqn
The constant $C_\Theta$ is the same as in Lemma {\ref{lm:res-aux-bds}}.
\end{theorem}

The parameter $\e$ measures the projection of the polydisc along the codimension 3 subspace of
irrelevant interactions (which are contracted by a factor $\leq\frac12$ under
application of $\ren$), while $\delta$ and $\eta$ measure its projection
to a dimension 3 center manifold of marginal perturbations.
With every application of $\ren$, the infrared cutoff parametrized by $\sig$ is scaled by
a factor $\rho^{-1}$.
The interaction kernels $\h_1$ behave like strictly marginal operators, i.e. their size remains
constant under repeated applications of $\ren$,
as long as $\sig<1$, see Section {\ref{resc-def-subsubsect-1}}.
The main new techniques in this paper address this regime, i.e., $\sig\leq1$.
When $\sig>1$, the kernels $\h_1$
behave like irrelevant operators, i.e. their size converges to zero at an exponential
rate under repeated applications of $\ren$.
In this case, they can then be completely controlled with the results of \cite{bcfs2}.

\subsection{Strong induction argument}
\label{subsec:SInd-1}

The upper bounds provided by (\ref{eq:Polparhat-leq1}) are clearly insufficient to
control the growth of $\delta$ and $\eta$ under repeated
applications of the renormalization map.
The next main step in the analysis is to prove that nevertheless, the size of
$\delta$ and $\eta$ does not increase under any number
of applications of $\ren$. To achieve this result, we let $\delta_n$, $\eta_n$ denote
the constants in the above bounds after $n$ iterations of $\ren$, and invoke
the {\em strong induction principle}.
This means that for every step $n\rightarrow n+1$, we study the entire orbit $\h^{(k)}\in
\Polyds(\e_k,\delta_k,\eta_k,\lTnl_k,\sig_k)$, for $0\leq k\leq n$,
with initial condition $\h^{(0)}\in\Polyds(\e_0,\delta_0,\eta_0,\lTnl_0,\sig_0)$
(provided by the "first Feshbach decimation step", see Section {\ref{subsec:first-F-step-1}}
and \cite{bcfs2}, where $\sigzero$ denotes the initial infrared cutoff in the fiber Hamiltonian $\Hpszero$).

We make the following key
observations:
\begin{itemize}
\item
By (\ref{eq:Polparhat-leq1}), uniform boundedness of $\delta_n$ in $n$ automatically
implies uniform boundedness of $\eta_n$.
\item
After $n$ applications of the renormalization map, one arrives at $\sig_n=\rho^{-n}\sigzero$. Thus,
$\sig_n\leq1$ if $n\leq N(\sigzero)$, and $\sig_n>1$ if $n>N(\sigzero)$ for
\eqn
    N(\sigzero)=\left\lceil\frac{\log\frac{1}{\sigzero}}{\log\frac1\rho}\right\rceil \;.
    \label{eq:Nsigzero-def-1}
\eeqn
Hence,  (\ref{eq:Polparhat-leq1}) and (\ref{eq:Polparhat-geq1}) imply, under the condition that
$\delta_n$ is  uniformly bounded in $n$, that
the interaction $\h_1$ undergoes a transition from strictly marginal behavior to
irrelevant behavior at $n=N(\sigzero)$.
This means that the upper bound in $\|\h_1\|_\xi<\eta_n$ is essentially independent of
$n$ in the regime $n\leq N(\sigzero)$, but in the regime $n>N(\sigzero)$, $\eta_n$
decreases by a factor at least $\frac12$ under every application of $\ren$.
\item
In the regime $n>N(\sigzero)$, it can be easily inferred from the estimates  (\ref{eq:Polparhat-geq1}) in
Theorem {\ref{Codimcontrsubsect}}  that
$\delta_n<\delta_{N(\sigzero)}+2\eta_{N(\sigzero)}$  uniformly in $n$, and that
$\e_n$, $\eta_n<2^{(n-N(\sigzero))_+}\eta_{N(\sigzero)}$
converge to zero at a $\sigzero$-independent exponential rate as $n\rightarrow\infty$.
For these large values of $n$, it is not necessary to invoke Theorem {\ref{thm:strong-induct}}.
\end{itemize}


The key goal of this part of the analysis is to prove that $\delta_n$ is uniformly bounded in $n$
and $\sigzero$. The main result can be stated as follows.

\begin{theorem}
\label{thm:strong-induct}
Let $\sigzero\ll1$ (the infrared cutoff
in the fiber Hamiltonian $\Hpszero$) be arbitrary but fixed.
Invoking Theorem   {\ref{thm:codim2contrthm}}, we assume that
$\ren$ is codimension 3 contractive on the polydisc
$\Polyds(\e_0,\delta_0+2\eta_0,\eta_0,\lTnl_0,\sig_0)$ for
$\e_0<\eta_0<c\g\ll\rho^3$ sufficiently small, $\delta_0<c\gs$, and $\lTnl_0=\frac12$.

Let $\h^{(0)}\in\Polyds(\e_0,\delta_0,\eta_0,\lTnl_0,\sig_0)$.
By $\sind[n]$, we denote the strong induction assumption that
for $0\leq k\leq n$, one has
\eqn
        \h^{(k)}\in\Polyds(\e_k,\delta_k,\eta_k,\lTnl_k,\sig_k) \;,
\eeqn
where
\eqn
        \h^{(k)} \; = \; \ren[\h^{(k-1)}]\;\;\;\;\;{\rm for \; }1\leq k\leq n \;,
\eeqn
and
\eqn
        \e_k&\leq& \eta_k \; < \; c \, \g
        \nonumber\\
        \delta_k&\leq&\cdstrong\gs
        \nonumber\\
        \eta_k&\leq& \cetaextw \, \g \, \xi^{-1}  (1+|\vp|+\cdstrong\gs)
        \nonumber\\
        \lTnl_k&=&\rho^k\lTnl_0 \;\;\;,\;\;\;\lTnl_0=\frac12
        \nonumber\\
        \sig_k&=&\rho^{-k}\sigzero \;.
\eeqn
Then, for $\gs<\gs_0$ with $\gs_0$ sufficiently small (independent of $\sigzero$)
and any $n\geq0$,
$\sind[n]$ implies $\sind[n+1]$.
The constant $\cdstrong$  is independent of $n$, $\gs$, and $\sig_0$,
and is determined in Proposition {\ref{prop:sind-step-prf-1}}.
\end{theorem}

From Theorem {\ref{thm:strong-induct}}, we obtain the desired uniform bounds on
$\delta_n$ with respect to $n$, and also on $\e_n$, $\eta_n$.

Theorem {\ref{thm:strong-induct}} is the key tool that
allows us to establish pure marginality of the interaction.
For its proof, we use  (\ref{QHQ-id-basic-1}), which is
a version of the identity of Lemma {\ref{first-last-scale-lemma-1}}
in \cite{bcfs2}; it allows to "collapse" intermediate scales between
effective Hamiltonians on non-successive scales.

The marginal operators in the model considered here are given by
$H_f$ and $\pfp[\z;\vp]\Ppar$ in $T$ (see (\ref{eq:T-def-structure-2}) and (\ref{eq:T-def-structure-3})),
and $\h_1=(w_{1,0},w_{0,1})$ (counted as only one marginal direction because $w_{1,0}=w_{0,1}^*$).
The key application of the soft photon sum rules is given in the proof of
Theorem  {\ref{thm:codim2contrthm}}; they are used to relate
$\pfp[\z;\vp]$  to  $\h_1$ (by gauge invariance), whereby the number of independent marginal operators
is reduced from three to two.

By definition of the renormalization map, the coefficient of the operator $H_f$ in $T$
has the constant value $1$. It thus remains to prove that the size of $\pfp[\z;\vp]$ is independent of $n$.
A main difficulty here is that the marginal operators
$H_f$ and $\pfp[\z;\vp]\Ppar$ in $T$ are {\em not} related via gauge invariance,
and it is at this point where the identity (\ref{QHQ-id-basic-1}) mentioned above enters.

\section{Proof of Theorem {\ref{thm:codim2contrthm}}}
\label{sect:codim2-contr-1}

In this section, we prove the codimension-3 contractivity of $\ren$
asserted in Theorem {\ref{thm:codim2contrthm}}.
For details omitted here, we refer to the proof of Theorem 6.6 in \cite{bcfs2}.

\subsection{Wick ordering}
\label{gen-wick-prf-subsect-1}

We adopt the following notation from  \cite{bcfs2}.
For fixed $L\in\N$, let
\eqn
        \underline{m,p,n,q}:=(m_1,p_1,n_1,q_1,\dots,m_L,p_L,n_L,q_L)\in\N_0^{4L} \;
\eeqn
and
\eqn
        M \; := \; |\underline{m}| \; = \; m_1+\dots+m_L \; \; , \; \;
        N \; := \; |\underline{n}| \; = \; n_1+\dots+n_L \;,
\eeqn
and we recall the definitions (\ref{eq:tilde-W-def-1}) and (\ref{eq:tilde-X-def-1}).
We let
\eqn
        &&\widetilde V^{(L)}_{\underline{m,p,n,q}}\big[\h\,\big|\,\spvar;K^{(M,N)}\big]
        \label{eq:tildVLmpnq-def-1}
        \\
        &&\hspace{2cm}:=\;\Bra\, \vac \, , \,
        \prod_{\ell=1}^L \, \Big\{ \, \widetilde W_\ell
        [\z;\rho(\spvar+\spvar_\ell);\rho K^{(m_\ell,n_\ell)}_\ell]
        \, F_\ell[\spvar] \, \Big\} \, \vac \, \Ket \;,
        \nonumber
\eeqn
and
\eqn
        &&V^{(L)}_{\underline{m,p,n,q}}\big[\h\,\big|\,\spvar;K^{(M,N)}\big]
        \;:=\;F_0[\spvar]\,
        \widetilde V^{(L)}_{\underline{m,p,n,q}}\big[\h\,\big|\,\spvar;K^{(M,N)}\big]
        \label{VLmpnq-def-1}
\eeqn
where
\eqn
        F_0[\spvar] \; := \; \piop_{\rho}\big[ \, \z \, ; \, \rho( \spvar+\widetilde \spvar_{0}) \, \big]
        \;\;\;,\;\;\;
        F_L[\spvar] \; := \; \piop_{\rho}\big[ \, \z \, ; \, \rho( \spvar+\widetilde \spvar_L) \, \big]
        \label{F0FLdef}
\eeqn
and
\eqn
        F_\ell[\spvar] \; := \;
        \frac{(\bar\chi_\rho^2\chi_1^2)[H_f+\rho(X_0+\widetilde X_{\ell,0})]}
        { E[\z]+\hfp[\z]H_f+\rho(X_0+\widetilde X_{\ell,0})+
        \bar\chi_\rho^2\widetilde T[\z;\opp+\rho(\spvar+\widetilde \spvar_\ell)] }
        \nonumber\\
        \label{Felldef}
\eeqn
for $\ell=1,\dots,L-1$, with $T[\z;\spvar]= X_0+\chi_1^2[X_0]\widetilde T [\z;\spvar])$.

Then, for $\zet\in\I$ and $\z:=\Ez^{-1}[\zet]$,
\eqn
        \widehat w_{M,N}[\zet;\spvar;K^{(M,N)}] \; = \;
        \frac{1}{\hfp[\z;\vp]} \; \widetilde w_{M,N}[\zet;\spvar;K^{(M,N)}] \;,
\eeqn
where
\eqn
        &&\widetilde w_{M,N}[\zet;\spvar;K^{(M,N)}] \; = \;
        \rho^{M+N-1} \, \sum_{L=1}^\infty (-1)^{L-1}
        \sum_{m_1+\cdots+m_L=M \atop n_1+\cdots + n_L=N}
        \label{hatwMNdefform}
        \\
        &&\hspace{1cm}
        \sum_{p_1,q_1,\dots,p_L,q_L: \atop m_\ell+p_\ell+n_\ell+q_\ell\geq1}
        \Big[\prod_{\ell=1}^L \,
        {m_\ell+p_\ell \choose p_\ell}
        {n_\ell+q_\ell \choose q_\ell} \,\Big] \,
        V^{(L)}_{\underline{m,p,n,q}}\big[\h\,\big|\,\spvar;K^{(M,N)}\big] \;,
        \nonumber
\eeqn
where the factors $\rho^{M+N-1}$ are due to the rescaling transformation, see (\ref{rescwMNdef}).

\begin{lemma}
\label{VLboundlm}
For $L\geq1$ fixed, and $\underline{m,p,n,q}\in\N_0^{4L}$ with $|\underline{m}|=M$
and $|\underline{n}|=N$, one has
$V^{(L)}_{\underline{m,p,n,q}}\in\Wspace_{M,N}^\sharp$. Furthermore,
\eqn
        \|F_0\|_{\Tspace} \; , \; \|F_L\|_{\Tspace} &<&C_\Theta
        \label{eq:F-CThet-1}
\eeqn
and
\eqn
        \|F_{\ell}\|_{\Tspace} &<& \frac{C_\Theta}{\rho} \;,
        \label{eq:F-CThet-2}
\eeqn
where the constant $C_\Theta$ is the same as in Lemma {\ref{lm:res-aux-bds}}
(it depends only on the choice of the smooth cutoff function $\Theta$ in (\ref{Thetadef})).
Moreover,
\eqn
        &&\rho^{M+N-1} \,
        \| \partial_{|\vp|} \, \partial_{|\vk|}^a\, V^{(L)}_{\underline{m,p,n,q}} \|_{M,N}
        \; \; , \; \;
        \rho^{M+N-1} \,
        \| \partial_{\zet}^a \,  V^{(L)}_{\underline{m,p,n,q}} \|_{M,N}
       \\
        &&\hspace{0.5cm}\leq \; (L+1)^2 \, C_\Theta^{L+1} \, \rho^{M+N+1+a-L} \,
        \prod_{l=1}^L\frac{\| w_{m_l+p_l,n_l+q_l}[\z] \|_{m_\ell+p_\ell,n_\ell+q_\ell}^\sharp}
        {p_l^{p_l/2}q_l^{q_l/2}} \;,\;\;\;
        \nonumber
\eeqn
for $a=0,1$ and any $\vk\in k^{(M,N)}$. Furthermore,
\eqn
        &&\rho^{M+N-1} \,
        \| \partial_{\spvar}^{\ua} \, V^{(L)}_{\underline{m,p,n,q}} \|_{M,N}
        \label{partspvaraindVL}\\
        &&\hspace{0.5cm}\leq \; (L+1)^2 \, C_\Theta^{L+2} \,
        \rho^{ M+N +|\ua|-L} \,
        \prod_{l=1}^L\frac{\| w_{m_l+p_l,n_l+q_l}[\z] \|_{m_\ell+p_\ell,n_\ell+q_\ell}^\sharp}
        {p_l^{p_l/2}q_l^{q_l/2}} \;,\;\;\;
        \nonumber
\eeqn
for $1\leq|\ua|\leq2$.
For $|\ua|=1$,
\eqn
        &&\rho^{M+N-1}
        \| \partial_{|\vp|} \,
        \partial_{\spvar}^{\ua} \, V^{(L)}_{\underline{m,p,n,q}} \|_{M,N}
        \label{partppartspvaraindVL}\\
        &&\hspace{0.5cm}\leq \; (L+1)^2 \, C_\Theta^{L+2}
        \, \rho^{M+N+|\ua|-L} \,
        \prod_{l=1}^L\frac{\| w_{m_l+p_l,n_l+q_l}[\z] \|_{m_\ell+p_\ell,n_\ell+q_\ell}^\sharp}
        {p_l^{p_l/2}q_l^{q_l/2}} \;.\;\;\;
        \nonumber
\eeqn
Consequently,
\eqn
        &&\rho^{M+N-1} \| V^{(L)}_{\underline{m,p,n,q}} \|_{M,N}^\sharp
        \\
        &&\hspace{0.5cm}\leq \; (L+1)^2 \, C_\Theta^{L+2} \, \rho^{M+N-L} \,
        \prod_{l=1}^L\frac{\| w_{m_l+p_l,n_l+q_l}[\z] \|_{m_\ell+p_\ell,n_\ell+q_\ell}^\sharp}
        {p_l^{p_l/2}q_l^{q_l/2}} \;,\;\;\;
        \nonumber
\eeqn
using the convention $p^p=1$ for $p=0$.
\end{lemma}

\prf
We only demonstrate the argument for the term involving a derivative in $\zet$ because it
has no counterpart in \cite{bcfs2} (where derivatives in the spectral parameter are
controlled by analyticity). For the other cases, we refer to \cite{bcfs2}.

We have  for $\z=\Erho^{-1}[\zet]$
\eqn
        &&\partial_{\zet} V_{\underline{m,p,n,q}}^{(L)}[\h\,\big|\,\spvar;K^{(M,N)}]
        \nonumber\\
        &&\hspace{1cm} =\;\sum_{j=0}^L
        \Bra   \vac \,\Big|\,
        \Big[\prod_{\ell=1}^{j-1}
        F_{\ell-1}[\spvar]\widetilde W_{\ell}[\z;\rho(\spvar+\spvar_\ell);
        \rho K_\ell^{(m_\ell,n_\ell)}]
        \Big]
        \label{eq:parzet-V-term1-1}
       \\
        &&\hspace{2.5cm}
        \;\Big(\partial_{\zet} F_j[\spvar]\Big)
        \Big[\prod_{\ell=j+1}^L  \widetilde W_{\ell}[\z;\rho(\spvar+\spvar_\ell);
        \rho K_\ell^{(m_\ell,n_\ell)}]
        F_\ell[\spvar]\Big] \vac \, \Ket
        \nonumber\\
        &&\hspace{1.5cm}+\;\sum_{j=1}^L
        \Bra \,  \vac \,\Big|\, F_0[\spvar]
        \Big[\prod_{\ell=1}^{j-1}
        \widetilde W_{\ell}[\z;\rho(\spvar+\spvar_\ell);
        \rho K_\ell^{(m_\ell,n_\ell)}] F_{\ell}[\spvar]
        \Big]
        \nonumber\\
        &&\hspace{2.5cm}
        \; \Big(\partial_{\zet}
        \widetilde W_{j}[\z;\rho(\spvar+\spvar_\ell);
        \rho K_j^{(m_j,n_j)}]\Big)
        \label{eq:parzet-V-term1-2}\\
        &&\hspace{2.5cm}\;
        \Big[\prod_{\ell=j+1}^L
        F_\ell[\spvar]\widetilde W_{\ell}[\z;\rho(\spvar+\spvar_\ell);
        \rho K_\ell^{(m_\ell,n_\ell)}]\Big]
        F_L[\spvar] \,\vac \, \Ket \;.
        \nonumber
\eeqn
Using (\ref{eq:F-CThet-1}) and (\ref{eq:F-CThet-2}) to bound
$\|F_\ell\|_{op}$ and $\|\partial_{\z}F_\ell\|_{op}$,
\eqn
        \|(\ref{eq:parzet-V-term1-1})\|_{M,N}&\leq& \rho (1+c\eta)
        \sum_{j=0}^L \|\partial_{\z}  F_j\|_{op}
        \Big[\prod_{\ell=0\atop \ell\neq j}^L
        \|F_\ell[\spvar]\|_{op} \Big]
        \nonumber\\
        && \hspace{2.5cm}\;
        \prod_{\ell=1}^L
        \Big\|\widetilde W_\ell[\rho(\spvar+\spvar_\ell);\rho K_\ell^{(m_\ell,n_\ell)}]\Big\|_{op}
        \nonumber\\
        &\leq& (L+1)C_\Theta^{L+1}\rho^{-L+2}\prod_{\ell=1}^L
        \Big\|\widetilde W_\ell[\z;\rho(\spvar+\spvar_\ell);\rho K_\ell^{(m_\ell,n_\ell)}]\Big\|_{op}
\eeqn
and
\eqn
        \|(\ref{eq:parzet-V-term1-2})\|_{M,N}&\leq& \rho  \,
        \Big[\prod_{\ell=0}^L \|F_\ell[\spvar]\|_{op}\Big]
        \Big\{ \sum_{j=1}^L
        \Big\|\partial_{\z} \widetilde W_j
        [\z;\rho(\spvar+\spvar_j);\rho K_j^{(m_j,n_j)}]\Big\|_{op}
        \nonumber\\
        &&  \;\prod_{\ell=1 \atop \ell\neq j}^L
        \Big\|\widetilde W_\ell[\z;\rho(\spvar+\spvar_j);\rho K_j^{(m_j,n_j)}]\Big\|_{op}\Big\}
        \nonumber\\
        &\leq& L C_\Theta^{L+1} \rho^{-L+2}
        \Big\{ \sum_{j=1}^L
        \Big\|W_{p_j,q_j}^{m_j,n_j}
        \Big[\partial_{\z} \h[\z]\Big|\rho(\spvar+\spvar_j);
        \rho K_j^{(m_j,n_j)}\Big]\Big\|_{op}
        \nonumber\\
        && \;\prod_{\ell=1 \atop \ell\neq j}^L
        \Big\|\widetilde W_\ell[\z;\rho(\spvar+\spvar_\ell);
        \rho K_\ell^{(m_\ell,n_\ell)}]\Big\|_{op}\Big\} \;.
\eeqn
Here, we used that for $\z=\Erho^{-1}[\zet]$,
\eqn
        | \, \partial_{\zet}f[\z] \,| \; \leq \;
        \rho \, (1+c\eta) \, | \, (\partial_{\z}f)[\z] \, | \;,
\eeqn
see (\ref{eq:derp-Ezinv-2}) below.
The factor $(1+c\eta)<2$ has been absorbed into the definition of the
constant $C_\Theta$.

The remaining cases can be adapted straightforwardly from the proof of Lemma 7.1 in \cite{bcfs2}.
\endprf

\subsection{Preservation of the soft photon sum rules}
\label{spsrproofsssect}

\begin{lemma}
The renormalization map preserves the soft photon sum rules,
\eqn
    \ren:\sbsrm\mapsto{\bf SR}[\rho^{-1}\sig]
\eeqn
where \sbsr is defined in (\ref{sbsr}).
That is, given
$\h\in\Polyds\Polpar$, which satisfies \sbsr, it follows that $\widehat\h=\ren[\h]$
satisfies ${\bf SR}[\rho^{-1}\sig]$.

\end{lemma}

\prf
It is proved in \cite{bcfs2} for the scalar model (zero electron spin)
that the renormalization map $\ren$ preserves the soft photon sum rules.
The argument is purely algebraic, and it applies equally to the spin $\frac12$ model.
The fact that the generalized Wick kernels are here complex $2\times2$ matrices, and not
scalars, does not affect the proof.
\endprf

\subsection{Preservation of the symmetries}
\label{sec:Ren-symm-1}
In this section, we prove that the symmetries of the fiber Hamiltonian $H(\vp,\sig)$
described in Section {\ref{rotreflinvsect}} are inherited by the
effective Hamiltonians, in the sense of Definition {\ref{def:rotref-1}},
and preserved by the renormalization map.

\begin{lemma}
Assume that $\h\in\Polyds\Polpar$, and that
\eqn
    U H[\h[\z;R\vp]] U^* = H[\h[\z;\vp]] \;,
    \label{eq:UHU-inv-1}
\eeqn
where $U$ stands either for $U_h$ or for $U_{ref}$, and $R$ denotes either $R_h$ or $-\1$ in
the notation of Definition
{\ref{def:rotref-1}}.
Then,
\eqn
    U H[\widehat\h[\z;R\vp]] U^* = H[\widehat\h[\zet;\vp]]
\eeqn
for $\widehat \h=\ren[\h]$, with $\zet=\Ez[\z]$.
\end{lemma}

\prf
Let for brevity
\eqn
    \omega[\z;\vp] \; := \; H[\h[\z;R\vp]]\,-\,\hfp[\z;\vp]H_f \;,
\eeqn
where $\hfp[\z;\vp]$ is defined in (\ref{eq:hfp-def-1}), and
\eqn
    \bar R[\z;\vp] \; := \; (\hfp[\z;\vp]H_f+\bar\chi_\rho[H_f]\omega[\z;\vp]\bar\chi_\rho[H_f])^{-1}
\eeqn
on $\Ran[\bar\chi_\rho[H_f]]$.
From
\eqn
    U \, f[H_f] \, U^* \; = \; f[H_f] \;   ,
\eeqn
for any Borel function $f$, we find
\eqn
    \lefteqn{
    U F_{\chi_\rho[H_f]}(H[\h[\z;R\vp]]\,,\,\hfp[\z;R\vp]H_f)U^*
    }
    \nonumber\\
    &&=\;
    U\Big(\hfp[\z;R\vp]H_f+\chi_\rho[H_f]\omega[\z;R\vp]\chi_\rho[H_f]
    \nonumber\\
    &&\hspace{1cm}-\chi_\rho[H_f]\omega[\z;R\vp]
    \bar\chi_\rho[H_f]\bar R[R\vp]\bar\chi[H_f]\omega[\z;R\vp]\chi_\rho[H_f]\Big)U^*
    \nonumber\\
    &&=\; \hfp[\z;R\vp]H_f+\chi_\rho[H_f] U \omega[\z;R\vp] U^* \chi_\rho[H_f]
    \nonumber\\
    &&\hspace{1cm}-\chi_\rho[H_f] U \omega[\z;R\vp] U^*
    \bar\chi_\rho[H_f] U \bar R[\z;R\vp] U^* \bar\chi[H_f]U\omega[\z;R\vp]U^*\chi_\rho[H_f]
    \nonumber\\
    &&=\;
    F_{\chi_\rho[H_f]}(U H[\h[\z;R\vp]] U^* \,,\,\hfp[\z;R\vp]H_f) \;.
\eeqn
Therefore,
\eqn
    U \renop[H[\h[\z;R\vp]]\,] U^*
    &=&U\Ez\circ\resc[F_{\chi_\rho[H_f]}(H[\h[\z;R\vp]]\,,\,\hfp[\z;R\vp]H_f)]U^*
    \nonumber\\
    &=&\Ez\circ\resc[U\,F_{\chi_\rho[H_f]}(H[\h[\z;R\vp]]\,,\,\hfp[\z;R\vp]H_f)\,U^*]
    \nonumber\\
    &=&\Ez\circ\resc[ F_{\chi_\rho[H_f]}(U H[\h[\z;R\vp]] U^*\,,\,\hfp[\z;R\vp]H_f) ]
    \nonumber\\
    &=&\renop[ U H[\h[\z;R\vp]] U^*\,]\;,
\eeqn
which implies that given (\ref{eq:UHU-inv-1}),
\eqn
    U \renop[H[\h[\z;R\vp]]\,] U^* = \renop[ H[\h[\z;\vp]] \,]
\eeqn
or likewise,
\eqn
    U H[\ren[\h][\zet;R\vp]] U^* = H[\ren[\h][\zet; \vp]]
\eeqn
with $\zet=\Ez[\z]$.

This implies that $\ren$ preserves rotation and reflection symmetry.
\endprf

\subsection{Codimension three contractivity}
\label{codim2contrproofsubsubsect-1}

We now come to the core of the proof of Theorem {\ref{thm:codim2contrthm}},
and verify that
$$
    \ren:\Polyds\Polpar\rightarrow\Polyds\Polparhat
$$
with (\ref{eq:Polparhat-leq1}) and (\ref{eq:Polparhat-geq1}).
This implies that $\ren$ is contractive on a codimension
three subspace of $\Polyds\Polpar$ at a contraction rate which is {\em independent}
of $\sig$.


Our proof is organized as follows.

First, we show that from application of $\ren$, the kernels $\h_{\geq2}$ are
contracted by a factor $\leq\frac12$ by pure scaling, for $\rho$ sufficiently small.
This implies that they belong to a codimension-3 subspace of $\Polyds\Polpar$ of
irrelevant perturbations.

To control the {\em marginal} kernels $\h_1=(w_{0,1},w_{1,0})$,
we invoke the soft photon sum rules, and relate
$\h_1$ to the coefficient of the marginal
operator $\Ppar$ in the non-interacting Hamiltonian $T$.
Hereby, we can reduce the number of independent marginal operators by one,
and it remains to control the renormalization of marginal operators in $T$
under $\ren$. This is the main topic of Section {\ref{sect:sind-thm-prf-1}} below.

In \cite{bcfs2}, a similar argument has been used in the special case
$|\vp|=0$, to determine the renormalized electron
mass for $|\vp|=0$ in the limit $\sig\searrow0$.

We now give the detailed proof of Theorem {\ref{thm:codim2contrthm}}.

\subsubsection{Bounds on $\widehat E$ and $\widehat T$}
\label{subsubsec:T-bounds-1}

We begin with $M+N=0$, and first discuss the renormalization of
$E[\z]$ (see (\ref{eq:Polyd-def-Ez-1}) in the definition of the polydisc
$\Polyd\Polpar$).
Let
\eqn
        \z=\Ez^{-1}[\zet]\;\in\;\Ie_{\frac{3\rho}{\Iconsttw}}
        \;\;\;{\rm for} \;\;\;
        \zet\in\I \;,
\eeqn
where $\Ez[\z] = \frac{1}{\hfp[\z]\rho} E[\z]$ (see Lemma {\ref{Ezlemma}}).
Let us first prove
\eqn
        \sup_{\zet\in\I}\{ \,
        | \, \partial_{|\vp|}\hfp[\Ez^{-1}[\zet];\vp] \, |
        \; , \;  | \, \partial_{\zet}\hfp[\Ez^{-1}[\zet];\vp] \, | \, \}
        &\leq& \frac{c\eta^2}{\rho^3}
        \label{eq:partvp-Ez-aux-2}
\eeqn
(note that in contrast to Proposition {\ref{Polydlemma}}, the argument is here $\Ez^{-1}[\zet]$) and
\begin{align}
        \sup_{\zet\in\I}| \, \partial_{|\vp|}\Ez^{-1}[\zet] \, | \; < \; c  \eta \;,
        \label{eq:derp-Ezinv-1}
        \\
        | \, \partial_{\zet}\Ez^{-1}[\zet] \, - \rho \, |
        \; < \; c \, \rho \,  \eta  \; .
        \label{eq:derp-Ezinv-2}
\end{align}
To this end, we recall that
\eqn
        \sup_{\z\in \Ie_{\frac{3\rho}{\Iconsttw}}}\{ \, | \, \hfp[\z;\vp] \, - \, 1 \, | \; , \;
        | \, (\partial_{|\vp|}\hfp)[\z;\vp] \, | \; , \;
        | \, (\partial_{\z}\hfp)[\z;\vp] \, | \, \}&\leq&  \frac{c\eta^2}{\rho^3}
        \label{eq:der-hfp-bds-1}
\eeqn
from Proposition {\ref{Polydlemma}}, and
\eqn
        \sup_{|\z|<\frac{1}{\Iconst} }\{ \, | \, E[\z] \, - \, \z \, | \; , \;
        | \, \partial_{\z}  E[\z] \, - \, 1 \, | \; , \;
        | \, \partial_{|\vp|} E[\z] \, | \, \} \; < \; \eta\;\;\;
        \label{eq:der-E-z-bds-1}
\eeqn
from the definition of the polydisc in Section {\ref{polyd-subsect-1-1}}.

The estimate (\ref{eq:partvp-Ez-aux-2}) follows from
\eqn
        \sup_{|\zet|<\frac{1}{\Iconst}}|\partial_{|\vp|}\hfp[\Ez^{-1}[\zet];\vp]| &\leq&
        \sup_{|\z|<\frac{3\rho}{\Iconsttw} }
        |\partial_{|\vp|}\hfp[\z;\vp]|
        \nonumber\\
        &&+ \; \Big( \sup_{|\z|<\frac{3\rho}{\Iconsttw} }|\partial_{\z}\hfp[\z;\vp]| \Big) \,
        \sup_{|\zet|<\frac{1}{\Iconst}}|\partial_{|\vp|}\Ez^{-1}[\zet]|
        \label{eq:partvp-Ez-aux-1}
\eeqn
and
\eqn
        \sup_{|\zet|<\frac{1}{\Iconst}}|\partial_{\zet}\hfp[\Ez^{-1}[\zet];\vp]|
        &\leq& \Big( \sup_{|\z|<\frac{3\rho}{\Iconsttw} } |\partial_{\z}\hfp[\z;\vp]| \Big) \,
        \sup_{|\zet|<\frac{1}{\Iconst}}|\partial_{|\zet|}\Ez^{-1}[\zet]| \;,
        \label{eq:partvp-Ez-aux-3}
\eeqn
and from using (\ref{eq:der-hfp-bds-1}), (\ref{eq:partvp-Ez-aux-2}).

To prove (\ref{eq:derp-Ezinv-1}), we observe that
$\partial_{|\vp|}\Ez[\Ez^{-1}[\zet]]=\partial_{|\vp|}\zet=0$ implies that
\eqn
        \sup_{|\zet|<\frac{1}{\Iconst}}|\partial_{|\vp|}\Ez^{-1}[\zet]| & \leq &
        \sup_{|\z|< \frac{3\rho}{\Iconsttw} }
        \frac{1}{|(\partial_{\z}\Ez)[\z]|} \, |(\partial_{|\vp|}\Ez)[\z]|\;.
\eeqn
We have
\eqn
        \sup_{\z\in \Ie_{\frac{3\rho}{\Iconsttw}}}|(\partial_{|\vp|}\Ez)[\z]|&\leq&
        \sup_{\z\in \Ie_{\frac{3\rho}{\Iconsttw}}}\frac{1}{|\hfp[\z]| \rho }
        \Big(\,\frac{|\partial_{|\vp|}\hfp[\z]|}{|\hfp[\z]|}|E[\z]| \,
        + \, |\partial_{|\vp|}E[\z]|\,\Big)
        \nonumber\\
        &\leq&\sup_{\z\in \Ie_{\frac{3\rho}{\Iconsttw}}}\frac{2}{\rho} \Big(\frac{ \eta^2 |\z|}{\rho^3} + \eta \Big)
        \nonumber\\
        &<&\frac{3\eta}{\rho} \;,
        \label{eq:der-z-E-1}
\eeqn
for $\eta\ll\rho$. On the other hand,
\begin{align}
        |(\partial_{\z}\Ez)[\z]|  \; \geq \; \frac{1}{|\hfp[\z]|\rho}\Big( |\partial_{\z}E[\z]|-
        \frac{|\partial_{\z}\hfp[\z]|}{|\hfp[\z]|}|E[\z]| \Big)
        \; \geq \; \frac{1-c'\eta}{(1+c\eta^2 )\rho} \; .
        \label{eq:partz-Tz-lowbd-1}
\end{align}
One thus obtains (\ref{eq:derp-Ezinv-1}).

The estimate (\ref{eq:derp-Ezinv-2}) follows immediately from
\eqn
        |\partial_{\zet}\Ez^{-1}[\zet]|\;=\;\frac{1}{|(\partial_z \Ez)[\Ez^{-1}[\zet]]|} \;,
\eeqn
together with
\begin{align}
        |(\partial_{\z}\Ez)[\z]| \; \leq \; \frac{1}{|\hfp[\z]|\rho}\Big( |\partial_{\z}E[\z]|+
        \frac{|\partial_{\z}\hfp[\z]|}{|\hfp[\z]|}|E[\z]| \Big)
        \; \leq \; \frac{1+c'\eta}{(1-c\eta^2)\rho} \; ,
        \label{eq:partz-Tz-uppbd-1}
\end{align}
and (\ref{eq:partz-Tz-lowbd-1}).

Let
\begin{align}
        \widehat E[\zet;\vp] \; := \; \widehat w_{0,0}[\z;\unull;\vp]
        \; = \; \zet \, + \, \hfp[\z]^{-1}\rho^{-1}
        \Delta \widetilde w_{0,0}[\Ez^{-1}[\zet];\unull;\vp] \;,
        \label{eq:hatE-def-1}
\end{align}
where
\begin{align}
        \Delta\widetilde w_{0,0}[\zet;\spvar;\vp] \; := \;
        \sum_{L=2}^\infty(-1)^{L-1}\sum_{p_1,q_1,\dots,p_L,q_L:\atop
        p_\ell+q_\ell\geq1} \widetilde V^{(L)}_{\underline{0,p,0,q}}[\z;\spvar;\vp] \;,
\end{align}
see (\ref{eq:tildVLmpnq-def-1}).

By the arguments presented in Section {\ref{sec:Ren-symm-1}}, the operator
$\Delta\widetilde w_{0,0}[\zet;\opp;\vp]$ is rotation- and reflection symmetric.
Thus, by Lemma {\ref{rotreflinvlemma}}, it is a {\em scalar}  (its vector part
is identically zero).

We note that
\eqn
        \rho^{-1}\|\widetilde V^{(L)}_{\underline{0,p,0,q}}\|^\sharp_{0,0}\leq10\,(L+1)^2
        C_\Theta \, \rho^{-L} \, \prod_{\ell=1}^\infty \, \frac{\| w_{p_\ell,q_\ell}[\z]
        \|_{p_\ell,q_\ell}^\sharp }
        {p_\ell^{p_\ell/2}q_\ell^{q_\ell/2}} \;,
        \label{VL0p0qest}
\eeqn
which corresponds to the bounds in Lemma {\ref{VLboundlm}} for $V^{(L)}_{\underline{0,p,0,q}}$,
but with one power of $C_\Theta$ less here because there is a factor $F_0$ less.
Hence, we find
\eqn
        \rho^{-1} \| \Delta\widetilde w_{0,0} \|_{\Tspace}&\leq&
        \rho^{-1} \,
        \sup_{\zet\in\I} \, \sum_{L\geq2}\sum_{p_1,q_1,\dots,p_L,q_L:\atop
        p_\ell+q_\ell\geq1} \,
        \Big[ \,  \sup_{\spvar}| \, \partial_{\zet} \,
        \widetilde V^{(L)}_{\underline{0,p,0,q}}[\spvar] \, |
        \nonumber\\
        && + \,  \sum_{|\ua|=1}\sup_{\spvar}| \, \partial_{|p|} \, \partial_{\spvar}^{\ua} \,
        \widetilde V^{(L)}_{\underline{0,p,0,q}}[\spvar] \, |
        \nonumber\\
        &&\hspace{3cm}\, + \, \sum_{0\leq|\ua|\leq2 }
        \sup_{\spvar}| \, \partial_{\spvar}^{\ua} \,
        \widetilde V^{(L)}_{\underline{0,p,0,q}}[\spvar] \, |
        \Big]
        \nonumber\\
        &\leq& C_\Theta \, \sum_{L=2}^\infty (L+1)^2
        \Big( \frac{C_\Theta}{\rho} \Big)^L
        \Big( \sum_{p+q\geq1} \sup_{\z \in\I} \| w_{p,q}[\z] \|_{p,q}^\sharp \Big)^L
        \nonumber\\
        &\leq& C_\Theta \sum_{L=2}^\infty (L+1)^2
        \Big(\frac{C_\Theta}{\rho} \Big)^L
        \Big(\xi\sum_{p+q\geq1}\xi^{-p-q}\sup_{\z \in\I}\| w_{p,q}[\z] \|_{p,q}^\sharp \Big)^L
        \nonumber\\
        &\leq& C_\Theta \, \sum_{L=2}^\infty (L+1)^2
        \Big( \frac{ C_\Theta\xi}{\rho} \Big)^L
        \Big( \|\h_{\geq1}\|_{\xi } \Big)^L
        \nonumber\\
        &\leq& 12\, C_\Theta \, \Big( \frac{C_\Theta\xi}{\rho}
        \| \h_{\geq1} \|_{\xi } \Big)^2
        \; \leq \;
        \frac{C_\Theta^3 \xi^2 }{\rho^2}(2\eta)^2
        \; \leq \;  \frac{\eta}{10} \;,
        \label{DiffThatTest}
\eeqn
for
\eqn
        \eta&\ll&  \rho^3
        \nonumber\\
        \xi&\leq& \frac14
        \label{parm-choice-1}
\eeqn
(see also (\ref{eq:const-def-1})).

Thus, (\ref{eq:hatE-def-1}) yields
\eqn
        \sup_{\zet\in\I}| \widehat E[\zet;\vp]-\zet |&\leq&\sup_{|\z|<\frac{3\rho}{\Iconsttw}} \,
        | \, \hfp[\z;\vp] \, |^{-1} \, | \, \Delta \widetilde w_{0,0}[\z;\opp=\unull;\vp] \, |
        \nonumber\\
        &\leq&
        ( 1 \, + \, c\eta^2 ) \| \Delta \widetilde w_{0,0} \|_{\Tspace}
        \nonumber\\
        &<& \frac{\eta}{2}\;,
\eeqn
for $a=0,1$ and (\ref{parm-choice-1}).
Moreover,
\eqn
        \sup_{\zet\in\I}| \, \partial_{|\vp|} \widehat E[\zet;\vp] \, |&\leq&\sup_{\zet\in\I}
        (  1 \, + \, c\eta^2  ) \,
        | \, (\partial_{|\vp|}  \Delta \widetilde w_{0,0})[\Ez^{-1}[\zet];\unull;\vp] \, |
        \nonumber\\
        &+& \sup_{\zet\in\I}
        (  1 \, + \, c\eta^2  ) \,
        | \, ( \partial_{\z }\Delta \widetilde w_{0,0})[\Ez^{-1}[\zet];\unull;\vp] \, |
        \, | \, \partial_{|\vp|} \Ez^{-1}[\zet] \, |
        \nonumber\\
        &\leq&
        (  1 \, + \, c\eta^2  ) \| \Delta \widetilde w_{0,0} \|_{\Tspace}
        \, (1 \, + \, c\eta )
        \nonumber\\
        &\leq& \frac\eta2
        \label{partpEest}
\eeqn
This completes the discussion of the renormalization of $E[\z]$.

Next, we discuss $T[\z;\opp;\vp]$, and determine the
renormalized expressions for the conditions
(\ref{eq:Polyd-def-lTnl-1}) - (\ref{eq:Polyd-def-pfp-1}) in the definition of
$\Polyd\Polpar$.
To this end, we again let $\z=\Ez^{-1}[\zet]$, and consider
\eqn
        \widehat w_{0,0}[\zet;\spvar;\vp]&=&\hfp[\z]^{-1}
        \Big[ \hfp[\z;\vp] H_f \, + \, E[\z]\chi_1^2[H_f] \,
        \nonumber\\
        &&\hspace{1cm}+ \,
        \chi_1^2[H_f]\Big((1-\hfp[\z])H_f \, + \, \pfp[\z;\vp]\Ppar
        \, + \, \rho\lTnl \Pf^2
        \nonumber\\
        &&\hspace{3cm}
        \, + \, \rho^{-1}\remT[\z;\rho\opp;\vp]
        \nonumber\\
        &&\hspace{4cm}
        +\,\Delta\widetilde w_{0,0}[ \z;\opp;\vp ]\Big) \,F_0[\z;\opp;\vp]
        \;\Big]
        \nonumber\\
        &=&\widehat E[\zet]\chi_1^2[H_f] \, + \, \widehat T[\zet;\opp;\vp]
\eeqn
with
\eqn
\begin{split}
        \widehat T[\zet;\opp;\vp] \; = \; H_f + \, \chi_1^2[H_f] \, \big( \; \widehat\pfp[\zet;\vp] \Ppar
        \, + \, & \widehat \lTnl \Pf^2
        \\
        \, + \, & \widehat{\remT}[\zet;\opp;\vp] \; \big) \,
        F_0[\z;\opp;\vp] \;.
        \label{hatw00def}
\end{split}
\eeqn
The terms in (\ref{hatw00def}) are determined by the Taylor expansion of $\Delta\widetilde w_{0,0}$
in $\opp$ up to a quadratic remainder term.

The operator $\widehat T[\zet;\opp;\vp]$ is rotation- and reflection symmetric,
and is therefore a scalar (see Lemma {\ref{rotreflinvlemma}}).

We note that there is no term proportional to $H_f$ in the brackets in (\ref{hatw00def}) because
the defining condition for $\hfp[\z]$,
\eqn
        1 \, - \, \hfp[\z] \, + \, (\partial_{X_0}\Delta\widetilde w_{0,0})
        [\z;\unull;\vp] \; = \; 0 \;,
\eeqn
suppresses the creation of a term proportional to $\chi_1^2[H_f]H_f$ by $\ren$, see
Proposition {\ref{Polydlemma}}, Remark  {\ref{rem:hfp-choice-1}}, and Remark {\ref{rem:hfp-reason}}.

Furthermore,
\eqn
        \widehat\pfp[\zet;\vp] \; = \; \hfp[\z;\vp]^{-1}\big(\pfp[\z;\vp]
        \,
        + \, (\partial_{X^\parallel}\Delta\widetilde w_{0,0})
        [\z;\unull;\vp]\big)
\eeqn
and
\eqn
        \widehat\lTnl \; = \; \rho\lTnl \;.
\eeqn
The operator
\eqn
        \widehat{\remT}[\zet;\opp;\vp]&=&\hfp[\z]^{-1}\Big[ \, \rho^{-1}
        \, \remT[\z;\rho\opp;\vp]
        \, + \, (1-\hfp[\z] ) \, \rho \, \lTnl \, \Pf^2
        \nonumber\\
        &&\hspace{2cm}+ \;\, \rho^{-1} \,
        \Delta\widetilde w_{0,0}[\z;\spvar;\vp]
        \nonumber\\
        &&\hspace{3cm}- \, \rho^{-1} \sum_{0\leq|\ua|\leq1}(\partial_{\spvar}^{\ua}\Delta\widetilde w_{0,0})
        [\z;\unull;\vp] \, \Big]
\eeqn
is of order $O(|\opp|^2)$ as $\opp\rightarrow\unull$, and contains the quadratic Taylor remainder
term of $\Delta\widetilde w_{0,0}$.

We first recall that
\eqn
        \sup_{\zet\in\I}| \, (\partial_{\z}\pfp)[\Ez^{-1}[\zet];\vp] \, | \; \leq \; \eta
        \label{eq:der-pfp-z-1}
\eeqn
from the definition of $\Polyds\Polpar$.
Therefore,
\eqn
        \sup_{\zet\in\I}| \, \partial_{\zet}\pfp[\Ez^{-1}[\zet];\vp] \, |
        &\leq&\eta \, \sup_{\zet\in\I}| \, \partial_{\zet} \Ez^{-1}[\zet]  \, |
        \nonumber\\
        &\leq& \rho \, \eta \, (1 \, + \, c\eta) \; \leq \; \frac\eta9
        \label{eq:der-zet-pfp-est-1}
\eeqn
from (\ref{eq:derp-Ezinv-2}), and $\rho\leq\frac{1}{10}$ ($\rho$ is determined
in (\ref{eq:rho-xi-parmchoice-1}) below).
For $a=0,1$,
\eqn
        |\, \partial_{|\vp|}^a (\widehat\pfp[\zet;\vp]-\pfp[\Erho^{-1}[\zet];\vp] )\,|
        &=& \big| \, \partial_{|\vp|}^a \big(  (\hfp[\Erho^{-1}[\zet];\vp]^{-1}-1  )
        \pfp[\Erho^{-1}[\zet];\vp] \big)
        \nonumber\\
        &&\hspace{1cm}+\,
        \partial_{|\vp|}^a\partial_{X^\parallel}\Delta\widetilde w_{0,0}
        [\Erho^{-1}[\zet];\unull;\vp] \, \big|
        \nonumber\\
        &<&\frac{c\eta^2}{\rho^3} \, + \, \frac{\eta}{10}\;.
        \label{eq:hatpfp-pfp-est-1}
\eeqn
This follows straightforwardly using
\eqn
        \sup_{\z\in\I}|\, \partial_{|\vp|}^a(\pfp[\z;\vp]+
        |\vp|) \,|&<&\frac\delta2 \;\;\;,\;\;\;a=0,1\;,
        \label{eq:pfp-indass-recall-1}
\eeqn
from the definition of the polydisc $\Polyd\Polpar$ (see Section {\ref{polyd-subsect-1-1}}), and
\eqn
    \sup_{\zet\in\I}\big|   \partial_{|\vp|}(\pfp [\Erho^{-1}[\zet];\vp] +|\vp|)\big|
    &=& \sup_{\zet\in\I}\Big[\big|  (\partial_{|\vp|}\pfp )[\Erho^{-1}[\zet];\vp]+1\big| \;
    \nonumber\\
    &&\hspace{2cm}
    +\,\big|   (\partial_{\z}\pfp)[\Erho^{-1}[\zet];\vp] \big|
    \,\big| \partial_{|\vp|}\Erho^{-1}[\zet] \big|\Big]
    \nonumber\\
    &<&\frac\delta2 \, + \, \frac{2\delta\eta}{\rho}
\eeqn
with (\ref{eq:der-z-E-1}) and (\ref{eq:der-pfp-z-1}).
Moreover,
\eqn
        \sup_{\zet\in\I}|\,( \partial_{|\vp|}^a \, \partial_{X^\parallel}
        \, \Delta\widetilde w_{0,0})
        [\Erho^{-1}[\zet];\unull;\vp] \, |&<&\frac{\eta}{10}
        \nonumber\\
        \sup_{\zet\in\I}|\,( \partial_{\zet} \, \Delta\widetilde w_{0,0})
        [\Erho^{-1}[\zet];\unull;\vp]\,|&<&\frac{\eta}{10}
\eeqn
from (\ref{DiffThatTest}), and
\eqn
    | \, \partial_{|\vp|}^a(1-\hfp[\Erho^{-1}[\zet]] )\,| \; < \; \frac{c\eta^2}{\rho^3}
    \; \; \; , \; \; \; a=0,1 \;,
\eeqn
see (\ref{eq:hfp-der-1}).
We thus find for $a=0,1$ that
\eqn
        \sup_{\zet\in\I}|\,\partial_{\z} (\widehat\pfp[\zet;\vp]+|\vp|)\,|&<&
        \frac\eta9+\frac{\eta}{10}
        \;<\; \frac\eta2
        \nonumber\\
        \sup_{\zet\in\I}|\,\partial_{|\vp|}^a(\widehat\pfp[\zet;\vp]+|\vp|)\,|&<&
        \frac\delta2+\frac{2 \delta \eta}{\rho}+\frac{\eta}{10}+\frac{c\eta^2}{\rho^3}
        \;<\;\frac\delta2+\frac\eta2
        \label{eq:hatdelta-ren-est-1}
\eeqn
from (\ref{eq:hatpfp-pfp-est-1}) and (\ref{eq:pfp-indass-recall-1}), for $\delta,\eta\ll\rho^3$
sufficiently small.

Next, we discuss the renormalization of $\delta T$.
Note that since
\eqn
    (\partial_{\spvar}^{\ua}\widehat{\delta T})[\z;\unull;\vp] \; = \; 0 \;,
\eeqn
we have
\eqnn
    \sup_{|\vX|\leq X_0}|\,\partial_{\spvar}^{\ua}\partial_{|\vp|}^b
    \widehat{\delta T}[\Erho^{-1}[\zet];\spvar;\vp]\,|
    \;\leq\;\sum_{|\ua'|=2}\sup_{|\vX|\leq X_0}
    |\,\partial_{\spvar}^{\ua'}\partial_{|\vp|}^b \widehat{\delta T}[\Erho^{-1}[\zet];\spvar;\vp]\,|
\eeqnn
for $0\leq |\ua|\leq2$ and $b=0,1$. Consequently,
\eqn
    \|\widehat{\delta T}\|_{\Tspace}&\leq& 32 \,
    \Big( \frac{|\partial_{|\vp|}\hfp[\Erho^{-1}[\zet];\vp]|}{|\hfp[\Erho^{-1}[\zet];\vp]|} +1\Big)
    \nonumber\\
    &&\hspace{1cm}
    \Big[ \, \rho \|\delta T\|_{\Tspace} \, + \,
    \big(|\partial_{|\vp|}\hfp[\Erho^{-1}[\zet];\vp]|
    +|\hfp[\Erho^{-1}[\zet];\vp]-1|\big)\rho\lTnl \|\vX^2\|_{\Tspace}
    \nonumber\\
    &&\hspace{5cm}
    +\|\rho^{-1}\Delta\widetilde w_{0,0}\|_{\Tspace} \, \Big]
    \nonumber\\
    &\leq&32 \, (1+c\eta^2) \, \rho \|\delta T\|_{\Tspace} + \frac{c\eta^2}{\rho^3} \;.
\eeqn
using (\ref{eq:partvp-Ez-aux-2}). Since by assumption, $\|\delta T\|_{\Tspace}< \delta$,
we find
\eqn
        \|\widehat{\delta T}\|_{\Tspace} \; < \;  \frac{\delta}{2}+ c\eta^2
\eeqn
for
\eqn
        \rho \; \leq \; \frac{1}{100} \;,
        \label{eq:rho-T-bd-aux-1}
\eeqn
which is determined in (\ref{eq:rho-xi-parmchoice-1}) below.
Therefore, $\widehat\delta$ is determined by (\ref{eq:hatdelta-ren-est-1}).

To carry out the induction step for (\ref{eq:Polyd-def-piop-1}), we recall from
(\ref{eq:piop-ref-def-1}) and  (\ref{F0FLdef}) that
\eqn
        \piop_{\chi_1}^{(\vp;\widehat\lTnl)} \; = \; \Gamma_\rho
        \piop_{\chi_\rho[H_f]}\big(T_0^{(\vp;\lTnl)}[\z;\opp] \, , \, H_f\big)\Big|_{\Ran(\chi_\rho[H_f])}
        \Gamma_\rho^* \;,
\eeqn
where $\Gamma_\rho$ is the unitary dilation operator, see
(\ref{eq:resc-def-1}), and
\eqnn
        F_0[\z;\opp;\vp] \; = \; \Gamma_\rho\piop_{\chi_\rho[H_f]}
        \big( (E[\z] \chi_1^2[H_f]+T[\z;\opp;\vp]) \, , \, \hfp[\z]H_f \big)
        \Big|_{\Ran(\chi_\rho[H_f])}\Gamma_\rho^* \;.
\eeqnn
We note that
\eqn
        T_1&:=&T_0^{(\vp;\lTnl)}[\z;\opp]\Big|_{\Ran(\chi_\rho[H_f])}
        \; = \; E[\z] \, + \, |\vp| \Ppar \, + \, \lTnl \Pf^2
\eeqn
and
\eqn
        T_2&:=&(E[\z] \chi_1^2[H_f]+T[\z;\opp;\vp])\Big|_{\Ran(\chi_\rho[H_f])}
        \nonumber\\
        & =& E[\z] \, + \, H_f \, + \, \pfp[\z]\Ppar \, + \, \lTnl \Pf^2
        \, + \, \remT[\z;\opp;\vp]
\eeqn
using
\eqn
        \chi_\rho[H_f]\chi_1[H_f]=\chi_\rho[H_f]
        \;\;\;{\rm and} \;\;\;
        \chi_\rho[H_f]\widetilde \piop[\z;\opp;\vp]=0 \;,
\eeqn
see (\ref{eq:chirho-tildpiop-annih-1}).
It is clear that
\eqn
        \chi_\rho[H_f]F_0[\z;\opp;\vp] \; = \; 0 \;,
\eeqn
by the definition (\ref{F0FLdef}). Moreover, using (\ref{eq:piop-diff-ed-1}) and
$\tau_1:=H_f$, $\tau_2:=\hfp[\z]H_f$,
\eqn
        \|F_0-\piop_{\chi_1}^{(\vp;\widehat\lTnl)}\|_{\Tspace}&<&
        \|\piop_{\chi_\rho[H_f]}(T_1,H_f)-\piop_{\chi_\rho[H_f]}(T_2,\hfp[\z]H_f)\|_{\Tspace,\rho}
        \nonumber\\
        &\leq&\|\bar\chi_\rho[H_f]\bar R_0(T_2,\tau_1)\piop_{\chi_\rho[H_f]}(T_2,\tau_1)\|_{\Tspace,\rho}
        \|T_2-T_1\|_{\Tspace,\rho}
        \\
        &&\hspace{2cm}+ \|\bar\chi_\rho[H_f] T_1\bar R_0(T_1,\tau_1)\|_{\Tspace,\rho}
        \|\tau_2-\tau_1\|_{\Tspace,\rho}
        \nonumber
\eeqn
where $\|\,\cdot\,\|_{\Tspace,\rho}$ is defined as $\|\,\cdot\,\|_{\Tspace}$ in (\ref{Tnormsharpdef}),
but with the supremum over $X_0\in[0,1]$ replaced by the supremum over $X_0\in[0,\rho]$.
We find
\eqn
        \|\bar\chi_\rho[H_f]\bar R_0(T_2,\tau_1)\piop_{\chi_\rho[H_f]}(T_2,\tau_1)\|_{\Tspace,\rho}
        &\leq&\frac{C_\Theta}{\rho}
\eeqn
and
\eqn
        \|\bar\chi_\rho[H_f] T_1\bar R_0(T_1,\tau_1)\|_{\Tspace,\rho} &\leq&C_\Theta \;.
\eeqn
Therefore
\eqn
        \|F_0-\piop_{\chi_1}^{(\vp;\widehat\lTnl)}\|_{\Tspace}
        &\leq&\frac{C_\Theta}{\rho}  \Big( \big\|  (E[\z]+ X_0+\pfp[\z] X^\parallel
        \, + \, \widehat\lTnl \vX^2+\remT[\z; X;\vp]\big)
        \nonumber\\
        &&\hspace{3cm}
        \, - \, \big(E[\z]+X_0 -|\vp|X^\parallel+\widehat\lTnl \vX^2 \big) \, \big\|_{\Tspace}
        \nonumber\\
        &&\hspace{2cm}
        +\|\hfp[\z] X_0 - X_0 \|_{\Tspace} \Big)
        \nonumber\\
        &\leq& \frac{C_\Theta}{\rho} \Big( \|(\hfp[\z]-1)X_0 \|_{\Tspace} \, + \,
        \| (\pfp[\z]+|\vp|)X^\parallel\|_{\Tspace}
        \nonumber\\
        &&\hspace{2cm}\,+ \| \remT[\z;\rho X;\vp]\|_{\Tspace}\Big)
        \nonumber\\
        &\leq&  \frac{C_\Theta}{\rho} \, \Big(4\delta+\frac{c\eta^2}{\rho^3}\Big)
    \; \leq  \; K_\Theta \, \widehat\delta
        \label{eq:F0-piop0-est-1}
\eeqn
for the choice (\ref{eq:rho-xi-parmchoice-1}) of $\rho$. The
constant $K_\Theta$ only depends on the smooth cutoff function $\Theta$
in (\ref{Thetadef}), and defines the value of the constant $K_\Theta$
in the definition of the polydisc $\Polyd\Polpar$, see (\ref{eq:Polyd-def-piop-1}).
Moreover, $\widehat\delta$ is as in (\ref{ccontr-def-1}).

In particular, we have
\eqn
        \big\| \widehat T-T_0^{(\vp;\widehat\lTnl)} \, \big\|_{\Tspace}
        &\leq&\big\| (\widehat\pfp+|\vp|)X^\parallel+\widehat{\remT}\big\|_{\Tspace}\big\|F_0\big\|_{\Tspace}
        \nonumber\\
        &&+ \, \big\|\,X_0+|\vp|X^\parallel+\widehat\lTnl X^2\,\big\|_{\Tspace}
        \|F_0-\piop_{\chi_1}^{(\vp;\widehat\lTnl)}\|_{\Tspace}
        \nonumber\\
        &<&  K_\Theta' (\widehat\delta+\widehat\eta) \;,
        \label{renT0part-1}
\eeqn
for a constant $K_\Theta'$ which only depends on $\Theta$.

\subsubsection{Irrelevant kernels: Bounds on $\|\widehat\h_{\geq2}\|_\xi$}
\label{subsubsec:irell-wMN-1}

Recalling
\eqn
        \widehat w_{M,N}[\zet;\spvar;\rvar \vn, \lambda] \; = \;
        \frac{1}{\hfp[\Ez^{-1}[\zet];\vp]} \, \widetilde w_{M,N}[\zet;\spvar;\rvar \vn, \lambda]
\eeqn
from  Lemma {\ref{Wickorderthm}},
\eqn
        \| \widehat w_{M,N} \|_{M,N}^\sharp &\leq&
        \Big(1+\Big|\frac{\partial_{|\vp|}\hfp[\Ez^{-1}[\zet];\vp]}{\hfp[\Ez^{-1}[\zet]z;\vp]}\Big|\Big)
        \frac{1}{|\hfp[\Ez^{-1}[\zet];\vp]|}\|\widetilde w_{M,N}\|_{M,N}^\sharp
        \nonumber\\
        &\leq&\Big(1+\frac{c\eta^2}{\rho^3}\Big) \|\widetilde w_{M,N}\|_{M,N}^\sharp \;,
        \label{eq:hatw-tildew-bd}
\eeqn
by Proposition {\ref{Polydlemma}}.
Using Lemma {\ref{hatwMNformalseriesthm}}, Lemma {\ref{VLboundlm}}, and
${ m +p \choose p  }\leq 2^{m+p}$,
we find
\eqn
        &&\| \widetilde w_{M,N}[\zet] \|_{M,N}^\sharp \; \leq \;
        \sum_{L=1}^\infty  \, C_\Theta^2 (L+1)^2
        \Big(\frac{C_\Theta}{\rho}\Big)^L(2\rho)^{M+N}
        \label{hatwMNbd-1}
        \\
        &&\hspace{0.5cm}\,\sum_{m_1+\cdots+m_l=M,\atop n_1+\cdots+n_L=N}
        \sum_{p_1,q_1,\dots,p_L,q_L:\atop
        m_\ell+p_\ell+n_\ell+q_\ell\geq1}
        \prod_{\ell=1}^L\Big\{\Big(\frac{2}{\sqrt{p_\ell}}\Big)^{p_\ell}
        \Big(\frac{2}{\sqrt{q_\ell}}\Big)^{q_\ell}
        \|w_{M_\ell,N_\ell}[\z]\|_{M_\ell,N_\ell}^\sharp
        \Big\}
        \nonumber
\eeqn
where
\eqn
        M_\ell:= m_\ell+p_\ell
        \;\;\;,\;\;\;
        N_\ell:=n_\ell+q_\ell \;.
\eeqn
Summing over $\underline{m,p,n,q}$, we get
\eqn
        \| \widehat \h_{\geq2} [\zet] \|^\sharp_{\xi}
        &\leq& 2 \, C_\Theta^2\rho^{2 }\sum_{M+N\geq2}\xi^{-M-N} \sum_{L=1}^\infty (L+1)^2
        \Big(\frac{C_\Theta}{\rho}\Big)^L
        \nonumber\\
        &&\,
        \sum_{m_1+\cdots+m_l=M,\atop n_1+\cdots+n_L=N}
        \sum_{p_1,q_1,\dots,p_L,q_L:\atop
        m_\ell+p_\ell+n_\ell+q_\ell\geq1}\prod_{\ell=1}^L\xi^{m_\ell+n_\ell}
        \\
        && \,\prod_{\ell=1}^L
        \Big\{
        \Big(\frac{2\xi}{\sqrt{p_\ell}}\Big)^{p_\ell}
        \Big(\frac{2\xi}{\sqrt{q_\ell}}\Big)^{q_\ell}
        \xi^{-(m_\ell+p_\ell+n_\ell+q_\ell)}
        \|w_{M_\ell,N_\ell}[\z]\|_{M_\ell,N_\ell}^\sharp
        \Big\}
        \nonumber
 \eeqn
 and using the definition of the norm $\|\,\cdot\,\|_\xi^\sharp$,
 \eqn
        \| \widehat \h_{\geq2} [\zet] \|^\sharp_{\xi}
        &\leq&2 \, C_\Theta^2\rho^{2}  \,
        \frac{C_\Theta}{\rho}\sum_{M+N\geq2}
        \xi^{-(M+N)}
        \|w_{M,N}[\z]\|_{M,N}^\sharp
        \nonumber
        \\
        &&\hspace{2cm}
        \Big(\sum_{p=0}^M\Big(\frac{2\xi}{\sqrt{p}}\Big)^{p}\Big)
        \Big(\sum_{q=0}^N\Big(\frac{2\xi}{\sqrt{q}}\Big)^{q}\Big)
        \label{hatwMNgeq1est1-L1}\\
        &&+\; 2 \, C_\Theta^2\rho^{2} \sum_{L=2}^\infty (L+1)^2
        \Big(\frac{C_\Theta}{\rho}\Big)^L
        \label{hatwMNgeq1est1}
        \\
        && \,\Big\{\sum_{M+N\geq1}
        \Big(\sum_{p=0}^M\Big(\frac{2\xi}{\sqrt{p}}\Big)^{p}\Big)
        \Big(\sum_{q=0}^N\Big(\frac{2\xi}{\sqrt{q}}\Big)^{q}\Big)
        \xi^{-(M+N)}
        \|w_{M,N}[\z]\|_{M,N}^\sharp
        \Big\}^L \;,
        \nonumber
\eeqn
where (\ref{hatwMNgeq1est1-L1}) accounts for $L=1$, and
(\ref{hatwMNgeq1est1}) for the rest. Hence,
\eqn
        \| \widehat \h_{\geq2} \|_{\xi}
        \; \leq \; 2 \, C_\Theta^3\rho   \,  A^2 \,
        \|w_{\geq2}\|_{\xi}
        + \, 2 \, C_\Theta^2\rho^{2} \sum_{L=2}^\infty (L+1)^2
        \Big(\frac{C_\Theta}{\rho}\Big)^L
        A^{2L}(\|\h_{\geq1}\|_{\xi})^L\nonumber
\eeqn
with
\eqn
        A \; := \; \sum_{p=0}^\infty\Big(\frac{2\xi}{\sqrt{p}}\Big)^{p}
        \; \leq \; \sum_{p=0}^\infty(2\xi)^p =\frac{1}{1-2\xi}
        \; \leq \; 2\;,
        \label{Aconstdef}
\eeqn
assuming that $\xi\leq\frac14$.

Letting
\eqn
        \label{eq:B-aux-def-1}
        B \; := \; \frac{C_\Theta}{\rho(1-2\xi)^2 }\|\h_{\geq1}\|_{\xi}
        \; \leq \;  \frac{4C_\Theta }{\rho}\|\h_{\geq1}\|_{\xi} \;,
\eeqn
one gets $\sum_{L=2}^\infty (L+1)^2 B^L \leq  12B^2$,
provided that $B<\frac{1}{10}$ (recalling that
$\|\h_{\geq k}\|_\xi=\sup_{\z\in\I}\|\h_{\geq k}[\z]\|_\xi^\sharp$).

Since $\|\h_{\geq2}\|_\xi<\e$ and
$\|\h_{\geq1}\|_\xi<\eta+\e<2\eta$ by definition of $\Polyds\Polpar$, one finds from
$\| \widehat \h_{\geq2} \|_{\xi}
\, \leq \, (1+c\eta) \, \| \widetilde \h_{\geq2}  \|_{\xi}$
that
\eqn
        \|\widehat \h_{\geq2}\|_{\xi}
        \; \leq \; 5 \rho C_\Theta^3 \e + 96 C_\Theta^3 \eta^2
        \; \leq \; \frac{\e}{4} + \frac\eta4
        \label{hatwMNbd-2}
\eeqn
with
\eqn
        \rho \; = \; \frac{1}{20 C_\Theta^3} \;,
        \label{eq:rho-irr-bd-aux-1}
\eeqn
and  $\eta\ll1$ sufficiently small.

\subsubsection{Marginal kernels: Bounds on $\|\widehat \h_1\|_\xi$}
\label{subsubsec:marg-kernels-1}

In the case $M+N=1$,
we use the soft photon sum rules \sbsr. That is, for any
arbitrary unit vector $\vn\subset\R^3$,
\eqn
        \lim_{\rvar\rightarrow0}  w_{1,0}[\z;\spvar;\rvar \vn, \lambda]
        \; = \; \g \, \mu(\sig) \, \pol(\vn,\lambda)\cdot\nabla_{\vX}  T[\z;\spvar;\vp]  \;,
\eeqn
and likewise for $w_{0,1}$.
Since the soft photon sum rules are preserved by $\ren$, they imply that the
renormalized quantities $\widehat T$, $\widehat w_{1,0}$ and $\widehat w_{0,1}$ likewise satisfy
\eqn
        \lim_{\rvar\rightarrow0}  \widehat w_{1,0}[\zet;\spvar;\rvar \vn, \lambda]
        \; = \; \g \,\mu(\widehat\sig) \, \pol(\vn,\lambda)\cdot\nabla_{\vX}  \widehat T[\zet;\spvar;\vp] \;,
        \label{w-deg1-zero-0}
\eeqn
where $\zet=\Ez[\z]$, $\widehat\sig=\rho^{-1}\sig$, and $\mu(\widehat\sig)=\rho^{ \chi(\sig>1)}\mu(\sig)$
(see (\ref{sbsr})).
With
\eqn
        \widehat\pfp[\zet;\vp]\;=\;\nabla_{\vX}\widehat T[\zet;\spvar=\unull;\vp] \;,
\eeqn
we have
\eqn
        \lim_{X\rightarrow0}\lim_{\rvar\rightarrow0}
        \widehat w_{1,0}[\zet;\spvar;\rvar \vn, \lambda]
        \; = \; \g \, \mu(\widehat\sig) \, \pol(\vn,\lambda)\cdot\widehat\pfp[\zet;\vp]\;.
        \label{w-deg1-zero-1}
\eeqn
For $\sig<1$, we have $\mu(\widehat\sig)=1$. For $\sig>1$, we have $\mu(\widehat\sig)=\rho\mu(\sig)$,
and thus gain a factor $\rho<\frac12$ from the application of $\ren$.

\noindent\underline{{\em (i) The case $\sig\leq1$.}}
We recall from (\ref{eq:hatdelta-ren-est-1}) that
\eqnn
        \sup_{\zet\in\I}| \, \partial_{\zet}\widehat\pfp[\Ez^{-1}[\zet];\vp] \, |
        &\leq&  \frac\eta2
        \nonumber\\
        | \, \partial_{|\vp|}^a\big( \widehat\pfp[\zet;\vp] \, + \, |\vp| \big) \, | & < &
        \frac\delta2 \, + \, \frac\eta2 \;\;\;,\;\;\; a=0,1\;.
\eeqnn
%
Since (\ref{w-deg1-zero-1}) only depends on $\z$ and $\vp$, but neither on $\spvar$ nor $|\vk|$, we find
\eqn
    \lefteqn{
    \sum_{a=0,1}\|\partial_{|\vp|}^a\widehat w_{1,0}[\zet;\spvar;\rvar \vn, \lambda]\|_{1,0}
    }
    \nonumber\\
    &\leq&
    \sum_{a = 0,1 }
    | \, \g\pol(\vn,\lambda)\cdot\partial_{|\vp|}^a \, \widehat \pfp[\zet;\vp] \, |
    \, + \, \sum_{|\ua|=1}\sum_{b=0,1}
    \|  \partial_{\spvar}^\ua \, \partial_{|\vp|}^b \, \widehat w_{1,0}\|_{1,0}
    \nonumber\\
    &&\hspace{6cm}
    + \, \sum_{a=0,1}\|\partial_{|\vp|}^{a} \, \partial_{|\vk|} \, \widehat w_{1,0}\|_{1,0}
    \label{eq:w10-norm-auxest-1}\\
    &\leq&\g \, \big( 1 \, + \, |\vp| \, + \, \widehat\delta \big)
    \, + \, \sum_{|\ua|=1}\sum_{b=0,1}
    \|  \partial_{\spvar}^\ua  \, \partial_{|\vp|}^b \, \widehat w_{1,0}\|_{1,0}
    \, + \, \sum_{a=0,1}\| \partial_{|\vp|}^{a} \, \partial_{|\vk|} \, \widehat w_{1,0}\|_{1,0}
    \nonumber
\eeqn
by Taylor's theorem.
All terms involving derivatives in $\spvar$ and $|\vk|$
obtain a factor $\rho$ from rescaling, as shown below.
Hence, (\ref{w-deg1-zero-1}) is the marginal part of $w_{1,0}$.

The key observation here is that by use of the soft photon sum rules,
the marginal parts of $\widehat w_{1,0}$ and $\widehat w_{0,1}$ are entirely determined by $\widehat T$.
Moreover, since $T$ is {\em scalar} (it has no vector part), only the scalar components of $w_{1,0}$
and $w_{0,1}$ are marginal, while the vector parts scale like {\em irrelevant} operators.
This implies that the term proportional to $\Bf$ in the fiber Hamiltonian $\Hps$ in (\ref{eq:Hps-def-1})
is an {\em irrelevant} operator.

Using (\ref{eq:w10-norm-auxest-1}), we find that
\eqn
        \| \widehat w_{1,0} \|^\sharp_{1,0}
        &=&\sum_{a=0,1}\| \partial_{\zet}^a \, \widehat w_{1,0} \|_{1,0}
        \, + \| \partial_{|\vp|} \, \widehat w_{1,0} \|_{1,0}
        \, + \, \sum_{1\leq|\ua|\leq2}\| \partial_{\spvar}^\ua \, \widehat w_{1,0} \|_{1,0}
        \nonumber\\
        &&\hspace{1cm}
        \, + \, \sum_{|\ua|=1}\| \partial_{|\vp|} \, \partial_{\spvar}^\ua \, \widehat w_{1,0} \,  \|_{1,0}
        \, + \, \sum_{a=0,1}\| \partial_{|\vp|}^a \, \partial_{|\vk|} \, \widehat w_{1,0} \|_{1,0}
        \nonumber\\
        &\leq&\g( 1 \, + \, |\vp| \, + \, \widehat\delta )
        \, + \, 2\Big( \; \sum_{1\leq|\ua|\leq2}\| \partial_{\spvar}^\ua \, \widehat w_{1,0} \|_{1,0}
        \, + \| \partial_{\zet} \, \widehat w_{1,0} \|_{1,0}
        \nonumber\\
        &&\hspace{1cm}
        \, + \, \sum_{|\ua|=1}\| \partial_{|\vp|} \, \partial_{\spvar}^\ua \, \widehat w_{1,0} \|_{1,0}
        \, + \, \sum_{a=0,1}\| \partial_{|\vp|}^a \, \partial_{|\vk|} \, \widehat w_{1,0} \|_{1,0} \; \Big) 
        \;. \; \; \; \; \; \; \; \; \; 
        \label{eq:hatw-auxbd-2}
\eeqn
To bound the sums in the bracket in (\ref{eq:hatw-auxbd-2}), we note that
similarly as in (\ref{eq:hatw-tildew-bd}),
\eqn
        \| \partial_{|\vp|}^a \, \partial_{Y} \, \widehat w_{1,0} \|_{1,0}
        &\leq&(1+c\eta)
        \| \partial_{|\vp|}^a \, \partial_{Y} \, \widetilde w_{1,0} \|_{1,0}
\eeqn
for $Y=|\vk|$ or a component of $\spvar$.

The leading term in $\widetilde w_{1,0}$ corresponding to $L=1$ (where $\underline p$ and
$\underline q$ are zero) is given by
\eqn
        V^{(L=1)}_{\underline{1,0,0,0}}[\h|\spvar;K] \; = \; \Bra\vac\,,\,
        F_0[\spvar+\uk] \, \widetilde W_{1}[\z;\rho\spvar;\rho K] \, F_1[\spvar] \, \vac \, \Ket
\eeqn
so that
\eqn
        \| V^{(L=1)}_{\underline{1,0,0,0}}[\z] \|_{1,0}^\sharp
        \; \leq \; \| w_{1,0}[\z;\rho \spvar;\rho K] \|^\sharp_{1,0} \; \| F_0 \|^\sharp_{\Tspace} \;
        \| F_1 \|^\sharp_{\Tspace}  \;,
\eeqn
by the Leibnitz rule, and recalling the definition of the norms
$\|\,\cdot\,\|^\sharp\equiv\|\,\cdot\,\|_{M,N}^\sharp$ in
(\ref{WMNnormdefinition}) and $\|\,\cdot\,\|^\sharp$ in (\ref{Tnormsharpdef}).
A similar calculation is explained in detail in \cite{bcfs2}.
By (\ref {w-deg1-zero-0}) and similar considerations as in (\ref{eq:hatw-auxbd-2}),
\eqn
        \| w_{1,0}[\z;\rho \spvar;\rho K] \|^\sharp_{1,0} &\leq&
        \g \, ( 1 \, + \, |\vp| \, + \, \delta )
        \nonumber\\
        &&
        + \; 2 \, \rho \, \Big(\sum_{1\leq|\ua|\leq2}\| \partial_{\spvar}^\ua \,  w_{1,0} \|_{1,0}
        \, + \, \sum_{|\ua|=1}\| \partial_{|\vp|} \, \partial_{\spvar}^\ua \, w_{1,0} \|_{1,0}
        \nonumber\\
        &&\hspace{2cm}
        \, + \| \partial_{\zet} \, w_{1,0} \|_{1,0}
        \, + \, \sum_{a=0,1}  \| \partial_{|\vp|}^a \, \partial_{|\vk|} \, w_{1,0} \|_{1,0} \Big)
        \nonumber\\
        &\leq&\g \, ( 1 \, + \, |\vp| \, + \, \widehat\delta )
        \, + \, \frac{\g \, \eta}{2}
        \, + \, 2 \, \rho \| w_{1,0} \|_{1,0}^\sharp
        \;,
\eeqn
where the factor $\rho$ enters through the derivatives with respect to $\spvar$, $\zet$ and $|\vk|$.
Moreover,
\eqn
        \| F_0 \|^\sharp_{\Tspace}\,,\,\| F_1 \|^\sharp_{\Tspace}&\leq& C_\Theta  \;.
\eeqn
Consequently,
\begin{align}
        \| V^{(L=1)}_{\underline{1,0,0,0}}[\z] \|_{1,0}^\sharp
        \; \leq \;  C_\Theta^2 \, \Big( \g \, ( 1 \, + \, |\vp| \, + \, \widehat\delta ) \, + \,
        \frac{\g\eta}{2} \, + \, 2 \, \rho \| w_{1,0} \|_{1,0}^\sharp \Big)  \;.
\end{align}
The case for $\widehat w_{0,1}$ is identical.

The sum of terms contributing to $\widehat w_{1,0}$  for $L\geq2$ can be bounded by
\eqn
        \lefteqn{
        2\,C_\Theta^2 \sum_{L=2}^\infty    (L+1)^2
        \Big(\frac{C_\Theta}{\rho}\Big)^L(2\rho )^{M+N}
        \sum_{m_1+\cdots+m_l=M,\atop n_1+\cdots+n_L=N}
        \sum_{p_1,q_1,\dots,p_L,q_L:\atop
        m_\ell+p_\ell+n_\ell+q_\ell\geq1}
        }
        \nonumber\\
        &&\hspace{0.5cm}\,
        \prod_{\ell=1}^L\Big\{\Big(\frac{2}{\sqrt{p_\ell}}\Big)^{p_\ell}
        \Big(\frac{2}{\sqrt{q_\ell}}\Big)^{q_\ell}
        \| w_{M_\ell,N_\ell}[\z] \|_{M_\ell,N_\ell}^\sharp
        \Big\}
        \nonumber\\
        &\leq&2\,C_\Theta^2 \, \rho \, \xi\sum_{L=2}^\infty    (L+1)^2 \, B^L \,
        \nonumber\\
        &\leq&384 \frac{C_\Theta^4}{\rho} \, \xi\,\| \h_1 \|_{\xi}^2 \;,
        \label{hat-w-pnull-bd-1}
\eeqn
with $B$ defined in (\ref{eq:B-aux-def-1}),
similarly as in the discussion of (\ref{hatwMNbd-2}).

In conclusion,
\eqn
        \|\widehat \h_1\|_{\xi}&\leq& \xi^{-1} \| \widehat w_{1,0} \|_{1,0}
        \, + \, \xi^{-1} \| \widehat w_{0,1} \|_{0,1}
        \nonumber\\
        &\leq&\cetaex \,  \g \, \xi^{-1} \, ( 1 \, + \, |\vp| \, + \, \widehat\delta )
        \nonumber\\
        &&\hspace{2cm}+
        \big( 10 C_\Theta^2 \, + \, 1536 \frac{C_\Theta^4 \, \xi \, \eta}{\rho^2} \,  \big) \, \rho \,
        \| \h_{\geq1} \|_{\xi}
        \nonumber\\
        &\leq&\cetaex \, \g \, \xi^{-1} \, (1+|\vp|+\widehat\delta) \, + \, \frac \eta2 \;,
        \label{hatwMNest}
\eeqn
independently of $\sigma$, for $\xi\leq \frac{1}{4}$, with
\eqn
        \eta&\ll&  \rho^3 \;,
        \nonumber\\
        \rho &=&\frac{1}{150 C_\Theta^2} \;,
        \label{eq:const-def-1}
\eeqn
and using $\|\h_{\geq1}\|_\xi<\eta+\e<2\eta$.

\noindent\underline{\em (ii) The case $\sig\geq1$.}
Combined with the additional scaling factor $\rho$ from $\mu(\widehat\sig)=\rho\mu(\sig)$
in (\ref{w-deg1-zero-1}), the arguments used for $M+N\geq2$ in Section {\ref{subsubsec:irell-wMN-1}}
straightforwardly imply
\eqn
        \| \widehat \h_1 \|_{\xi}&\leq&\frac\eta2
\eeqn
if $\sig>1$.

\subsection{Concluding the proof of  Theorem {\ref{thm:codim2contrthm}}}
For
\begin{align}
        \rho \; = \; \min\Big\{\frac{1}{K_\Theta} \, , \,
        \frac{1}{20 C_\Theta^3} \, , \, \frac{1}{150 C_\Theta^2} \, , \, \frac{1}{100}\Big\}
        \; \; \; , \; \; \;
        \xi \; = \; \frac{1}{10} \;,
        \label{eq:rho-xi-parmchoice-1}
\end{align}
(see (\ref{eq:Polyd-def-piop-1}), (\ref{parm-choice-1}), (\ref{eq:rho-T-bd-aux-1}),
(\ref{eq:rho-irr-bd-aux-1}))
and
\eqn
        \gs \; , \; \eta \; , \; \delta \; \ll \; \rho^3 \;,
\eeqn
we conclude that
\eqn
        \ren:\Polyds\Polpar\rightarrow \Polyds\Polparhat
\eeqn
with
\eqn
        \widehat\eta&=&\left\{
        \begin{array}{cl}
        \cetaex \, \g \, \xi^{-1} \, (1+|\vp|+\widehat\delta) \, + \,  \eta/2&{\rm if\;}\sig\leq1 \\
        \eta/2&{\rm if\;}\sig>1
        \end{array}\right.
        \nonumber\\
        \widehat\e&\leq&\frac{\e}{4}+\frac\eta4
        \nonumber\\
        \widehat\delta&\leq&\delta+ \frac\eta2
        \nonumber\\
        \widehat\lTnl&=&\rho \, \lTnl
        \nonumber\\
        \widehat\sigma&=&\rho^{-1}\sigma\;.
        \label{ccontr-def-1}
\eeqn
All constants only depend on the smooth cutoff function $\Theta$
(given that $\rho$ is fixed by (\ref{eq:rho-xi-parmchoice-1})).
This completes the proof of  Theorem {\ref{thm:codim2contrthm}}.
\qedprf

\section{Proof of Theorem {\ref{thm:strong-induct}} }
\label{sect:sind-thm-prf-1}

In this section, we establish the strong induction step
\eqn
        \sind[n-1]\,\Rightarrow\,\sind[n]
        \label{eq:sInd-step-1}
\eeqn
for $n\geq1$ to prove Theorem {\ref{thm:strong-induct}}.
In order to verify (\ref{eq:sInd-step-1}),
we combine Theorem {\ref{thm:codim2contrthm}} with algebraic identities
satisfied by the smooth Feshbach map.

Let $\sigzero$ denote the infrared cutoff in the
original fiber Hamiltonian $\Hpszero$, and
\eqn
    N(\sigzero) \; := \; \left\lceil\frac{\log\frac{1}{\sigzero}}{\log\frac1\rho}\right\rceil \;.
\eeqn
For the range of scales $n\leq N(\sigzero)$, one has $\sig_n=\rho^{-n}\sigzero\leq1$.
As has been noted before,  (\ref{eq:Polparhat-leq1})
in Theorem {\ref{thm:codim2contrthm}} is insufficient to control the growth of the
parameters $\delta_n$ and $\eta_n$.

For the range of scales $n>N(\sigzero)$ where $\sig_n=\rho^{-n}\sigzero>1$,
part (\ref{eq:Polparhat-geq1}) of
Theorem {\ref{thm:codim2contrthm}} implies that $\delta_n$ and $\eta_n$ decay
exponentially. Hence, given $\sind[N(\sigzero)]$,
Theorem {\ref{thm:codim2contrthm}} immediately implies (\ref{eq:sInd-step-1})
for all $n>N(\sigzero)$.

\subsection{Base case: The first decimation step}
\label{subsec:first-F-step-1}

We associate the fiber Hamiltonian $\Hpszero$ with the scale $-1$.
In the first decimation step, the spectrally shifted
fiber Hamiltonian $\Hpszero+E^{(-1)}[\z_{-1}]$ (with $\z_{-1}\in\I(\frac{\vp^2}{2})$) is mapped
to an element
\eqn
        \h^{(0)} \; \in \; \Polyds(\e_0,\delta_0,\eta_0,\lTnl_0,\sigzero) \;.
\eeqn
The parameters $\e_0<\eta_0$ and $\delta_0$ are independent of $\sigzero$, and satisfy
\eqn
    \e_0&\leq&\eta_0 \; < \; c\,\g
    \nonumber\\
    \delta_0&=&C_0\gs
    \nonumber\\
    \eta_0&=&\cetaex \, \g\,\xi^{-1} \, (1+|\vp|+C_0\gs)
    \nonumber\\
    \lTnl_0&=&\frac12 \;,
\eeqn
imposing
\eqn
    \gs \; \ll \; \rho^6
\eeqn
on the finestructure constant, see (\ref{eq:rho-xi-parmchoice-1}).
These results are proved in Section 11 of \cite{bcfs2}.

\subsection{Strong induction step}
\label{ssec:strong-ind-1}
From here on, the parameters $\rho$ and $\xi$ are assigned the fixed values in (\ref{eq:rho-xi-parmchoice-1}).

The strong induction assumption $\sind[n-1]$
states that for all $0\leq k\leq n-1$,
\eqn
        \h^{(k)}\in\Polyds(\e_k,\delta_k,\eta_k,\lTnl_k,\sig_k)
        \label{eq:hk-n-1-sInd-1}
\eeqn
with
\eqn
            \h^{(k)} \; = \; \ren[\h^{(k-1)}] \;\;\;\;{\rm for \; }1\leq k\leq n-1 \;,
\eeqn
and
\eqn
        \begin{array}{rcl}
        \e_k&\leq& \eta_k
        \\
        \delta_k&\leq&\cdstrong   \gs
        \\
        \eta_k&=&  \cetaextw \, \g \, \xi^{-1}  (1+|\vp|+\cdstrong\gs)
        \\
        \lTnl_k&=&\rho^k\lTnl_0 \;\;\;\;\;\;{\rm with}\;\lTnl_0=\frac12
        \\
        \sig_k&=&\rho^{-k}\sigzero \;,
        \end{array}
        \label{eq:sind-nmin1-1}
\eeqn
see Theorem {\ref{thm:strong-induct}}.
The constant  $\cdstrong$ is independent of $n$ and $\sig_0$, and will be determined
in Proposition {\ref{prop:sind-step-prf-1}} below.

To prove Theorem {\ref{thm:strong-induct}}, we assume $\sind[n-1]$,
and infer from Theorem {\ref{thm:codim2contrthm}} that
\eqn
        \h^{(n)} \; = \;
        \ren[\h^{(n-1)}] \; \in \; \Polyds(\e_{n},\delta_{n},\eta_{n},\lTnl_{n},\sig_{n}) \;,
\eeqn
where
\eqn
        \begin{array}{rcl}
        \delta_{n}&\leq&\cdstrong\gs + \frac{\eta_{n-1}}2
        \\
        \eta_{n}&=&
        \cetaex  \, \g \, \xi^{-1} \, (1+|\vp|+\cdstrong\gs)\, + \, \frac{\eta_{n-1}}{2}
        \\
        \lTnl_{n}&=&\rho^{n} \, \lTnl_0
        \\
        \sig_{n}&=&\rho^{-n} \, \sig_0 \;.
        \end{array}
        \label{eq:ind-step-recall-1}
\eeqn
and
\eqn
        \e_{n}\;\leq\;\frac{\e_{n-1}}{4} \, + \, \frac{\eta_{n-1}}4 \;\leq\;  \frac{\eta_{n-1}}2
        \; \leq \;\eta_n\;.
\eeqn
To establish $\sind[n]$, and to determine the constant $C_0$, we use the following
"bootstrap" argument:
We assume that $\sind[n-1]$ holds for an unspecified finite constant $C_0$.
Using this assumption, we
prove (in Propositions {\ref{prop:pfp-bd-1}} and {\ref{prp:derp-pfp-bd-1}} below)
that for all $\gs<\gs_0$ with $C_0\gs_0\ll1$ sufficiently small,
there exists an explicitly computable constant $C_0'$ {\em independent of $n$ and $\gs$} such that
\eqn
        \delta_{k} \; \leq \; C_0'\gs
        \label{eq:delta-n-sInd-bd-1}
\eeqn
for all $k$ with  $0\leq k\leq n$.
Together with $\sind[n-1]$ and Theorem {\ref{thm:codim2contrthm}}, we then find
\eqn
        \eta_{k} \; \leq \; \cetaex  \, \g \, \xi^{-1} \, (\, 1 \, + \, |\vp| \, + \, C_0' \, \gs \,)
        \,+\,\frac{\eta_{k-1}}{2} \;
        \label{eq:eta-n-sInd-bd-1}
\eeqn
for $1\leq k\leq n$, from which one infers
\eqn
        \eta_{n} \; \leq \; \cetaextw \, \g \, \xi^{-1} \, (\, 1 \, + \, |\vp| \, + \, C_0' \, \gs \,) \;.
\eeqn
This implies that in $\sind[n]$, we can choose
\eqn
        C_0 \; = \; C_0' \;,
\eeqn
and since $n$ is arbitrary, this is valid for all $n$.
The details left out here are presented in Proposition {\ref{prop:sind-step-prf-1}}.

Let $\z_{n} \in \I$ denote the spectral parameter corresponding to $\h^{(n)}[\z_n]$.
The spectral parameters $\z_k$ associated to $\h^{(k)}[\z_k]$, for $0\leq k< n$,
are recursively defined by
\eqn
        J_{(k)}:\z_k\,\mapsto\,\z_{k+1} \; = \; \Ez^{(k)}[\z_{k}] \;,
\eeqn
see (\ref{eq:Erho-def-1}), and
\eqn
        \z_{k} \; = \; J_{(k,n)}^{-1}[\z_n] \; := \; (J_{(k)})^{-1}\circ\cdots\circ(J_{(n-1)})^{-1}[\z_{n}]
        \label{eq:Jmn-def-1}
\eeqn
for $-1\leq k<n$. Furthermore, $\z_{0}=J_{(-1)}[\z_{-1}]$ is obtained in the first
decimation step. For a detailed discussion of this part, we refer to \cite{bcfs1,bcfs2}.


For notational convenience, we write
\eqn
        \hfp_{(k)}&:=&\hfp^{(k)}[\z_k;\vp]
        \nonumber\\
        H_{(k)}&:=&H[\h^{(k)}[\z_k;\vp]]
\eeqn
and
\eqn
        Q_{(k)} &:=&Q_{\chi_\rho[H_f]}
        (H_{(n)},\hfp_{(k)}H_f)
        \nonumber\\
        Q_{(k)}^{\sharp}&:=&Q_{\chi_\rho[H_f]}^{\sharp}
        (H_{(n)},\hfp_{(k)}H_f)
        \;.
\eeqn
For $\vu\in S^2$, let $\psi_\vu\in\C^2$ with $\|\psi_\vu\|_{\C^2}=1$,
$\bra \, \psi_\vu \, , \vec\pauli \, \psi_\vu \, \ket = \vu$, and we define
\eqn
    \vacpsi \; := \; \psi_\vu\,\otimes\,\vac \;.
\eeqn
To establish (\ref{eq:delta-n-sInd-bd-1}),
we prove that the coefficient
\eqn
        \pfp_{(n)} &\equiv& \pfp^{(n)}[\z_n;\vp]\;:=\;  \bra \vacpsi\,,\,\partial_{\Ppar}H_{(n)}\vacpsi\ket
\eeqn
of the marginal operator $\Ppar$ in the non-interacting part of $H[\h^{(n)}[\z_n]]$ satisfies
\eqn
    |\,  \pfp_{(n)}+|\vp| \, | \,,\,| \, \partial_{|\vp|}\pfp_{(n)}\,+\,1 \, |\;<\; c\gs \;,
    \label{eq:pfp-bds-statement-1}
\eeqn
where the constant  is independent of $n$ and $\sigzero$.
This in turn directly implies (\ref{eq:eta-n-sInd-bd-1})
via  the soft photon sum rules, as explained in Section
{\ref{subsubsec:marg-kernels-1}}.

To prove (\ref{eq:pfp-bds-statement-1}), we invoke the following identities which are
provided by Lemma 15.2 in \cite{bcfs2}.

\begin{lemma}
\label{first-last-scale-lemma-1}
For $n> m\geq0$, let
\eqn
        Q_{(m,n)}&:=&Q_{(m)}\Gamma_\rho^* Q_{(m+1)}\Gamma_\rho^*
        \cdots Q_{(n-1)}\Gamma_\rho^*\;,
        \nonumber\\
        Q_{(m,n)}^\sharp&:=&\Gamma_\rho
        Q_{(n-1)}^\sharp\Gamma_\rho \cdots Q_{(m+1)}^\sharp\Gamma_\rho
        Q_{(m)}^\sharp
\eeqn
and
\eqn
        Q_{(-1,n)}&:=&Q_{(-1)}Q_{(0,n)}\;,
        \nonumber\\
        Q_{(-1,n)}^\sharp&:=& Q_{(0,n)}^\sharp Q_{(-1)}^\sharp\;.
\eeqn
Then, the identities
\eqn
        H_{(m)}Q_{(m,n)}&=&\Big[\prod_{k=m}^{n-1}\hfp_{(k)}\Big]
        \rho^{n-m_+} (\Gamma_\rho^*)^{n-m_+} \chi_1[H_f] H_{(n)} \;,
        \nonumber\\
        Q_{(m,n)}^\sharp H_{(m)}&=&
        \Big[\prod_{k=m}^{n-1}\hfp_{(k)}\Big]\rho^{n-m_+}  H_{(n)}\chi_1[H_f] (\Gamma_\rho)^{n-m_+} \;,
        \label{HQ-id-min1-n-1}
\eeqn
and
\eqn
        Q^\sharp_{(m,n)}H_{(m)}Q_{(m,n)}
        \; =\;\Big[\prod_{k=m}^{n-1}\hfp_{(k)}\Big] \rho^{n-m_+}
        \Big[H_{(n)} - H_{(n)}\bar\chi_1[H_f]H_f^{-1}\bar\chi_1[H_f]H_{(n)}\Big]
        \nonumber\\
        \label{QHQ-id-basic-1}
\eeqn
hold for all $m$ with $-1\leq m < n$ and $m_+:=\max\{m,0\}$.
\end{lemma}

Some basic properties of the vectors
$Q_{(-1,n)}\vacpsi\in\C^2\otimes\Fo$ and $Q_{(m,n)}\vacpsi\in\H_{red}$
are summarized in the following proposition.

\begin{proposition}
\label{prop:derHf-H-1}
Assume that $\sind[n-1]$ holds for $n\geq0$. Then,
\begin{align}
        \Bra \,\vacpsi \, ,\, Q_{(-1,n)}^\sharp Q_{(-1,n)}\vacpsi\,\Ket
        \; = \; \Big\|\,Q_{(-1,n)}\vacpsi \,\Big\|^2
    \; = \; \Big[\prod_{l=-1}^{n-1}\hfp_{(l)}\Big]
        (1-\err_n^{(1)}) \; ,
        \label{eq:derHfQHQ-aux-2}
\end{align}
where $\err_n^{(1)}$ is defined in (\ref{eq:errn-def-1}),
and
\eqn
    | \, \err_n^{(1)} \, | \; , \; | \, \partial_{|\vp|}\err_n^{(1)} \, | \; < \; c \, \eta_n^2 \;.
\eeqn
In particular,
\eqn
    \Big|\prod_{k=-1}^n\hfp_{(k)}\Big| \; < \; \exp\Big[c \, \sum_{k=0}^n \eta_k^2\Big]
    \; < \; \exp\Big[c \, \gs \min\{n, N(\sigzero)\}\Big]
    \label{eq:prodhfp-bd-1}
\eeqn
for constants $c$ which are independent of $n$.

Moreover,
\eqn
          1 \; < \; \Big\|\,Q_{(m,n)}\vacpsi\,\Big\|
          \; \leq \;
         \Big\|\,Q_{(-1,n)}\vacpsi\,\Big\|     \;,
         \label{eq:prodhfp-id-1}
\eeqn
for any $m$  with  $-1\leq m<n$, and
\eqn
        \Big\| H_f^{\frac s2}  \, Q_{(-1,n)} \vacpsi\, \Big\|^2 \; < \;
        \frac{c \, \gs}{s} \, \Big\| Q_{(-1,n)} \vacpsi \, \Big\|^2 \;,
        \label{eq:Hfpsi-est-1}
\eeqn
for any $0<s\leq1$, where the constant $c$ is independent of $n$, $\sigzero$,
$\gs$, and $s$.
\end{proposition}

\prf
We first of all note that since the spectral parameters $\z_k$, $-1\leq k \leq n$,
are real-valued, $ Q_{(m,n)}^\sharp$ is the adjoint of $Q_{(m,n)}$, and we immediately have
\eqn
    \Bra \, \vacpsi \, ,\, Q_{(m,n)}^\sharp \, Q_{(m,n)} \, \vacpsi \, \Ket & = &
    \Big\| Q_{(m,n)}\vacpsi \, \Big\|^2
    \label{eq:Psi-real-norm-1}
\eeqn
for all $-1\leq m \leq n$.

The following result can be straightforwardly
adopted from Lemma 15.3 in \cite{bcfs2}.
For $0\leq|\vp|<\puppbd$, and any choice of $m$ with $-1\leq m<n$, we have
\eqn
        &&\Bra\vacpsi\,,\,\partial_{H_f}H_{(n)}\vacpsi\Ket
        \label{derHf-Hn-id-1}
        \\
        &&\hspace{1cm}=\,\Big[\prod_{k=-1}^{n-1}\hfp_{(k)}^{-1}\Big]
        \,\Bra\vacpsi \,,\, Q_{(-1,n)}^\sharp\,(\partial_{H_f}H_{(-1)})\,Q_{(-1,n)}\vacpsi\Ket+\err_n  \;,
        \nonumber
\eeqn
where the error term is defined by
\eqn
        \err_n^{(1)} \; = \;  (I_1) \, + (I_2) \, + \, (II)
        \label{eq:errn-def-1}
\eeqn
with
\eqn
        (I_1)&=&
        \Big[\prod_{j=-1}^{n-1}\hfp_{(j)}^{-1}\Big]\, \Bra \vacpsi\,,\,Q_{(-1,n)}^\sharp H_{(-1)}
        \partial_{H_f}Q_{(-1,n)}\vacpsi\Ket
        \nonumber\\
        (I_2)&=&
        \rho^{-n}\Big[\prod_{j=-1}^{n-1}\hfp_{(j)}^{-1}\Big]\,
        \Bra \vacpsi \,,\,(\partial_{H_f}Q_{(-1,n)}^\sharp) H_{(-1)} Q_{(-1,n)}\vacpsi\Ket
        \; = \; (I_1)^*
        \; \; \; \; \;
\eeqn
and
\eqn
        (II) \; = \; \Bra\vacpsi\,,
        \,\partial_{H_f}\Big(H_{(n)}\bar\chi_1[H_f]H_f^{-1}
        \bar\chi_1[H_f]H_{(n)}\Big)\vacpsi\Ket \;.
\eeqn
Using (\ref{HQ-id-min1-n-1}), and
\eqn
        \partial_{H_f}\Gamma_\rho \; = \; \rho\,\Gamma_\rho\partial_{H_f}
        \;\;\;,\;\;\;
        \partial_{H_f}\Gamma_\rho^* \; = \; \rho^{-1}\Gamma_\rho^*\partial_{H_f} \;,
\eeqn
we find
\eqn
        (I_1)&=&\rho^{n}\Bra \vacpsi\,,\,H_{(n)}\chi_1[H_f] \Gamma_\rho^{n}
        \partial_{H_f}Q_{(-1,n)}\vacpsi\Ket
        \nonumber\\
        &=&\sum_{j=-1}^{n-1}\rho^{n-j} \, \Bra \vacpsi\,,\,H_{(n)}\Gamma_\rho^{n}
        \nonumber\\
        &&\hspace{3cm}\chi_1[\rho^{-n}H_f]
        Q_{(-1,j-1)} (\partial_{H_f}Q_{(j)})\Gamma_\rho^*Q_{(j+1,n)}\vacpsi\Ket
        \nonumber\\
        &=&\rho \, \Bra  \vacpsi\,,\,H_{(n)}\chi_1[H_f]\Gamma_\rho
        \partial_{H_f}Q_{(n-1)}  \vacpsi\Ket \;.
\eeqn
Here, we used
\eqn
        \chi_1[\rho^{-n}H_f]Q_{(-1,j-1)}&=&\chi_1[\rho^{-n}H_f]Q_{(-1)}Q_{(0)}\Gamma_\rho^*
        Q_{(1)}\Gamma_\rho^*\cdots Q_{(j-1)}\Gamma_\rho^*
        \nonumber\\
        &=&\chi_1[\rho^{-n}H_f] Q_{(0)}\Gamma_\rho^*
        Q_{(1)}\Gamma_\rho^*\cdots Q_{(j-1)}\Gamma_\rho^*
        \nonumber\\
        &=&\Gamma_\rho^*\chi_1[\rho^{-n+1}H_f]
        Q_{(1)}\Gamma_\rho^*\cdots Q_{(j-1)}\Gamma_\rho^*
        \nonumber\\
        &=&
        \cdots \; = \; (\Gamma_\rho^*)^k\chi_1[\rho^{-n+k}H_f]
        Q_{(k)}\Gamma_\rho^*\cdots Q_{(j-1)}\Gamma_\rho^*
        \nonumber\\
        &=&(\Gamma_\rho^*)^{j}\chi_1[\rho^{-n+j}H_f]\;,
    \label{eq:chir-Qk-aux-red-1}
\eeqn
since for all $r>1$,
\eqn
        \chi_1[\rho^{-r}H_f]Q_{(k)}&=&
        \chi_1[\rho^{-r}H_f](\chi_\rho[H_f]
        \nonumber\\
        &&\hspace{2cm}-\bar\chi_\rho[H_f]\bar R_{(k)}\bar\chi_\rho[H_f]
        (H_{(k)}-\tau_{(k)}H_f)\chi_\rho[H_f])
        \nonumber\\
        &=&
        \chi_1[\rho^{-r}H_f]\chi_\rho[H_f]
        \nonumber\\
        &=&\chi_1[\rho^{-r}H_f]  \;,
    \label{eq:chir-Qk-aux-red-2}
\eeqn
see also (\ref{FeshComp-2}), and
Lemma 15.3 in \cite{bcfs2}. By $\sind[n-1]$, we conclude that
\eqn
        |(I_1)| \; \leq \; \|W_{(n)}\|_{op} \|\partial_{H_f}Q_{(n-1)}\vacpsi\|
        \;\leq\; c \, \eta_n^2
\eeqn
for a constant which is independent of $n$ and $\sigzero$ (the constant only depends on
$\rho$, which is fixed by  (\ref{eq:rho-xi-parmchoice-1}) in this part of the analysis).
The term $(I_2)$ can be treated in
the same way.
Moreover, it is easy to see that
\eqn
        |(II)| \; \leq \; c \, \big(\|W_{(n)}\|_{op} \, + \|\partial_{H_f}W_{(n)}\|_{op}\big)^2
        \;\leq\; c\,\eta_n^2 \;.
\eeqn
Thus, $\err_n^{(1)}$ depends only on the effective Hamiltonian
on the last scale $n$, and from $\sind[n-1]$ follows that
\eqn
        |\err_n^{(1)}|  \; < \; c\, \eta_n^2 \;.
\eeqn
We note that
\eqn
       |\partial_{|\vp|}\err_n^{(1)}| \; < \; c\, \eta_n^2
\eeqn
is obtained from a similar analysis.

Since $\partial_{H_f}H_{(-1)}=1$, we find
\eqn
        \Bra \vacpsi \, ,\, Q_{(-1,n)}^\sharp Q_{(-1,n)}\vacpsi\,\Ket
        \; = \; \Big[\prod_{l=-1}^{n-1}\hfp_{(l)}\Big]
        (1-\err_n^{(1)}) \; ,
\eeqn
which establishes (\ref{eq:derHfQHQ-aux-2}).

To prove (\ref{eq:prodhfp-bd-1}), we recall from (\ref{eq:Ezlemma-eq-2}) that
$|\hfp_{(k)}-1|< c \eta_k^2$, one gets
\eqn
        \Big|\prod_{k=-1}^{N(\sigzero)}\hfp_{(k)}\Big|
        \; < \; \exp\Big[ N(\sigzero)\sup_{0\leq k\leq N(\sigzero)}\eta_k^2  \Big]
\eeqn
and
\eqn
        \Big|\prod_{k=N(\sigzero)+1}^{\infty}\hfp_{(k)}\Big|
        & < & \exp\Big[c \, \eta_{N(\sigzero)}^2\sum_{k>N(\sigzero)} 2^{-2(k-N(\sigzero))}\Big]
        \nonumber\\
        & < & \exp\Big[c \, \eta_{N(\sigzero)}^2 \Big]\;.
\eeqn
Since  by $\sind[n-1]$,
\eqn
        \sup_{0\leq k < n}\eta_k \; < \;  c \,\xi^{-1}\gs \;,
        \label{eq:supkn-aux-1}
\eeqn
holds for $n=N(\sigzero)$,
and $\xi$ is independent of $\gs$ and $\sigzero$, the claim follows.

To prove (\ref{eq:prodhfp-id-1}), let $P_1[H_f]=\chi[H_f<1]$. We have
\eqn
    \| Q_{(m,n)}\vacpsi\|\|Q_{(-1,n)}\vacpsi\|
    &=&
    \|(\Gamma_\rho^*)^m P_1[H_f]Q_{(m,n)}\vacpsi\|
    \|Q_{(-1,n)}\vacpsi\|
    \nonumber\\
    &\geq&\Big|\Bra (\Gamma_\rho^*)^m P_1[H_f]Q_{(m,n)}\vacpsi \,,\,Q_{(-1,n)}\vacpsi\Ket\Big|
    \nonumber\\
    &=&\Big|\Bra Q_{(m,n)}\vacpsi \,,\,\Gamma_\rho^m P_1[\rho^{-m}H_f]Q_{(-1,n)}\vacpsi\Ket\Big|
    \nonumber\\
    &=&\Big|\Bra Q_{(m,n)}\vacpsi \,,\,\Gamma_\rho^m (\Gamma_\rho^*)^m P_1[H_f]Q_{(m,n)}\vacpsi\Ket\Big|
    \nonumber\\
    &=&\|Q_{(m,n)}\vacpsi\|^2 \;,
\eeqn
which follows from the same considerations as in (\ref{eq:chir-Qk-aux-red-1}) and (\ref{eq:chir-Qk-aux-red-2}).
Thus,
\eqn
    \| Q_{(m,n)}\vacpsi\| \; \leq \; \|Q_{(-1,n)}\vacpsi\| \;.
\eeqn
Moreover, one easily sees that
\eqn
        \bra\vacpsi\,,\,Q_{(m,n)}\vacpsi\ket \; = \; \bra\vacpsi\,,\,\vacpsi\ket \; = \; 1 \;,
\eeqn
hence
\eqn
        \Big\|Q_{(m,n)}\vacpsi\Big\| \; \geq \; 1
        \label{eq:derHfQHQ-aux-3}
\eeqn
for any $-1\leq m<n\in\N_0$.
This implies (\ref{eq:prodhfp-id-1}).

To prove (\ref{eq:Hfpsi-est-1}), we observe that
\eqn
        \Big\| H_f^{\frac s2}  \, Q_{(m,n)} \vacpsi\Big\| &\leq&
        \Big\|H_f^{\frac s2}  \,\chi_\rho[H_f] \Gamma_\rho^* Q_{(m+1,n)} \vacpsi \Big\|
        \nonumber\\
        &&\hspace{1cm} + \,
        \Big\|H_f^{\frac s2} Q_{(m)}' \,\Gamma_\rho^* Q_{(m+1,n)} \vacpsi \Big\|
\eeqn
where for $0\leq m <n$,
\eqn
        Q_{(m)}' \; := \; \bar\chi_\rho[H_f] \bar R_{(m)}\bar\chi_\rho[H_f]
        (H_{(m)}-\tau_{(m)}H_f)\chi_\rho[H_f] \;,
\eeqn
on $\H_{red}$, and $Q_{(m)}=\chi_\rho[H_f]-Q_{(m)}'$. For $m=-1$,
\eqn
        Q_{(-1)}' \; := \; \bar\chi_1[H_f] \bar R_{(-1)}\bar\chi_1[H_f]
        (H_{(-1)}-\tau_{(-1)}H_f)\chi_1[H_f]
\eeqn
on $\C^2\otimes\Fo$ and $Q_{(-1)}=\chi_1[H_f]-Q_{(-1)}'$.

Next, we use the estimate
\eqn
        \Big\|Q_{(m)}' \Gamma_\rho^* Q_{(m+1)}\Big\| \; < \;
        \frac{c(\eta_{m}+\eta_{m+1})}{\rho^3} \;< \; \frac{c\eta_m}{\rho^3}
        \label{eq:Q-Q-prod-est-1}
\eeqn
from Lemma 12.2 in \cite{bcfs2}.
It implies that
\eqn
        \Big\| H_f^{\frac s2}  \, Q_{(m,n)} \vacpsi\Big\| &\leq&
        \rho^{\frac s2}\Big\|H_f^{\frac s2}    Q_{(m+1,n)} \vacpsi \Big\|
        \nonumber\\
        &&\hspace{1cm}+ \, \frac{c\eta_m}{\rho^3}
        \Big\|\chi_\rho[H_f] Q_{(m+2,n)} \vacpsi \Big\|
        \nonumber\\
        &<&
        \rho^{\frac s2}\Big\|H_f^{\frac s2}    Q_{(m+1,n)} \vacpsi \Big\|
        \nonumber\\
        &&\hspace{1cm}+ \, \frac{c\eta_m}{\rho^3}(1-c\eta_m)
        \Big\|\chi_\rho[H_f] Q_{(-1,n)} \vacpsi \Big\| \;,
\eeqn
using Lemma {\ref{prop:derHf-H-1}}.
Thus, iterating,
\eqn
        \Big\|H_f^{\frac12}  \, Q_{(0,n)}\Big\|&<&\sup_{0\leq m<n}
        \Big\{\frac{c\eta_m}{(1-\rho^{\frac s2})\rho^3}(1-c\eta_m)\Big\}
        \Big\|\chi_\rho[H_f] Q_{(-1,n)} \vacpsi \Big\|
        \nonumber\\
        &<& \frac{c\g}{(1-\rho^{\frac s2})\rho^3}
        \Big\|\chi_\rho[H_f] Q_{(-1,n)} \vacpsi \Big\| \;,
\eeqn
by use of $\sup_{0\leq m<n}\eta_m<c\g$, which follows from $\sind[n]$.

Hence, from $1-\rho^{\frac s2}\sim s$ as $s\searrow0$,
\eqn
        \Big\| H_f^{\frac12}  \, Q_{(-1,n)} \vacpsi\Big\|
        &\leq&\Big\|H_f^{\frac s2}  \, Q_{(0,n)}\Big\|+\|H_f^{\frac s2}Q_{(-1)}'Q_{(0,n)}\vacpsi\Big\|
        \nonumber\\
        &\leq&\Big\|H_f^{\frac s2}  \, Q_{(0,n)}\Big\| + \, \frac{c \, \g}{s} \, \Big\|Q_{(-1,n)}\vacpsi\Big\|
\eeqn
using
\eqn
        \|H_f^{\frac s2}Q_{(-1)}'Q_{(0,n)}\vacpsi\Big\|&\leq&\Big\|H_f^{\frac s2}Q_{(-1)}'Q_{(0)}\Big\|_{op} \,
        \Big\|Q_{(1,n)}\vacpsi\Big\|
        \nonumber\\
        &\leq&\frac{c \, \g}{s} \, \Big\|Q_{(1,n)}\vacpsi\Big\|
        \nonumber\\
        &\leq&\frac{c \, \g}{s} \, \Big\|Q_{(-1,n)}\vacpsi\Big\|  \;,
\eeqn
and
\eqn
        \Big\|H_f^{\frac s2}Q_{(-1)}'Q_{(0)}\Big\|_{op}
        &\leq&\Big\|H_f^{\frac s2}|\bar R_{(-1)}|^{\frac12}\Big\|_{op}
        \nonumber\\
        &&
        \,\Big\||R_{(-1)}|^{\frac12}\bar\chi_1[H_f] (H_{(-1)}-\tau_{(-1)}H_f)Q_{(0)}\Big\|_{op}
        \label{eq:Hf-srho-H0-est-aux-1}
\eeqn
with
\eqn
        \Big\|H_f^{\frac s2}|\bar R_{(-1)}|^{\frac12}\Big\|_{op}
        &<&c
        \nonumber\\
        \Big\||R_{(-1)}|^{\frac12}\bar\chi_1[H_f] (H_{(-1)}-\tau_{(-1)}H_f)Q_{(0)}\Big\|_{op}
        &<& c \, \g \;.
        \label{eq:Hf-srho-H0-est-aux-2}
\eeqn
(\ref{eq:Hf-srho-H0-est-aux-1}) and (\ref{eq:Hf-srho-H0-est-aux-2})
are obtained straightforwardly from
results in Section 13.1.1 of \cite{bcfs2}.
This implies  (\ref{eq:Hfpsi-est-1}).
\endprf





\begin{remark}
\label{rem:real-spec-var}
A key reason we are using
spectral parameters in $\R$, as opposed to $\C$ in \cite{bcfs2},
is because then, (\ref{eq:Psi-real-norm-1}) is available.
In \cite{bcfs2}, the fact that the interaction is irrelevant
makes an application of  (\ref{eq:Psi-real-norm-1}) unnecessary.
\end{remark}

\begin{proposition}
\label{prop:pfp-bd-1}
For $n\geq0$,
\eqn
        \pfp_{(n)} \; = \; \frac{\bra \, \vacpsi\,,\,Q_{(-1,n)}^\sharp
        (\partial_{\Ppar}H_{(-1)}) Q_{(-1,n)} \, \vacpsi \, \ket}
        {\bra \, \vacpsi \, , \, Q_{(-1,n)}^\sharp Q_{(-1,n)} \, \vacpsi \, \ket} (1- \err_n^{(1)})+\err_n^{(2)}  \;,
        \label{eq:pfp-n-id-1}
\eeqn
where $\err_n^{(2)}=O(\eta_n^2)$ is defined in (\ref{eq:pfp-I-II-def-1}).

For $\gs<\gs_0$ with $C_0\gs\ll1$ sufficiently small (see (\ref{eq:sind-nmin1-1})),
\eqn
        |\,\pfp_{(n)} \, | \; < \; c_0 \, \gs \;
        \label{eq:pfp-n-pnull-est-1}
\eeqn
where the constant $c_0$ is independent of $n$ and $\gs$.
\end{proposition}

\prf
From (\ref{QHQ-id-basic-1}), we find
\eqn
        \pfp_{(n)} \; = \; \bra \vacpsi \,,\,H_{(n)}\vacpsi \ket
        \; = \; \Big[\prod_{k=-1}^n\hfp_{(k)}^{-1}\Big]\big({\rm main}_n + \err_n^{(2)}\big)
        \label{eq:pfp-main-err-1}
\eeqn
where
\eqn
        {\rm main}_n &=&\rho^{-n}
        \Bra\vacpsi\,,\,\partial_{\Ppar}(Q_{(-1,n)}^\sharp  H_{(-1)}  Q_{(-1,n)})\vacpsi\Ket
        \nonumber\\
        \err_n^{(2)}&=&-\Bra\vacpsi\,,\,\partial_{\Ppar}(H_{(n)}
        \bar\chi_\rho[H_f]H_f^{-1} \bar\chi_\rho[H_f] H_{(n)} )\vacpsi\Ket \;.
        \label{eq:pfp-I-II-def-1}
\eeqn
It is easy to verify that
\eqn
        |\err_n^{(2)}| \; , \; |\partial_{|\vp|}\err_n^{(2)}| \; < \; c \, \eta_{n}^2 \;.
        \label{eq:beta-II-errbd-aux-1}
\eeqn
From Lemma 15.5 in \cite{bcfs2},
\eqn
        {\rm main}_n \; = \; \Bra\vacpsi\,,\,Q_{(-1,n)}^\sharp (\partial_{\Ppar}H_{(-1)})
        Q_{(-1,n)}\vacpsi\Ket \;.
\eeqn
The factor $\rho^{-n}$ is eliminated by pulling the differentiation operator $\partial_{\Pf}$
through the $n$ rescaling operators $\Gamma_\rho$ in $Q_{(-1,n)}^\sharp$ from the left.

Using Proposition {\ref{prop:derHf-H-1}}, we thus find
\eqn
        \pfp_{(n)} \; = \; \frac{\bra \vacpsi\,,\,Q_{(-1,n)}^\sharp
        (\partial_{\Ppar}H_{(-1)}) Q_{(-1,n)}\vacpsi\ket}
        {\bra \vacpsi\,,\, Q_{(-1,n)}^\sharp Q_{(-1,n)}\vacpsi \ket} (1- \err_n^{(1)})+
        \err_n^{(3)} \;,
\eeqn
where $\err_n^{(1)}$ is defined in (\ref{eq:errn-def-1}). This establishes (\ref{eq:pfp-n-id-1}).
From
\eqn
        \partial_{\Ppar} H_{(-1)} \; = \;\partial_{\Ppar}\Hpszero
        \; = \; -|\vp|+\Ppar+\g\Apar \;,
\eeqn
one finds
\eqn
        \pfp_{(n)} \; = \; -|\vp|\,+\,\err_n^{(3)}
\eeqn
with
\eqn
    \err_n^{(3)} & := & |\vp| \, \err_n^{(1)} \,+\, \err_n^{(2)}
    \nonumber\\
    &&+\,\frac{\bra \vacpsi \,,\, Q_{(-1,n)}^\sharp (\Ppar+\g\Apar) Q_{(-1,n)}\vacpsi\ket}
        {\bra\vacpsi\,,\, Q_{(-1,n)}^\sharp Q_{(-1,n)} \vacpsi\ket}(1- \err_n^{(1)})
        \;.
    \label{eq:tildeerr-def-1}
\eeqn
This establishes (\ref{eq:pfp-n-id-1}).

Let $\Af=\Af^{+}+\Af^{-}$, where $\Af^{-}$ is the term involving annihilation operators.
From the Schwarz inequality,
\eqn
        \big\| \Af^{-} \, Q_{(-1,n)} \, \vacpsi \, \big\| \; < \; c \,
        \big\| H_f^{\frac12} \,  Q_{(-1,n)} \, \vacpsi \, \big\| \;.
\eeqn
Moreover, $|\Pf|\leq H_f$. Thus, using Proposition {\ref{prop:derHf-H-1}},
\eqn
        \lefteqn{
        \Big| \, \frac{\bra \vacpsi \, , \, Q_{(-1,n)}^\sharp \, (\Ppar+\g\Apar) \, Q_{(-1,n)}\vacpsi\ket}
        {\bra\vacpsi\,,\, Q_{(-1,n)}^\sharp \, Q_{(-1,n)} \vacpsi\ket } \, \Big|
        }
        \nonumber\\
        &<&\frac{ \| H_f^{\frac12} Q_{(-1,n)} \, \vacpsi \|^2
        \, + \, \g \| Q_{(-1,n)} \, \vacpsi \|
        \| H_f^{\frac12} Q_{(-1,n)}\vacpsi\|}
        {\| Q_{(-1,n)} \, \vacpsi \|^2 }
        \nonumber\\
        &<&c \, \gs   \;,
        \label{eq:pfp-err-est-aux-1}
\eeqn
uniformly in $n$.
\endprf




\begin{proposition}
\label{prp:derp-pfp-bd-1}
Assume that $\sind[n-1]$ holds for $n\geq0$. Then,
\eqn
        \partial_{|\vp|} \pfp_{(n)} \; = \; \Big[-1+2\frac{\bra\vacpsi\,,\,(\partial_{|\vp|}Q_{(-1,n)}^\sharp)
        H_{(-1)}(\partial_{|\vp|}  Q_{(-1,n)})\vacpsi\ket}
        {\bra\vacpsi\,,\, Q_{(-1,n)}^\sharp  Q_{(-1,n)}\vacpsi\ket}
        \Big](1-\err_n^{(1)})
        \nonumber\\
         + \,  \err_n^{(3)} \;, \; \; \; \; \;  
        \label{eq:derppfp-id-err-1}
\eeqn
where $| \err_n^{(3)}| \; < \; c \, \eta_n^2$.

Moreover, for $|\vp|<\puppbd$ and $C_0\gs\ll1$ sufficiently small (see (\ref{eq:sind-nmin1-1})),
\eqn
        |\, \partial_{|\vp|}\pfp_{(n)} + 1 \, | \; < \; c_0  \gs \; ,
        \label{eq:derp-pfp-unifbd-1}
\eeqn
where the constant $c_0$ is independent of $n$ and $\gs$.
\end{proposition}

\prf
To prove (\ref{eq:derppfp-id-err-1}), let
\eqn
        \Psi_{(-1,n)} \; := \; Q_{(-1,n)}\vacpsi \;.
\eeqn
We recall (\ref{eq:pfp-main-err-1}) whereby
\eqn
        \partial_{|\vp|}\pfp_{(n)}&=&\partial_{|\vp|}\Big[
        \frac{\bra \Psi_{(-1,n)}  \,,\,(\partial_{\Ppar}H_{(-1)})\Psi_{(-1,n)} \rangle}
        {\bra \Psi_{(-1,n)} \,,\, \Psi_{(-1,n)} \ket}(1-\err_n^{(1)})+\err_n^{(2)}\Big]
        \nonumber\\
        &=&\Big[-\frac{\bra\partial_{|\vp|}\Psi_{(-1,n)} \,,\, \Psi_{(-1,n)} \ket
        +\bra \Psi_{(-1,n)} \,,\, \partial_{|\vp|} \Psi_{(-1,n)} \ket}
        {\bra \Psi_{(-1,n)}  \,,\, \Psi_{(-1,n)} \ket}
        \nonumber\\
        &&\hspace{3cm}\cdot\;\frac{\bra \Psi_{(-1,n)}  \,,\,(\partial_{\Ppar}H_{(-1)})\Psi_{(-1,n)} \rangle}
        {\bra \Psi_{(-1,n)} \,,\, \Psi_{(-1,n)} \ket}
        \nonumber\\
        &&+ \, \frac{\bra \partial_{|\vp|}\Psi_{(-1,n)} \,,\,
        (\partial_{\Ppar}H_{(-1)})\Psi_{(-1,n)} \ket }{\bra \Psi_{(-1,n)} \,,\, \Psi_{(-1,n)} \ket}
        \\
        &&+ \, \frac{ \bra \Psi_{(-1,n)} \,,\,
        (\partial_{\Ppar}H_{(-1)})\partial_{|\vp|}\Psi_{(-1,n)} \ket}{\bra \Psi_{(-1,n)} \,,\, \Psi_{(-1,n)} \ket}
        \; - \; 1 \, \Big] \, ( 1 \, - \, \err_n^{(1)} )
        \nonumber\\
        &&- \, \frac{\bra \Psi_{(-1,n)}  \,,\,(\partial_{\Ppar}H_{(-1)})\Psi_{(-1,n)} \rangle}
        {\bra \Psi_{(-1,n)} \,,\, \Psi_{(-1,n)} \ket}\;\partial_{|\vp|}\err_n^{(1)}
        \;+\;\partial_{|\vp|}\err_n^{(2)} \;,
        \nonumber
\eeqn
using $\partial_{|\vp|}\partial_{\Ppar}H_{(-1)}=1$. The error terms
$\err_n^{(1)}$ and $\err_n^{(2)}$ are defined in (\ref{eq:errn-def-1}) and (\ref{eq:pfp-I-II-def-1}),
respectively.

From (\ref{HQ-id-min1-n-1}), we find
\eqn
        \lefteqn{
        \Big| \Bra \partial_{|\vp|}\Psi_{(-1,n)}\,,\,\partial_{|\vp|}(H_{(-1)}\Psi_{(-1,n)})\Ket \Big|
        }
        \nonumber\\
        &\leq&\rho^{n}\Big[ \, \sum_{\ell=-1}^{n-1} | \, \partial_{|\vp|} \, \hfp_{(\ell)}\, |
        \prod_{j=-1 \atop j\neq\ell}^{n-1}|\hfp_{(j)}| \, \Big]
        \Big| \, \Bra \, \partial_{|\vp|} \, \Psi_{(-1,n)} \, , \, \Gamma_\rho^{-n} \,
        \chi_1[H_f] \, H_{(n)} \, \vacpsi \, \Ket \, \Big|
        \nonumber\\
        && \hspace{1cm}+ \; \rho^{n} \, \Big[ \, \prod_{j=-1}^{n-1}| \, \hfp_{(j)} \, | \, \Big]
        \Big|\Bra \partial_{|\vp|} \Psi_{(-1,n)}\,,\,\Gamma_\rho^{-n}
        \chi_1[H_f] \, \partial_{|\vp|} \, H_{(n)} \, \vacpsi \, \Ket \, \Big|\;
        \label{eq:derp-Q-error-1}
\eeqn
using (\ref{eq:prodhfp-bd-1}). By (\ref{eq:hfp-der-1}), we have
\eqn
        \sum_{\ell=-1}^{n-1} | \, \partial_{|\vp|} \, \hfp_{(\ell)}\, |
        \prod_{j=-1 \atop j\neq\ell}^{n-1}|\hfp_{(j)}|
        &\leq& n\,\Big[ \, \sup_\ell\,| \, \partial_{|\vp|} \, \hfp_{(\ell)}\, | \, \Big]
        \,\prod_{j=-1 \atop j\neq\ell}^{n-1}(1 \, + \, c\,\eta_j^2)
        \nonumber\\
        &\leq& c \, n \, \eta_n^2 \, \exp\Big( c \, \sum_{j=-1}^{n-1} \eta_j^2 \Big)
        \nonumber\\
        & < & c \, n \, \gs \, \exp( c \, \gs \, n \,)
\eeqn
and using (\ref{FeshComp-2})
(see also (\ref{eq:Qj-chi-l-null-1}) and the subsequent discussion), we find
\eqnn
        (\ref{eq:derp-Q-error-1})
        &\leq&
        \big( 1 \, + \, c \, \gs \, n \big) \,
        \rho^{n} \, \exp( c \, \gs^2 \, n ) \,
        \sum_{a=0,1} \, \Big| \, \Bra \, \partial_{|\vp|} \, Q_{( n)} \, \vacpsi \, , \,
        \chi_1[H_f] \, \partial_{|\vp|}^a \, H_{(n)} \, \vacpsi \, \Ket \, \Big|
        \nonumber\\
        &\leq&
        \big( 1 \, + \, c \, \gs \, n \big) \,
        \rho^{n} \, \exp( c \, \eta_0^2 \, n \big) \, \Big\|
        \partial_{|\vp|}Q_{( n)} \, \vacpsi\, \Big\|
        \,\sum_{a=0,1} \, \Big\| \partial_{|\vp|}^a \, H_{(n)} \, \vacpsi \, \Big\|
       \nonumber\\
        &\leq&c \, \eta_n^2
\eeqnn
for $\rho\leq\frac{1}{100}$ (see (\ref{eq:rho-xi-parmchoice-1})).
Thus,
\eqn
         &&\bra \, \Psi_{(-1,n)} \, , \, ( \partial_{|\vp|} \, H_{(-1)} ) \, \Psi_{(-1,n)} \, \ket
         \\
         &&\hspace{3cm}= \; - \, \bra \, \Psi_{(-1,n)} \, , \,
         H_{(-1)} \, \partial_{|\vp|} \, \Psi_{(-1,n)} \, \ket\;
         + \; O( \eta_n^2 ) \;.
         \nonumber
\eeqn
Moreover, from
\eqn
        H_{(-1)} \; = \; T_{(-1)} \, + \, W_{(-1)} \, + \, J_{(-1,n)}^{-1}[\z_n]
\eeqn
(see (\ref{eq:Jmn-def-1}) for the definition of $J_{(m,n)}$), and
\eqn
        \partial_{|p|}H_{(-1)} \; = \; - \, \partial_{\Ppar}H_{(-1)} \,
        + \, \partial_{|\vp|} \, J_{(-1,n)}^{-1}[\z_n]
\eeqn
(from $\partial_{|\vp|}\Hpszero=-\,\partial_{\Ppar}\Hpszero$), we find
\begin{align}
        \partial_{|\vp|}J_{(-1,n)}^{-1}[\z_n] \; = \;
        \frac{\bra \Psi_{(-1,n)}  \,,\,(\partial_{\Ppar}H_{(-1)})\Psi_{(-1,n)} \rangle}
        {\bra \Psi_{(-1,n)} \,,\, \Psi_{(-1,n)} \ket}
    \, + \, O( \eta_n^2 )
\end{align}
(noting that $\|\Psi_{(-1,n)}\|\geq1$, since $\bra\vacpsi,\Psi_{(-1,n)}\ket=1$). Thus,
\eqnn
        \partial_{|\vp|}\pfp_{(n)} \; = \; \Big[-1 \, + \, 2\frac{ \bra \partial_{|\vp|}\Psi_{(-1,n)} \,,\,
        H_{(-1)} \partial_{|\vp|}\Psi_{(-1,n)} \ket}{\bra \Psi_{(-1,n)} \,,\, \Psi_{(-1,n)} \ket}\Big]
        (1-\err_n^{(1)})
        \,+\,\err_n^{(3)}
\eeqnn
where
\eqn
         \err_n^{(3)}&:=&-\,\frac{ \bra  \Psi_{(-1,n)} \,,\,
         (\partial_{\Ppar}H_{(-1)})  \Psi_{(-1,n)} \ket}{\bra \Psi_{(-1,n)} \,,\, \Psi_{(-1,n)} \ket}\;
         \partial_{|\vp|}\err_n^{(1)} \,
        +\,\partial_{|\vp|}\err_n^{(2)} \; + \; O( \eta_n^2 )
        \nonumber\\
        &=&-\,(\pfp_{(n)}-\err_n^{(n)})\,
        \frac{ \partial_{|\vp|}\err_n^{(1)} }{1-\err_n^{(n)}}
        \,+\,\partial_{|\vp|}\err_n^{(2)} \; + \; O( \eta_n^2 ) \;.
    \label{eq:derp-pfp-id-1}
\eeqn
To estimate $\err_n^{(3)}$, we note that by (\ref{eq:sind-nmin1-1}) and (\ref{eq:ind-step-recall-1})
(which are based on $\sind[n-1]$ and Theorem {\ref{thm:codim2contrthm}}),
\eqn
        | \, \pfp_{(n)} \, - \, |\vp| \, | &\leq& C_0\gs \, + \,\frac{\eta_{n-1}}{2}
        \nonumber\\
        &\leq&C_0\gs \, + \,  \cetaex  \, \g \, \xi^{-1} \,
        ( 1 \, + \, |\vp| \, + \, \cdstrong \, \gs )
        \; < \; \,  c  \, \g
\eeqn
Hence, by Proposition {\ref{prop:pfp-bd-1}} and (\ref{eq:beta-II-errbd-aux-1}),
\eqn
        |\err_n^{(3)}|&\leq&c \, \g \,\frac{|\partial_{|\vp|} \, \err_n^{(1)}|}{1 \, - \, |\err_n^{(1)}|}
        \,+\,|\partial_{|\vp|}\err_n^{(2)}| \,+\, c \, \eta_n^2
        \nonumber\\
        &\leq&\,\frac{c\g\, \eta_n^2 }{1 \, - \, c  \gs}
        \, +   c  \, \eta_n^2
        \nonumber\\
        &\leq&c \, \eta_n^2 \;,
\eeqn
where the constants are independent of $n$.

To prove (\ref{eq:derp-pfp-unifbd-1}), we use
\eqn
        \partial_{|\vp|}Q_{(m,n)} \; = \; (\partial_{|\vp|}Q_{(m)})\Gamma_\rho^*Q_{(m+1,n)}
        \, + \, Q_{(m)}\Gamma_\rho^*\partial_{|\vp|}Q_{(m+1,n)}
\eeqn
and
\eqn
        H_{(m)}Q_{(m)} \; = \; \hfp_{(m)} \, \rho \, \Gamma_\rho^* \, \chi_1[H_f] \, H_{(m+1)} \;.
\eeqn
Clearly,
\eqn
        \lefteqn{
        \bra\vacpsi\,,\,(\partial_{|\vp|}Q_{(-1,n)}^\sharp)
        H_{(-1)}(\partial_{|\vp|}  Q_{(-1,n)})\vacpsi\ket
        }
        \nonumber\\
        &=&\sum_{-1\leq j,k<n}\Bra \, \vacpsi \, , \, Q_{(j+1,n)}^\sharp
        \, \Gamma_\rho \, (\partial_{|\vp|}Q_{(j)}^\sharp)
        \nonumber\\
        &&\hspace{3cm}
        Q_{(-1,j-1)} \,
        H_{(-1)} \, Q_{(-1,k-1)}
        \label{eq:dpQHdpQ-aux-1}
        \\
        &&\hspace{5cm}
        (\partial_{|\vp|}  Q_{(k)})
        \, \Gamma_\rho^* \, Q_{(k+1,n)} \, \vacpsi \, \Ket \;.
        \nonumber
\eeqn
Let us consider the case $j\leq k$.

We first show that the terms with $j<k$ vanish. One has
\eqn
        Q_{(-1,j-1)} H_{(-1)}Q_{(-1,k-1)}
        &=&Q_{(-1,j-1)} H_{(-1)}Q_{(-1,l-1)} Q_{(l,k-1)}
        \nonumber\\
        &=&\Big[\prod_{l=-1}^{j-1}\hfp_{(l)}\Big]\rho^{k-j}\Big\{H_{(j)}Q_{(j,k-1)}
        \\
        &&\hspace{1cm}-H_{(j)}
        \bar\chi_1[H_f]H_f^{-1}\bar\chi_1[H_f]H_{(j)}Q_{(j,k-1)}\Big\} \;,
        \nonumber
\eeqn
where the second term in the brackets vanishes unless $j=k$, since
\eqn
        \bar\chi_1[H_f]H_{(j)}Q_{(j,k-1)}
        &=&\rho^{k-j}\bar\chi_1[H_f]\chi_{\rho^{k-j}}[H_f]H_{(k-1)}
        \nonumber\\
        &=&0 \;\;{\rm if } \; j<k
\eeqn
(because $\bar\chi_1[H_f]\chi_{\rho^{l}}[H_f]=0$ for all $l>0$).

Thus, assuming that $j<k$, the corresponding term in (\ref{eq:dpQHdpQ-aux-1}) is given by
\eqn
        &&\rho^{k}\Bra\vacpsi\,,\,Q_{(j+1,n)}^\sharp\Gamma_\rho(\partial_{|\vp|}Q_{(j)}^\sharp)
        \nonumber\\
        &&\hspace{3cm}
        \chi_{\rho^{k-j} }[H_f]
        H_{(k)}
        \\
        &&\hspace{5cm}
        (\partial_{|\vp|}  Q_{(k)})
        \Gamma_\rho^* Q_{(k+1,n)}\vacpsi\Ket \;.
        \nonumber
\eeqn
However, for any $l>0$,
\eqn
        \partial_{|\vp|}Q_{(j)}^\sharp\chi_{\rho^{l} }[H_f]
        \; = \; 0 \;,
        \label{eq:Qj-chi-l-null-1}
\eeqn
since the kernel of $\partial_{|\vp|}Q_{(j)}^\sharp$ in $\H_{red}$
is contained in $\Ran(\bar\chi_\rho[H_f])$.

We thus conclude that
\eqn
        \Bra\vacpsi\,,\,(\partial_{|\vp|}Q_{(-1,n)}^\sharp)
        H_{(-1)}(\partial_{|\vp|}  Q_{(-1,n)})\vacpsi\Ket
        \; = \; \sum_{-1\leq  k<n} A_k
        \label{eq:dpQHdpQ-aux-2}
\eeqn
where
\eqn
        A_k&:=&\rho^{k-1}\Big[\prod_{l=-1}^{k-1}\hfp_{(l)}\Big]
        \Bra\vacpsi\,,\,Q_{(k+1,n)}^\sharp\Gamma_\rho(\partial_{|\vp|}Q_{(k)}^\sharp)
        \nonumber\\
        &&\hspace{3.5cm}
        \Big[H_{(k)}
        -H_{(k)} \bar\chi_1[H_f]H_f^{-1}\bar\chi_1[H_f]H_{(k)} \Big]
        \\
        &&\hspace{6cm}
        (\partial_{|\vp|}  Q_{(k)})
        \Gamma_\rho^* Q_{(k+1,n)}\vacpsi\Ket \;,
        \nonumber
\eeqn
with $|A_{-1}| \, < \, c\gs$ (uniformly in $\sigzero\geq0$) from \cite{bcfs2}.
Next, we use
\eqn
        \Big\|(\partial_{|\vp|}  Q_{(k)})\Gamma_\rho^* Q_{(k+1)}\Big\|_{op}
        \; \leq \;  c \, (\eta_k+\eta_{k+1})
        \; \leq \;  c \,  \eta_k   \;,
\eeqn
see Lemma 12.2 in \cite{bcfs2} where the constants are $O(\rho^{-3})$ (we
recall that $\rho$ has been fixed for
this part of our analysis, see  (\ref{eq:rho-xi-parmchoice-1})).
Moreover,
\eqn
        \Big\|Q_{(k+1,n)}\vacpsi\Big\| \; \leq \; (1+c\eta_k) \,
        \Big|\prod_{l=-1}^{k-1}\hfp_{(l)}^{-1}\Big| \, \Big\|Q_{(-1,n)}\vacpsi\Big\|\;.
\eeqn
and
\eqn
        \|H_{(k)}\|_{op} \; < \; c \;.
\eeqn
Hence,
\eqn
        0 \; \; < \; \; (\ref{eq:dpQHdpQ-aux-2})&<&| \, A_{-1} \, | \, + \, c \,
        \gs \, \Big[ \, \sum_{0\leq k<n} \, \rho^{k-1} \, \Big] \,
        \Big\| Q_{(-1,n)} \, \vacpsi \, \Big\|^2
        \nonumber\\
        &<&c \, \gs \, \Big\| Q_{(-1,n)} \, \vacpsi \, \Big\|^2
        \;,
\eeqn
where the constant $c$ is independent of $n$, $\sigzero$ and $\gs$.

Collecting our results, we have established that
\eqn
        | \, \partial_{|\vp|}\pfp_{(n)} \, + \, 1 \,| \; < \; c_0\, \gs
\eeqn
where the constant $c_0$ is independent of $n$ and $\gs$
(and also from  $\e_n\leq\eta_n<c\g$ and $\delta_n<C_0\gs$).
\endprf

\begin{proposition}
\label{prop:sind-step-prf-1}
Let $C_0$ denote the constant in the
definition of $\sind[n]$ in Theorem {\ref{thm:strong-induct}}, and
let $c_0$ denote the constant in  (\ref{eq:derp-pfp-unifbd-1}).
Then, for
\eqn
        C_0 \; = \; 2\,c_0 \;
\eeqn
the strong induction assumption $\sind[n-1]$ implies $\sind[n]$
for any $n\geq0$,
provided that $\gs\leq\gs_0$ with $c_0\gs_0\ll1$ sufficiently small (independently of $\sigzero$),.
\end{proposition}

\prf
We recall the discussion at the beginning of Section {\ref{ssec:strong-ind-1}}. 
We first assume for arbitrary $n\geq1$ that $\sind[n-1]$ holds for an unspecified,
finite constant $C_0$. Moreover, we assume that $\gs<\gs_0$ with  $C_0\gs_0\ll1$
sufficiently small such that (\ref{eq:pfp-n-pnull-est-1}) and (\ref{eq:derp-pfp-unifbd-1}) hold.

Then, Propositions {\ref{prop:pfp-bd-1}} and {\ref{prp:derp-pfp-bd-1}}
imply that there exists an explicitly computable constant $c_0$ independent of $n$ and $\gs$ such that
\eqn
    \sum_{a=0,1}| \, \partial_{|\vp|}^a (\pfp_{(n)} \, + \,|\vp|) \, | \; \leq \; 2\,c_0 \,\gs  \;.
\eeqn
Since $\delta_n$ is by definition an upper bound on the left hand side, we can choose
\eqn
        \delta_n \; \leq \; 2 \, c_0 \, \gs \;,
        \label{eq:deltn-sind-bd-1}
\eeqn
where $c_0$ is the same constant as in (\ref{eq:pfp-n-pnull-est-1}) and (\ref{eq:derp-pfp-unifbd-1}).
Likewise, the same argument implies for all $0\leq k\leq n$ that $\delta_k\leq 2c_0\gs$,
for the given, unspecified choice of $C_0$.

By Theorem {\ref{thm:codim2contrthm}}, this implies that
\eqn
    \eta_k \; \leq \; \cetaex \, \g \, ( 1 \, + \, |\vp| \, + \, 2 \, c_0 \, \gs ) \, + \, \frac{\eta_{k-1}}{2}
\eeqn
for all $1\leq k \leq n$.
Thus,
\eqn
    \eta_n & \leq & \sum_{k=0}^n \, \cetaex \, \g \, ( 1 \, + \, |\vp| \, + \, 2 \, c_0 \, \gs ) \, 2^{-k}
    \nonumber\\
    & = & \cetaextw \, \g \, ( 1 \, + \, |\vp| \, + \, 2 \, c_0 \, \gs ) \;.
    \label{eq:etan-sind-bd-1}
\eeqn
This establishes $\sind[n]$ with
\eqn
        C_0 \; = \; 2 \, c_0 \;.
        \label{eq:C0-marg-const-1}
\eeqn
Since $n$ was arbitrary, and $c_0$ is independent of $n$, this implies that $\sind[n]$
holds for $C_0=2\,c_0$ and all $n$,
provided that $\gs\leq \gs_0$ with $c_0\gs_0\ll1$ sufficiently small,
and $\gs_0$ independent of $\sigzero$.
\endprf



\section{Proof of Theorem  {\ref{mainthm-1}}}
\label{sec:main-thm-proof-1}

The  proof of Theorem  {\ref{mainthm-1}} can be straightforwardly completed
by use of Theorems {\ref{thm:codim2contrthm}} and {\ref{thm:strong-induct}}.
We will in fact demonstrate that as $n\rightarrow\infty$,
$$
        \pfp_{(n)} \; \longrightarrow \;   - \, \partial_{|\vp|}\Egrdzero
$$
and
$$
        \partial_{|\vp|} \pfp_{(n)} \; \longrightarrow \;-\, m_{ren}(\vp,\sigzero)^{-1} \; ,
$$
where the sequence of spectral parameters $(\z_n)_{n\geq0}$ in $\pfp_{(n)}$ is chosen suitably.

We have proved in the previous sections that $\sind[n]$ holds for all $n$, and that
\eqn
        | \, \pfp_{(n)} \, + \, |\vp| \, | \; \; \; , \; \; \;
        | \, \partial_{|\vp|}\pfp_{(n)} \, + \, 1 \, | \; < \; c_0 \, \gs
\eeqn
hold uniformly in $n$.
But this implies for the renormalized infrared mass that
\eqn
        | \, m_{ren}(\vp,\sigzero) \, - \, 1 \, | \; < \; c_0 \, \gs
\eeqn
uniformly in $\sigzero$.

For this part of the analysis, we will extensively apply
constructions and results from \cite{bcfs2} to abbreviate our discussion.

\subsection{Reconstruction of the ground state}

We determine the ground state eigenvalue $E(\vp,\sigzero)$ of $H(\vp,\sigzero)$ and its 2-dimensional
eigenspace. This is accomplished by combining Theorem {\ref{thm:codim2contrthm}} and
Theorem {\ref{thm:strong-induct}} with arguments from \cite{bcfs2}.

As proved in Section {\ref{sect:sind-thm-prf-1}},
the property $\sind[n]$ formulated in Theorem {\ref{thm:strong-induct}} holds for all $n\in\N_0$.
This implies the following:
\begin{itemize}
\item
For $n\leq N(\sigzero)$, we have
\eqn
    \delta_n&\leq&C_0 \, \gs
    \nonumber\\
    \e_n \;\; \leq \;\; \eta_n&\leq&C_1 \, \g
    \nonumber\\
    \sig_n&=&\rho^{-n}\sigzero
    \nonumber\\
    \lTnl_n&=&\rho^n\lTnl_0
    \label{eq:parm-flow-1}
\eeqn
for constants $C_0$, $C_1$ independent of $n$, $\sigzero$, and $\gs$.
\item
For $n>N(\sigzero)$,  we have
\eqn
    \delta_n&\leq&C_0 \, \gs
    \nonumber\\
    \e_n \; \leq \; \eta_n &\leq& 2^{-(n-N(\sigzero))_+} \, C_1 \, \g
    \nonumber\\
    \sig_n&=&\rho^{-n} \, \sigzero
    \nonumber\\
    \lTnl_n&=&\rho^n \, \lTnl_0 \;.
    \label{eq:parm-flow-2}
\eeqn
\end{itemize}

We let
\eqn
        E_{(n)}[\z;\vp] \; := \; w_{0,0}^{(n)}[\z;\unull;\vp]
\eeqn
and recall from Lemma {\ref{Ezlemma}} that
\eqn
        J_{(n)} \; : \; {\mathcal U}_{(n)}\rightarrow \I
        \; \; , \; \;
        \z \mapsto (\hfp[\z;\vp] \rho)^{-1} E_{(n)}[\z;\vp] \;,
\eeqn
where
$$
        {\mathcal U}_{(n)} \; := \; {\mathcal U}[\h^{(n)}]
        \; = \; \big\{\z\in\I\,\big|\, |E_{(n)}[\z;\vp]| \leq\frac{\rho}{\Iconst}\big\} \;.
$$
We define for $-1\leq n<m$
\eqn
        e_{(n,m)} \; := \; J_{(n)}^{-1}\circ\cdots\circ J_{(m)}^{-1}[0] \;\;\in\R\;.
\eeqn
By the same arguments as in the proof of Theorem 12.1 in \cite{bcfs2},
\eqn
        e_{(n,\infty)} \; := \; \lim_{m\rightarrow\infty}e_{(n,m)} \;\;\in\R
\eeqn
exists, and by construction,
\eqn
        J_{(n)}[e_{(n,\infty)}] \; = \; e_{(n+1,\infty)} \;.
\eeqn
Moreover,
\eqn
        |e_{(n,\infty)}| \; < \; 2^{-(n-N(\sigzero))_+}\eta_0 \;,
\eeqn
which tends to zero at an exponential rate as $n\rightarrow0$.

Let
\eqn
        \widetilde H_{(n)} & := & H[\h^{(n)}[e_{(n,\infty)};\vp]]
        \nonumber\\
        \widetilde\hfp_{(n)} & := & \hfp^{(n)}[e_{(n,\infty)};\vp]
    \nonumber\\
    \widetilde\pfp_{(n)}&:=&\pfp^{(n)}[e_{(n,\infty)};\vp]
\eeqn
and
\eqn
        \widetilde Q_{(n)}^{(\sharp)} \; := \; Q_{\chi_\rho[H_f]}^{(\sharp)}
        (\widetilde H_{(n)},\widetilde\hfp_{(n)}H_f) \;.
\eeqn
Moreover, for $-1\leq m < n$, let
\eqn
        \widetilde Q_{(m,n)}&:=&
        \widetilde Q_{(m)} \Gamma_\rho^* \widetilde Q_{(1)}
        \cdots \cdots \Gamma_\rho^* \widetilde Q_{(n-1)}\Gamma_\rho^*
        \nonumber\\
        \widetilde Q_{(m,n)}^\sharp&=&
        \Gamma_\rho\widetilde Q_{(n-1)}^\sharp\Gamma_\rho \cdots\cdots
        \widetilde Q_{(1)}^\sharp\Gamma_\rho
        \widetilde Q_{(m)}^\sharp   \;.
\eeqn
and
\eqn
        \widetilde Q_{(-1,n)}&:=& \widetilde Q_{(-1)}
        \widetilde Q_{(0,n)}
        \nonumber\\
        \widetilde Q_{(-1,n)}^\sharp&=&
        \widetilde Q_{(0,n)}^\sharp \widetilde Q_{(-1)}^\sharp \;.
\eeqn
We emphasize that as before, $\widetilde Q_{(m,n)}^\sharp$ is the adjoint of $\widetilde Q_{(m,n)}$,
since the spectral parameters $e_{(n,\infty)}$ are real-valued.

\begin{proposition}
\label{prop:eigen-eq-est-1}
There exists a constant $\gs_0>0$ (independent of $\sigzero$) such that
for any $\sigzero>0$, and all $\gs<\gs_0$,
the infimum $\Egrdzero$ of the spectrum of $\Hpszero$
is an eigenvalue of multiplicity two at the bottom of the essential
spectrum. For $\vu\in S^2$, let $\psi_\vu\in\C^2$ with $\|\psi_\vu\|_{\C^2}=1$,
$\bra \, \psi_\vu \, , \vec\pauli \, \psi_\vu \, \ket = \vu$, and
$\vacpsi = \psi_\vu \otimes \vac$.

Then, for any choice of $\psi_\vu$, the strong limit
\eqn
        \label{eq:Omgrdzeropsi-def-1}
        \Omgrdzeropsi \; := \; s-\lim_{n\rightarrow\infty} \, \widetilde Q_{(-1,n)} \, \vacpsi
\eeqn
exists in $\C^2\otimes\Fo$. Under the normalization condition
\eqn
        \Bra \, \vacpsi \,,\,\Omgrdzeropsi \, \Ket_{\C^2\otimes\Fo} \; = \; 1 \;,
\eeqn
we have
\eqn
    \Big\|\,\Omgrdzeropsi\,\Big\|_{\C^2\otimes\Fo}&\leq&\exp\big( c \, \gs \, N(\sigzero) \big)
    \; < \; \infty \;,
    \label{eq:Omgrd-norm-est-1}
\eeqn
and
\eqn
        \Hpszero \, \Omgrdzeropsi \; = \; \Egrdzero \, \Omgrdzeropsi \;,
        \label{eq:eigen-eq-1}
\eeqn
with $\bra \, \Omgrdzeropsi \, , \vec\pauli \, \Omgrdzeropsi \, \ket = \vu$.
Hence, $\Omgrdzeropsi$ is an element of the 2-dimensional eigenspace corresponding to
the ground state eigenvalue $\Egrdzero$ at the infimum of the spectrum of $\Hpszero$.


More generally, for any $n\geq0$ and any choice of $\psi_\vu$, the strong limit
\eqn
        \widetilde\Psi_{(n,\infty)} \; := \; s-\lim_{m\rightarrow\infty} \widetilde Q_{(n,m)} \vacpsi
        \label{eq:Psi-n-grd-ex-1}
\eeqn
exists in $\H_{red}$, and
\eqn
        \widetilde H_{(n)}\widetilde\Psi_{(n,\infty)} \; = \; 0 \;.
        \label{eq:Psi-n-grd-eq-1}
\eeqn
The vector $\widetilde\Psi_{(n,\infty)}$ belongs to the 2-dimensional eigenspace corresponding
to the ground state eigenvalue 0 of $\widetilde H_{(n)}=H[\h^{(n)}[e_{(n,\infty)}]]$.
\end{proposition}


\prf
Proposition {\ref{prop:eigen-eq-est-1}} corresponds to Theorem 12.1 in \cite{bcfs2}
(for spin 0 and $\gs<\gs_0(\sigzero)$).
Given (\ref{eq:parm-flow-1}) and (\ref{eq:parm-flow-2}),
which are uniform in $\sigzero$, we can straightforwardly adapt the proof of
Theorem 12.1 in \cite{bcfs2} to our situation.
Accordingly, we shall omit some of the details in our exposition,
and refer to \cite{bcfs2} instead.

We first establish the existence of the strong limit
\eqn
        \Omgrdzeropsi \; = \; s-\lim_{n\rightarrow\infty} \, \widetilde Q_{(-1,n)} \, \vacpsi
\eeqn
in $\C^2\otimes\Fo$.
To this end, we verify that the sequence $(\widetilde Q_{(-1,n)}\vacpsi)_{n\geq0}$
is Cauchy in $\C^2\otimes\Fo$.
We have for any $m>N(\sigzero)$
\eqn
        \Big\| \widetilde Q_{(-1,m)}\vacpsi\,-\,
        \widetilde Q_{(-1,m+1)}\vacpsi\, \Big\|_{\C^2\otimes\Fo}
        &\leq& \| \widetilde Q_{(-1,m)}\|_{op} \,
        \Big\| (\widetilde Q_{(m)}-\chi_\rho[H_f])\vacpsi\, \Big\|_{\H_{red}}
        \nonumber\\
        &\leq&\|\widetilde Q_{(-1)} \widetilde Q_{(0)}\|_{op} \,
        \Big[\prod_{j= 1 \atop j \; odd}^{m-1}
        \|\widetilde Q_{(j)}\Gamma_\rho^*\widetilde Q_{(j+1)}\|_{op}\Big]\,
        \nonumber\\
        &&\hspace{2cm}
        \|  (\widetilde Q_{(m)}-\chi_\rho[H_f])\vacpsi\|_{\H_{red}}
        \nonumber\\
        &\leq&
        c\,\eta_m \, \Big[\prod_{j=-1 \atop j \; odd}^{m-1} (1+c\eta_j+c\eta_{j+1})\Big]
        \nonumber\\
        &\leq&c\, \eta_m \, \exp\big(c\sum_{j=-1}^{m-1}\eta_j\big)
        \\
        & \leq &  c\,\g\,\exp\big(c \g N(\sigzero)\big)\,2^{-(m-N(\sigzero))_+} \;,
        \nonumber
\eeqn
see (\ref{eq:Q-Q-prod-est-1}),
(\ref{eq:parm-flow-1}), (\ref{eq:parm-flow-2}), combined with
$\|(\widetilde Q_{(m)}-\chi_\rho[H_f])\vacpsi\|<c \, \eta_m$, see \cite{bcfs2}.
Therefore, for any $n>m$,
\eqn
        \Big\| \widetilde Q_{(-1,m)}\vacpsi\,-\, \widetilde Q_{(-1,n)}\vacpsi\, \Big\|_{\C^2\otimes\Fo}
        &\leq&c\,\g\,\exp\big(c \g N(\sigzero)\big)\,\sum_{j=m}^{n-1}2^{-(j-N(\sigzero))_+}
        \nonumber\\
        &\leq&2 \, c\,\g\,\exp\big(c \g N(\sigzero)\big)\, 2^{-(m-N(\sigzero))_+}\;.
\eeqn
Since the upper bound converges to zero as $m\rightarrow\infty$,
$(\widetilde Q_{(-1,n)}\vacpsi)_{n\geq0}$ is a Cauchy sequence in $\C^2\otimes\Fo$.
For a detailed exposition, we refer to \cite{bcfs1, bcfs2}.
Moreover, we have
\eqn
    \lim_{n\rightarrow\infty}\Big\|\,\widetilde Q_{(-1,n)}\vacpsi\,\Big\|^2_{\C^2\otimes\Fo}
    &=& \lim_{n\rightarrow\infty}
    \Big[\prod_{j=-1}^n\widetilde\hfp_{(n)}\Big]\,(1-\widetilde\err_n^{(1)})
    \nonumber\\
    &\leq&\exp\big(c\gs N(\sigzero)\big)\;<\;\infty
\eeqn
from Proposition {\ref{prop:derHf-H-1}}, and
\eqn
    \Bra\,\vacpsi \, , \, \widetilde Q_{(-1,n)} \, \vacpsi \, \Ket_{\C^2\otimes\Fo} \; = \; 1 \;,
\eeqn
independently of $n$, which implies (\ref{eq:Omgrd-norm-est-1}).

Furthermore,
\eqn
        \widetilde H_{(-1)}\Omgrdzeropsi\;=\;0
\eeqn
follows from an iterated application of (\ref{HQ-id-min1-n-1}), and is equivalent to
(\ref{eq:eigen-eq-1}).

Since the choice of $\psi_\vu\in\C^2$ was arbitrary,
and $H_{(-1)}=\Hpszero-\Egrdzero$ is independent of $\psi$, (\ref{eq:eigen-eq-1})
implies that
\eqn
        \;\;\;\;\;\;\;
        \Big\{\, \Omgrdzero \; {\rm as \; in \; } (\ref{eq:Omgrdzeropsi-def-1})\,
        \Big| \, \psi_\vu \, \in \, \C^2 \; , \;
        \bra \, \psi_\vu \, , \, \vec \pauli \, \psi_\vu \, \ket \; = \; \vu \, \Big\}
        \; \subset\; \C^2\otimes\Fo
\eeqn
is the 2-dimensional eigenspace corresponding to $\Egrdzero$, the ground state eigenvalue
of $\Hpszero$,
which borders without a gap to  ${\rm ess\,spec}\Hpszero$.


For general $n$, (\ref{eq:Psi-n-grd-ex-1}) and (\ref{eq:Psi-n-grd-eq-1}) follow from the
same reasoning. Since $\widetilde H_{(n)}$ is independent of $\psi_\vu$, and
(\ref{eq:Psi-n-grd-eq-1}) holds for any choice of $\psi_\vu$,
the eigenspace associated to the eigenvalue 0 at the bottom of the
spectrum of $\widetilde H_{(n)}$ is given by
$\{\widetilde\Psi_{(n,\infty)} {\rm \; as \; in \; }(\ref{eq:Psi-n-grd-ex-1})
\,\big|\,\psi_\vu\in\C^2\,,\,\|\psi_\vu\|_{\C^2}=1\}\subset\H_{red}$.
For further details, we refer to \cite{bcfs1,bcfs2}.
\endprf

\subsection{Infrared mass renormalization}

Finally, we establish the uniform bounds on the renormalized electron mass
stated in Theorem {\ref{mainthm-1}}. To this end, we use the
Feynman-Hellman formula
\eqn
    \partial_{|\vp|}\Egrdzero \; = \; \frac{\bra \Omgrdzeropsi\,,\,
        (\partial_{|\vp|}\Hpszero) \Omgrdzeropsi \ket}
        {\bra \Omgrdzeropsi \,,\,\Omgrdzeropsi \ket} \;,
    \label{eq:Feyn-Hell-1}
\eeqn
and
\eqn
        \partial_{|\vp|}^2\Egrdzero  &=&
        1 \, - \, 2\frac{\bra \partial_{|\vp|}\Omgrdzeropsi\,,\,
        (\Hpszero-\Egrdzero) \, \partial_{|\vp|}\Omgrdzeropsi\ket}
        {\bra \Omgrdzeropsi \, , \, \Omgrdzeropsi\ket}
        \nonumber\\
        &<& 1 \;.
        \label{eq:Feyn-Hell-2}
\eeqn
A detailed discussion of these formulas is given in \cite{bcfs2}.

\begin{proposition}
For $\gs<\gs_0$ with $\gs_0$ sufficiently small, independently of $\sigzero$,
the renormalized electron mass is bounded by
\eqn
        1 \; < \; m_{ren}(\vp,\sigzero) \; < \; 1 \, + \, c_0\,\gs \;,
\eeqn
with a constant $c_0$ independent of $\sigzero\geq0$.

\end{proposition}

\prf
For $n\geq0$, $\vu\in S^2$, and any $\psi_\vu\in\C^2$ with $\bra \, \psi_\vu \, , \, \vec \pauli \,
\psi_\vu \, \ket = \vu$, we have that
\eqn
        \widetilde \pfp_{(n)} \; = \;  \bra \, \vacpsi \, , \, \partial_{\Pf}\widetilde H_{(n)}
        \, \vacpsi \, \ket \;,
\eeqn
with $\vacpsi=\psi_\vu\otimes\vac$, satisfies
\eqn
        | \, \widetilde \pfp_{(n)} \, + \, |\vp| \, | \; < \; c_0 \, \gs \;,
\eeqn
uniformly in $n$, where
\eqn
        \widetilde H_{(n)} \; = \; H[\h^{(n)}[e_{(n,\infty)}]] \;.
\eeqn
We have
\eqnn
        \widetilde\pfp_{(n)} \; = \;
        \frac{\bra \, \widetilde Q_{(-1,n)} \, \vacpsi \, , \,
        \big( \partial_{\Ppar}\Hpszero \big) \widetilde Q_{(-1,n)} \, \vacpsi \, \ket}
        {\bra \widetilde Q_{(-1,n)}\vacpsi\,,\, \widetilde Q_{(-1,n)}\vacpsi\ket}
        (1- \widetilde\err_n^{(1)} )
        \, + \, \widetilde\err_n^{(2)} \;,
\eeqnn
where
\eqn
        \widetilde{\err}_n^{(j)} \; := \; \err_n^{(j)}\Big|_{\z_n=e_{(n,\infty)}}
        \;\;\;,\;\;j=1,2,3\;,
\eeqn
and
\eqn
    |\widetilde\err_n^{(1)}| \,,\,|\widetilde\err_n^{(2)}|& \leq & c\eta_n^2 \; \searrow \; 0
    \;\;\;\;\;(n\rightarrow\infty) \;,
\eeqn
see (\ref{eq:errn-def-1}) and Proposition
{\ref{prop:derHf-H-1}}.
Hence, with $\Omgrdzeropsi$ as in (\ref{eq:Omgrdzeropsi-def-1}), we find
\eqnn
        \lim_{n\rightarrow\infty}\widetilde\pfp_{(n)} \; = \;
        - \, \frac{\bra \Omgrdzeropsi\,,\,
        \big( \partial_{|\vp|}\Hpszero \big) \Omgrdzeropsi \ket}
        {\bra \Omgrdzeropsi \,,\,\Omgrdzeropsi \ket}
        \; = \;
        - \, \partial_{|\vp|}\Egrdzero \;,
\eeqnn
which follows from
\eqn
    \partial_{|\vp|}\Hpszero \; = \; - \, \partial_{\Ppar}\Hpszero \;,
\eeqn
and the Feynman-Hellman formula (\ref{eq:Feyn-Hell-1}).
Hence, by Proposition {\ref{prop:derHf-H-1}},
\eqn
        | \, \partial_{|\vp|}\Egrdzero-|\vp| \, | \; < \; c\gs \;,
\eeqn
uniformly in $\sigzero$.

To estimate $\partial_{|\vp|}^2\Egrdzero$,
we consider
\eqnn
        \partial_{|\vp|}  \widetilde\pfp_{(n)} & = &
        (\partial_{|\vp|} \pfp_{(n)})\Big|_{\z_n=e_{(n,\infty)}}
        + \, (\partial_{\z_n}\pfp_{(n)})\Big|_{\z_n=e_{(n,\infty)}}\partial_{|\vp|}e_{(n,\infty)} \;.
\eeqnn
We recall from Propositions {\ref{prop:pfp-bd-1}} and {\ref{prp:derp-pfp-bd-1}} that
\eqn
        \sup_{\z_n\in \I}|\, \partial_{|\vp|} \, \widetilde\pfp_{(n)}  \, + \, 1\, |
        \;,\;
        \sup_{\z_n\in \I}|\, \partial_{\z_n} \, \widetilde\pfp_{(n)} \,|\; < \; c\gs \;,
\eeqn
where $c$ is independent of $n$ and $\gs$.

Moreover, we observe that
\eqn
        \lim_{n\rightarrow\infty}| \, \partial_{|\vp|}e_{(n,\infty)} \, | \; = \; 0\;.
    \label{eq:en-limn-zero-1}
\eeqn
To see this, we note that $\widetilde H_{(n)}\widetilde\Psi_{(n,\infty)}=0$ is equivalent to
\eqn
         \big( \widetilde T_{(n)} \, + \, \widetilde W_{(n)} \big) \, \widetilde \Psi_{(n,\infty)}
         \; = \; - \, e_{(n,\infty)} \, \chi_1^2[H_f] \, \widetilde \Psi_{(n,\infty)} \; ,
\eeqn
which follows from the definition of $\widetilde H_{(n)}$.
Therefore,
\eqn
        \partial_{|\vp|}e_{(n,\infty)} \; = \;- \,
        \frac{\bra\Psi_{(n,\infty)}\,,\,(\partial_{|\vp|}( \widetilde T_{(n)} \, + \,
        \widetilde W_{(n)} ) ) \, \Psi_{(n,\infty)}\ket}
        {\bra \, \Psi_{(n,\infty)}\,,\,\chi_1^2[H_f] \, \Psi_{(n,\infty)}\ket} \;.
\eeqn
Using
\eqn
        s-\lim_{n\rightarrow\infty}\widetilde\Psi_{(n,\infty)} \; = \; \vacpsi
\eeqn
and
\eqn
        \lefteqn{
        \lim_{n\rightarrow\infty}\bra \, \Psi_{(n,\infty)} \, , \,
        \big( \partial_{|\vp|}( \widetilde T_{(n)} \, + \, \widetilde W_{(n)} ) \big)
        \, \Psi_{(n,\infty)} \, \ket
        }
        \nonumber\\
        &=&\lim_{n\rightarrow\infty}\bra \, \vacpsi \, , \,
        \big( \partial_{|\vp|}( \widetilde T_{(n)} \, + \, \widetilde W_{(n)} ) \big)
        \, \vacpsi \, \ket
        \nonumber\\
        &=&\lim_{n\rightarrow\infty}\bra\vacpsi \, , \,
        ( \partial_{|\vp|}\widetilde W_{(n)} ) \, \vacpsi \, \ket
        \; < \; \lim_{n\rightarrow\infty} \| \partial_{|\vp|} \widetilde W_{(n)} \|_{op}
        \nonumber\\
        &\leq& \lim_{n\rightarrow\infty}\| \h^{(n)}_{\geq1} \|_{\xi}
        \; \leq \;
        \lim_{n\rightarrow\infty}\eta_n \; = \; 0 \;,
\eeqn
for every fixed value of the infrared cutoff, $\sigzero>0$.
Therefore,
\eqn
        \lefteqn{
        \partial_{|\vp|} \widetilde \pfp_{(n)}
        }
        \nonumber\\
        &=&\Big[- \, 1
        \, + \, 2\frac{\bra \partial_{|\vp|}\widetilde Q_{(-1,n)}\vacpsi\,,\,
        \big( \Hpszero-\Egrdzero \big) \partial_{|\vp|}\widetilde Q_{(-1,n)}\vacpsi\ket}
        {\bra \widetilde Q_{(-1,n)}\vacpsi\,,\,\widetilde Q_{(-1,n)}\vacpsi\ket }\Big]
        (1-\widetilde\err_n^{(1)})
        \nonumber\\
        && \hspace{4cm}\, + \, \widetilde\err_n^{(3)}
        \,+\, (\partial_{\z_n}\pfp_{(n)}\big)\Big|_{\z_n=e_{(n,\infty)}}
        \partial_{|\vp|}e_{(n,\infty)} \;,
\eeqn
as follows from Proposition {\ref{prp:derp-pfp-bd-1}}.

Hence, in combination with (\ref{eq:en-limn-zero-1}), we find
\eqn
        \lim_{n\rightarrow\infty}  \partial_{|\vp|} \widetilde \pfp_{(n)}
        &=&-1
        +2\frac{\bra \partial_{|\vp|}\Omgrdzeropsi\,,\,
        (\Hpszero-\Egrdzero) \partial_{|\vp|}\Omgrdzeropsi\ket}
        {\bra \Omgrdzeropsi \, , \, \Omgrdzeropsi\ket}
        \nonumber\\
        &=&-\partial_{|\vp|}^2\Egrdzero
\eeqn
by (\ref{eq:Feyn-Hell-2}),  since
\eqn
        | \, \widetilde\err_n^{(3)} \, | \; < \; c \, \eta_n^2 \; \searrow \; 0
        \;\;\;\;\; (n\rightarrow\infty) \;.
\eeqn
We thus obtain
\eqn
        1 \, - \, c_0 \, \gs \; < \; \partial_{|\vp|}^2 \, \Egrdzero \; < \; 1 \;,
\eeqn
uniformly in $\sigzero$.

Thus, we obtain for  the renormalized electron mass
\eqn
        m_{ren}(\vp,\sigzero) \; = \; \frac{1}{\partial_{|\vp|}^2\Egrdzero }
        \label{eq:mren-def-3}
\eeqn
that
\eqn
        1 \; < \; m_{ren}(\vp,\sigzero)   \; < \; 1 \, + \, c_0 \, \gs \;,
        \label{eq:mren-unifbd-1}
\eeqn
uniformly in $\sigzero\geq0$.
\endprf

This completes the proof of {\em\underline{A.} - \underline{C.}} in Theorem {\ref{mainthm-1}}.

\section{Existence of the renormalized mass}
\label{sec:mren-limits-1}

In this section, we prove the existence of the renormalized mass in the limit  
in which the infrared regularization is removed, in the form
as stated in part {\em\underline{D.}} of Theorem {\ref{mainthm-1}}.
This will conclude the analysis of this paper. 

\subsection{The limit $\sig\searrow0$ for fixed $\vp$}
\label{ssec:mren-lim-sig-1}

Because of the bounds (\ref{eq:mren-unifbd-1}) which are uniform in $\sigzero$ and $\vp$,
we find that for every $\vp$ with $0\leq|\vp|<\puppbd$, there exists 
a sequence $\{\sigma_n\}$ with $\sigma_n \searrow0$ as $n\rightarrow\infty$ such that
\eqn
	m_{ren}(\vp) \; := \; \lim_{n\rightarrow\infty}m_{ren}(\vp,\sigma_{n})
	\label{eq:mren-p-siglim-1}
\eeqn
exists.

\subsection{The joint limit $(\vp,\sig)\rightarrow(\vnull,0)$ for $\sigma>0$}
\label{ssec:mren-lim-sig-2}

It is established in \cite{bcfs2}, that the limit 
\eqn
	\widetilde m_{ren}(\vnull) \; := \; \lim_{\sigma\rightarrow0}m_{ren}(\vnull,\sigma)
\eeqn
exists, and a convergent algorithm is constructed to compute it to any
arbitrary given level of precision. 
In this subsection, we shall prove that the joint limit $(\vp,\sigma)\rightarrow(\vnull,0)$
of $m_{ren}(\vp,\sigma)$
exists on 
$$
	D_N \; := \; \{ \, (\vp,\sigma) \, | \, 0\leq|\vp|<\puppbd \, , \, \sigma\geq|\vp|^N \, \} \, ,
$$
for $\frac{\vp}{|\vp|}$ fixed, and $1\ll N<\infty$ arbitrarily large, but finite.
That is,
\eqn
	\widetilde m_{ren}(\vnull) \; = \; \lim_{D_N\ni(\vp,\sigma)\rightarrow(\vnull,0)}m_{ren}(\vp,\sigma) \,.
\eeqn 
To this end, we observe that 
\eqn
	E'''(\vp,\sigma)  \; = \;  6 \, \bra \,  \Psi_{\vu}'(\vp,\sigma) \, , \,
	(\,  H'(\vp,\sigma)-E'(\vp,\sigma) \, ) 
	\, \Psi_{\vu}'(\vp,\sigma) \, \ket \,,
\eeqn
given that $\| \Omgrd\|=1$, and where $(\,\cdot\,)'$ is shorthand for $\partial_{|\vp|}(\,\cdot\,)$.
Using Lemma {\ref{lm:derpPsi-est-1}} below, combined with the Schwarz inequality and $|E'(\vp,\sigma)|<c$,
one gets
\eqn
	| \, E'''(\vp,\sigma) \, | \; \leq \; c \, \log\frac1\sigma \,,
\eeqn
for a constant $c$ independent of $\vp$.
Accordingly, we find
\eqn
	| \, E''(\vp,\sigma) \, - \,   E''(\vnull,\sigma) \, |
	\; \leq \; c \, |\vp| \, \log\frac1\sigma \;,
\eeqn
and consequently, for all 
$$
	(\vp,\sigma) \; \in \; D_{N,\delta} \; := \; \{ \, (\vp,\sigma) \, \in \, 
	D_N \, | \, |(\vp,\sigma)| \, = \, \sqrt{|\vp|^2+\sigma^2} \, < \, \delta \, \} \, ,
$$
we have
\eqn
	| \, E''(\vp,\sigma) \, - \, E''(\vnull,\sigma) \, |
	& \leq & \sup_{(\vp,\sigma)\in D_{N,\delta}}  c  \, |\vp| \, \log\frac{1}{\sigma} 
	\nonumber\\
	& < &  C(N) \, \delta^{1-\eta} \; = \; o_\delta(1)\;,
\eeqn
for an arbitrary $\eta>0$.
Consequently,
\eqn
	\lim_{D_N\ni(\vp,\sigma)\rightarrow(\vnull,0)} E''(\vp,\sigma) 
	\; = \; \lim_{\sigma\rightarrow0} E''(\vnull,\sigma) \;.
\eeqn
But from the definition of the renormalized mass (\ref{eq:mren-def-3}), this is equivalent to
\eqn
	\lim_{D_N\ni(\vp,\sigma)\rightarrow(\vnull,0)}m_{ren}(\vp,\sigma)
	\; = \; \lim_{ \sigma \rightarrow 0}m_{ren}(\vnull,\sigma)
	\; = \; \widetilde m_{ren}(\vnull) \, ,
\eeqn
which is what we wanted to show.
 
\begin{lemma} 
\label{lm:derpPsi-est-1}
Assume that $\| \, \Psi_{\vu}(\vp,\sigma)  \, \|=1$. Then,
\eqn
	\| \,  H'(\vp,\sigma)  \, \Psi_{\vu}'(\vp,\sigma)  \, \|^2  \; , \;
	\| \, \Psi_{\vu}'(\vp,\sigma)  \, \|^2 \; < \; c \, \log \frac1\sigma \, .
\eeqn 
\end{lemma}

\begin{proof}
The estimate
\eqn 
	\| \, \Psi_{\vu}'(\vp,\sigma)  \, \|^2 \; < \; c \, \log \frac1\sigma \, 
\eeqn
follows from similar considerations as those explained in the proof of 
Proposition {\ref{prp:derp-pfp-bd-1}}. We shall not repeat the detailed argument here.

To prove the second asserted estimate, we recall that
\eqn 
	H(\vp,\sigma) \; = \; \frac12 \, ( \, H'(\vp,\sigma) \, )^2 \, + \, \g \, \vec\pauli \cdot \Bf \, + \, H_f \,.
\eeqn
Since 
\eqn
	| \, \Bf \, | \; \leq \; c \, \sqrt{1+H_f} \; \leq \; c \, ( \, 1 \, + \, H_f \, ) \, ,
\eeqn
it is clear that 
\eqn
	H_f \, + \, \g \, \vec\pauli \cdot \Bf \; \geq \; - \, c \, \g \, + \, ( \, 1 \, - \, c' \, \g \, ) \, H_f \, ,
\eeqn
where $1-c'\g>0$.
Therefore, we have
\eqn
	\| \,  H'(\vp,\sigma)  \, \Psi_{\vu}'(\vp,\sigma)  \, \|^2 
	& \leq & \bra \, \Psi_{\vu}'(\vp,\sigma)  \, , \, ( \, H(\vp,\sigma) - E(\vp,\sigma) \, ) \,
	\Psi_{\vu}'(\vp,\sigma)  \, \ket 
	\nonumber\\
	& & \hspace{2cm} + \, ( \, | \, E(\vp,\sigma) \, | \, + \, c \, \g \, )  \, \| \, \Psi_{\vu}'(\vp,\sigma) \, \|^2
	\nonumber\\
	& \leq & | \, m_{ren}(\vp,\sigma) \, - \, 1 \, | \, + \, c \, \| \, \Psi_{\vu}'(\vp,\sigma) \, \|^2
	\nonumber\\
	& \leq & c \, \gs \, + \, c' \, \log\frac1\sigma \, ,
\eeqn
as claimed.
\end{proof}

\subsection{The limit $|\vp|\rightarrow 0$ for $\sigma=0$}
\label{ssec:mren-lim-sig-3}

Finally, we prove that for $\sigma=0$, the limit $|\vp|\rightarrow 0$ agrees
with $\widetilde m_{ren}(\vnull)$, i.e., the order of taking the limits $|\vp|\rightarrow0$
and $\sigma\rightarrow0$
can be reversed.

To this end, let $\{\vp_j\}_{j\in\N}$ denote any sequence converging to $\vnull$
along a fixed direction $\frac{\vp_j}{|\vp_j|}$. For every $\vp_j$, 
let $\{\sigma_n(\vp_j)\}$ denote the sequence in (\ref{eq:mren-p-siglim-1}) 
corresponding to $\vp=\vp_j$.
From each such sequence, we may extract an element $\sigma_{n_j}(\vp_j)$  
such that the sequence $\{\sigma_{n_j}(\vp_j)\}_{j\in\N}$ converges to 0 as $j\rightarrow0$,
in such a way that $0<c_1<\frac{|\vp_j|}{\sigma_{n_j}(\vp_j)}<c_2$ holds for all $j$.  

Accordingly, we find that
\eqn
	| \, m_{ren}(\vp_j) \, - \, \widetilde m_{ren}(\vnull) \, | &\leq&
	| \, m_{ren}(\vp_j) \, - \, m_{ren}(\vp_j,\sigma_{n_j}(\vp_j)) \, |
	\\
	&&\hspace{2cm} \, + \, 
	| \, m_{ren}(\vp_j,\sigma_{n_j}) \, - \, \widetilde m_{ren}(\vnull)\, |
	\nonumber
\eeqn
Taking $j\rightarrow\infty$, it follows from the discussion in Section
{\ref{ssec:mren-lim-sig-1}} that the first term on the right hand side
converges to zero, and from the discussion in Section {\ref{ssec:mren-lim-sig-2}}  
that the second term also converges to zero. This implies
\eqn
	\lim_{j\rightarrow\infty} m_{ren}(\vp_j) \; = \; \widetilde m_{ren}(\vnull) \, ,
\eeqn
for {\em any} sequence $\{\vp_j\}$ converging to $\vnull$, with $\frac{\vp_j}{|\vp_j|}$ fixed.

This concludes the proof of part {\em\underline{D.}} of Theorem {\ref{mainthm-1}}.
\qedprf







\subsection*{Acknowledgements}
The contents of this paper are further developments of the work \cite{ch1} (unpublished).
All results and methods of \cite{ch1} have here been fundamentally improved, optimized,
and further extended.
I am profoundly grateful to J\"urg Fr\"ohlich, my advisor in \cite{ch1}, for his support,
advice, generosity, encouragement, friendship, and for everything he has taught me.
I also express my deep gratitude to Volker Bach and Israel Michael Sigal
for their generosity, support, and friendship.
This paper has greatly benefitted
from our collaborations in \cite{bcfs1,bcfs2},
and from what I learned through very helpful discussions with all of them.
I heartily thank Michael Aizenman and Elliott H. Lieb
for enjoyable and inspiring discussions, and for their generosity, interest, and kindness.
In particular, I am deeply grateful to my wife Isabelle for her patience,
tolerance, support, and endurance.
This work was supported by NSF grant DMS-0524909.

\parskip = 0 pt
\parindent = 0 pt

\end{document}